%% file: surface_ALE.tex
\documentclass[11pt]{article}
\usepackage{verbatim,amsmath,amssymb}
\usepackage{epsfig,float,color}
\usepackage[font=small,labelfont=bf]{caption}
\usepackage{geometry}
\usepackage{setspace}
\usepackage{natbib}
\usepackage{amsthm,mathrsfs,amsfonts,dsfont} 
\usepackage{hyperref}

\definecolor{darkblue}{rgb}{0,0,1}
\definecolor{col1}{rgb}{1,0,1}
\definecolor{col2}{rgb}{0,0.7,0}
\definecolor{col3}{rgb}{0.0,0.5,1}
\hypersetup{pdftex=true, colorlinks=true, breaklinks=true, linkcolor=darkblue, menucolor=darkblue, pagecolor=darkblue, citecolor=darkblue, urlcolor=darkblue}

\newtheoremstyle{rem}%name
{6pt}%Space above
{6pt}%Space below
{\small}%Body font
{}%Indent amount
{\bf}%Theorem head font
{:}%Punctuation after theorem head
{.5em}%Space after theorem head
{}%Theorem head spec(can be left empty, meaning ‘normal’)

\theoremstyle{rem}
\newtheorem{remark}{Remark}[section]

\newfloat{boxtable}{luc}{Box}
\newfloat{Box}{luc}{Box}

\input{neco.tex}
\begin{document}

\begin{center}

\Large{\bf{A curvilinear surface ALE formulation for self-evolving Navier- Stokes manifolds -- General theory and analytical solutions}} \\[4mm]

\end{center}

\renewcommand{\thefootnote}{\fnsymbol{footnote}}

\begin{center}
\large{Roger A. Sauer$^{\mra,\mrb,\mrc,}$\footnote[1]{corresponding author, email: roger.sauer@rub.de}}
\vspace{3mm}

\small{\textit{
$^\mra$Institute for Structural Mechanics, Ruhr University Bochum, 44801 Bochum, Germany \\[1mm]
$^\mrb$Department of Structural Mechanics, Gda\'{n}sk University of Technology, 80-233 Gda\'{n}sk, Poland \\[1mm]
$^\mrc$Mechanical Engineering, Indian Institute of Technology Guwahati, Assam 781039, India}}

\vspace{3mm}

\small{Published\footnote[2]{This pdf is the personal version of an article whose journal version is available at \href{https://doi.org/10.1017/jfm.2025.10327}{www.cambridge.org}} 
in \textit{J.~Fluid Mech.}, \href{https://doi.org/10.1017/jfm.2025.10327}{doi:10.1017/jfm.2025.10327} \\
Submitted on 28 January 2025; Revised on 31 May 2025; Accepted on 1 June 2025} 

\end{center}

\renewcommand{\thefootnote}{\arabic{footnote}}

\vspace{-4mm}

%\doublespacing

\rule{\linewidth}{.15mm}
{\bf Abstract:}
A new arbitrary Lagrangian-Eulerian (ALE) formulation for Navier-Stokes flow on self-evolving surfaces is presented. 
It is based on a general curvilinear surface parameterization that describes the motion of the ALE frame. 
Its in-plane part becomes fully arbitrary, while its out-of-plane part follows the material motion of the surface. 
This allows for the description of flows on deforming surfaces using only surface meshes. 
The unknown fields are the fluid density or pressure, the fluid velocity and the surface motion, where the latter two share the same normal velocity. 
The corresponding field equations are the continuity equation or area-incompressibility constraint, the surface Navier-Stokes equations, and suitable surface mesh equations.
Particularly advantageous are mesh equations based on membrane elasticity.
The presentation focuses on the coupled set of strong and weak form equations, and presents several manufactured steady and transient solutions.
These solutions are used together with numerical simulations to illustrate and discuss the properties of the proposed new ALE formulation.
They also serve as basis for the development and verification of corresponding computational methods.
The new formulation allows for a detailed study of fluidic membranes such as soap films, capillary menisci and lipid bilayers. 


\vspace{-5mm}
\rule{\linewidth}{.15mm}

\vspace{-1mm}

\section{Introduction}\label{s:intro}

The understanding of fluid flow on evolving surfaces is important in many phenomena such as liquid films, bubbles, foams and lipid bilayers.
Due to surface evolution, the flow has three spatial velocity components -- two in-plane and one out-of-plane.
The latter leads to shape changes, that are generally unknown.
While surface changes are more commonly described in a Lagrangian frame, fluid flows are more commonly described in an Eulerian frame, as this facilitates numerical descriptions, which are the motivation behind this work.
Evolving surface flows thus benefit from a combined Lagrangian-Eulerian description.
In classical computational fluid dynamics such a combination is often used in the framework of arbitrary Lagrangian-Eulerian (ALE) descriptions.
ALE formulations are often presented and understood as a numerical method, but they are much more than that: 
They are a general and flexible way to parameterize and describe evolving domains and their partial differential equations (PDEs).
What is often called the \textit{mesh} velocity, is actually the velocity of the ALE frame of reference.

The adaption of ALE formulations to evolving surfaces has only recently appeared, and is still not fully general.
The generality aimed at here, is one that is suitable for advanced computational methods, for example in the context of geometrically accurate surface finite element methods. 
We therefore focus here on the development of a general theory, its weak form and corresponding analytical solutions, as these are all required for the construction and verification of a computational formulation, which is then studied in %future work.
\citet{ALEcomp}.

ALE formulations for classical solid and fluid mechanics have a long history, cf.~\citet{donea04} and references therein.
For surface flows, the seminal work of \citet{scriven60} already introduces the distinction between Lagrangian and Eulerian surface descriptions -- also denoted as convected and surface-fixed coordinates, respectively.
A recent discussion on the two descriptions can be found in \citet{steigmann18}.
Many existing simulation methods for evolving surfaces use Lagrangian descriptions that can lead to large mesh distortion and require mesh stabilization or remeshing strategies, e.g.~see \citet{brakke92,ma08,elliott13,mikula14,droplet,liquidshell,dharmavaram21}.

The Eulerian and ALE settings can avoid large mesh distortions, but they have been applied to surface flows only in recent years.
\citet{elliott12} propose a surface ALE scheme for the surface advection-diffusion equation, where the material and mesh velocity are prescribed.
\citet{rahimi12} implement a surface ALE scheme for lipid bilayer flow with interlayer sliding in an axisymmetric setting.
The mesh velocity is penalized resulting in a near-Eulerian description in tangential direction.
\citet{torres19} develop a general 3D ALE formulation for cellular membranes. 
In their applications they advance the surface by a normal offset that is based on \citet{rangarajan15}.
This special ALE motion leads to a near-Eulerian parameterization that still requires remeshing in certain applications. 
\citet{ALE} develop a general curvilinear ALE framework for fluid flow on evolving surfaces and use it to study the stability of fluid films based on an in-plane Eulerian mesh description.
In-plane Eulerian mesh descriptions have also been used in the recent works of \citet{reuther20} and \citet{alizzi23}. 

These ALE works for surface Navier-Stokes, even though formulated generally, restrict their application to in-plane Eulerian or near-Eulerian surface parameterizations. 
Thus, a truly general surface ALE formulation for surface Navier-Stokes flow has not been fully investigated yet.
Especially not for in-plane mesh motion that is unconditionally stable.
The curvature-dependent mesh redistribution scheme of \citet{barrett08}, recently applied to surface flows \citep{krause23}, can be expected to become unstable under mesh refinement, as the surface curvature is invariant with respect to in-plane mesh motion.
The recent ALE mesh motion approach proposed by \citet{sahu24}, while successfully capturing tether formation and translation,
is based on viscosity and hence can be expected to lose stability for decaying velocities.
Another restriction of the ALE theory of \citet{ALE} and \citet{sahu24} is that its mesh motion is not fully general and
can become unsuitable at inflow boundaries. 

These limitations of existing theories motivate the development of a more general ALE formulation.
This is an important and timely topic, as the study of surface flows on known (fixed or prescribed) surfaces has received a lot of recent attention, using either a vorticity-stream function formulation (e.g.~\citet{nitschke12}), or a velocity-pressure formulation (e.g.~\citet{rangamani13}).
A detailed survey of existing computational approaches for surface flows %
will be conducted in future work. %
Here we focus on theoretical developments and solutions for Navier-Stokes flows on surfaces.
Three cases cases be distinguished: 
Flow on stationary surfaces, flow on evolving surfaces with prescribed surface motion, and flow on evolving surfaces with unknown surface motion. 
The third case can be further subdivided into surfaces advected by surrounding media, as in the case of free boundaries \citep{walkley05} and interfaces between phases \citep{bothe10}, and self-evolving surfaces, where the unknown surface motion is driven by the overall surface flow problem. 

The case of known surface motion has received much attention and there are many analytical (usually manufactured) flow solutions available for various surfaces, such as spheres \citep{nitschke12,reuther15,olshanskii18}, cylinders \citep{lederer20,suchde20}, wavy tubes \citep{fries18}, ellipsoids \citep{gross18}, toroids \citep{busuioc20} and other shapes \citep{gross20}.

For flows on fixed surfaces the unknown fluid velocity is characterized by two tangential velocity components that depend on the surface geometry. 
In the evolving surface case, the velocity has three unknown components that can be chosen based on a background Cartesian coordinate system instead of using tangential and normal surface components.
Even though this later description greatly facilitates the description of the governing equations and their discretization, many works use the tangential/normal decomposition for the flow equations.
This is probably due to the fact that evolving surface flows are regarded as an extension of fixed surface flows, following the notion that the in-plane surface flow equations are coupled to the out-of-plane moving surface equation, e.g.~see \citet{koba17,jankuhn18,miura18,reusken20,olshanskii22} for prescribed surface motions and \citet{reuther20} for self-evolving surfaces.
Instead of following the viewpoint of coupled component equa-tions, the entire system can be described and solved directly by a single 3D vector-valued~PDE, which is simply the equation of motion following from surface momentum balance, as was already noted by \citet{scriven60}. 
Thus the surface Navier-Stokes system can be expressed more compactly if viewed as the unity that it really is.
Here it is interesting to note that there are different versions of the surface Navier-Stokes equations in the literature \citep{brandner22}.

Self-evolving solid surfaces have been studied for a long time in the context of general membrane and shell formulations -- beginning with the works of \citet{oden67} and \citet{naghdi72}.
They are best described in a convected, Lagrangian parameterization, which is much more straightforward than an Eulerian parameterization.
Self-evolving fluid surfaces have therefore attracted much less and only recent attention.
Initial formulations have instead neglected tangential material flow and focussed on the shape change in a Lagrangian framework.
Computational examples have thus studied droplet contact \citep{brown80b,droplet}, 
lipid bilayer evolution \citep{feng06,arroyo09},
Willmore flow \citep{barrett08a,dziuk08},
phase evolution on surfaces \citep{elliott13,phaseshell},
vesicles immersed in a flow \citep{barrett15}, and
particles floating on liquid membranes \citep{dharmavaram22}.
As was mentioned already, in these works the Lagrangian description is often combined with mesh stabilization or remeshing strategies, to remedy large mesh distortion.
Eulerian descriptions for flow on evolving surfaces, on the other hand, are much rarer, and they have only been used in the already mentioned ALE works.
But those lack key aspects, which are addressed here.

The focus, here, is placed on area-incompressible surface flows, using a two-field, velocity-pressure description.
The ALE frame deformation and its velocity -- generally unknown -- add another two fields to the problem, such that the resulting formulation is a four-field problem.
The surface can be endowed with bending elasticity, as will be shown, but this is not a main focus of this work.
Further restrictions are
to consider mass-conserving surface flows, where there is no mass exchange with surrounding media, 
and to consider no thickness change of the surface, implying that there is no flow in thickness direction and that area-incompressibility follows from volume-incompressibility.
Otherwise the surface deformation would need to be decomposed into elastic and inelastic contributions \citep{FeFi}.
Despite these restrictions, the present formulation is still very general, and is written such that it allows for future extensions.
It contains free films as well as bounding interfaces, it applies to fixed as well as self-evolving surfaces, and it applies to closed surfaces, as well as open ones containing evolving inflow boundaries.
It also applies to arbitrary surface topologies. 

In summary, this work contains several important theoretical novelties: \\[-8mm]
\begin{itemize}
\item It presents and discusses a general surface ALE formulation in curvilinear coordinates, \\[-7mm]
\item that allows for truly arbitrary in-plane mesh motion, \\[-7mm]
\item including mesh motion defined from membrane elasticity. \\[-7mm]
\item It is used to formulate area-(in)compressible Navier-Stokes flow on self-evolving manifolds, \\[-7mm]
\item in vector form, without decomposition into in-plane and out-of-plane equations, \\[-7mm]
\item and solves this for several analytical and numerical benchmark examples, including non-laminar surface flows and expanding soap bubbles with evolving inflow boundaries.  \\[-7mm]
\item They demonstrate the capabilities and advantages of the proposed ALE formulation. \\[-7mm]
\end{itemize}

The remainder of this paper is organized as follows.
The curvilinear ALE frame is introduced in Sec.~\ref{s:surf} and used to describe surface deformation and flow.  
Secs.~\ref{s:bal} and \ref{s:consti} then proceed with formulating the field equations and constitutive equations in the ALE frame.
Their weak form is then provided and discussed in Sec.~\ref{s:wf}. 
Sec.~\ref{s:anex} provides several analytical and numerical solutions and uses them to study the proposed formulation.
Conclusions are drawn in Sec.~\ref{s:concl}.

\section{Surface description in the ALE frame}\label{s:surf}

This section presents the ALE formulation for evolving surfaces and their kinematical description.
The formulation follows \citet{ALE} but with a generalization of the possible frame motion, and with many additions, especially in Secs.~\ref{s:gradv}-\ref{s:J}.
The formulation is expressed in very general terms, such that it can capture arbitrarily large deformations and motions, including arbitrary rigid body translations and rotations.
The latter is confirmed by an example in Sec.~\ref{s:spin}.
Even though the formulation is based on a surface parameterization, it is emphasized that all quantities without free indices are independent of the ALE frame and hence observer-invariant quantities on the evolving surface.
This includes time derivatives.

\subsection{Different surface parameterizations}

The surface under consideration is denoted $\sS$ and its surface points are described by the parametrization
\eqb{l}
\bx = \bx(\zeta^\alpha,t)\,.
\label{e:bx}\eqe
Here, $\zeta^\alpha$ denotes a set of arbitrary surface coordinates ($\alpha = 1,\,2$) that is associated with the ALE frame of reference.
In a computational description, this frame of reference can be taken as the computational grid or mesh.
It can also be associated with an observer motion \citep{nitschke22}.

\begin{figure}[h]
\begin{center} \unitlength1cm
\begin{picture}(0,7.3)
\put(-6.2,-1.7){\includegraphics[width=123mm]{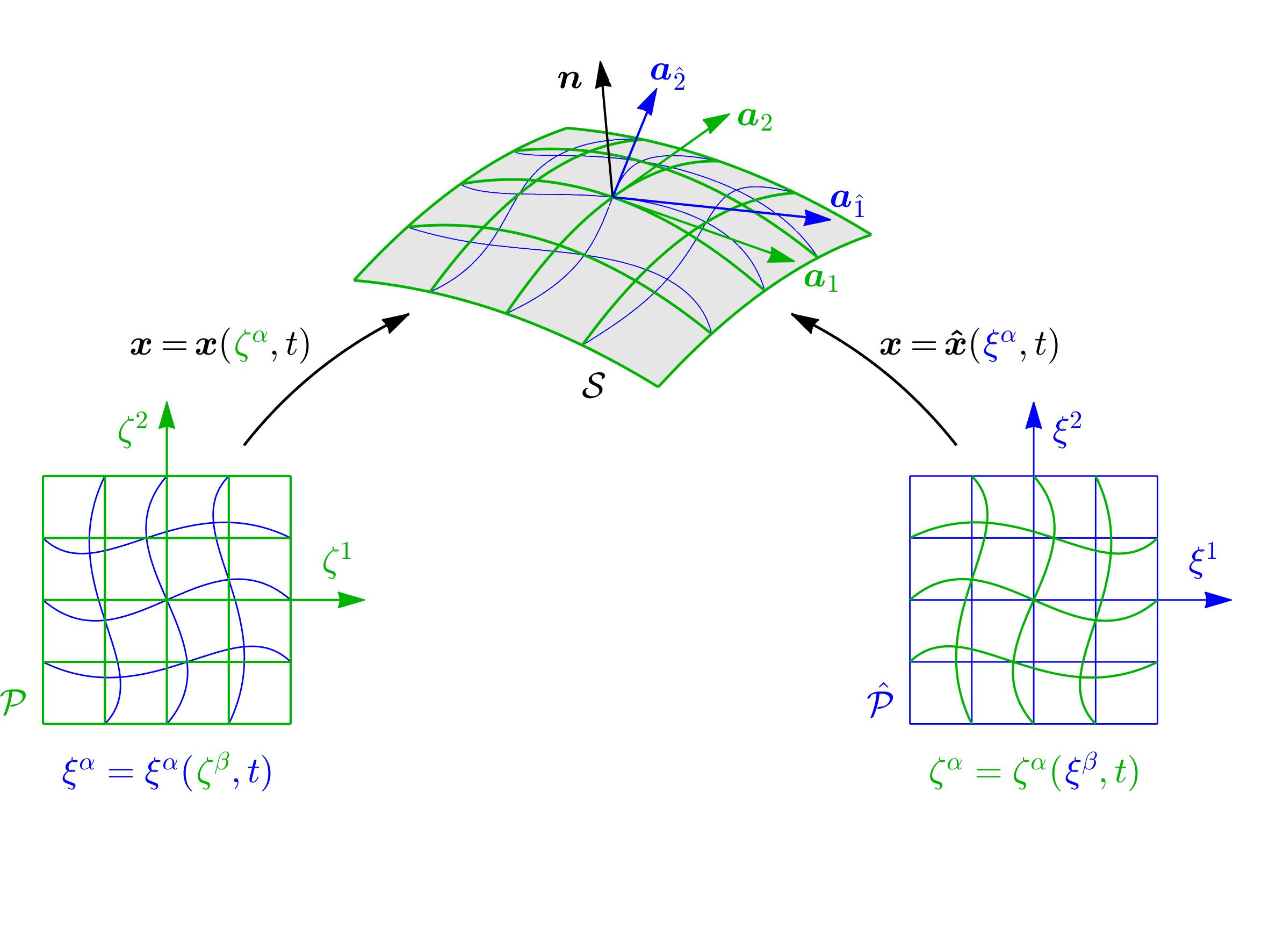}}
\end{picture}
\caption{ALE surface parameterization: 
The surface $\sS$ can be described by the ALE mapping $\bx=\bx(\zeta^\alpha,t)$ from the ALE parameter domain $\sP$, and the material mapping $\bx=\hat\bx(\xi^\alpha,t)$ from the material parameter domain $\hat\sP$.
From these mappings follow the tangent vectors $\ba_\alpha := \bx_{\!,\alpha}$ and $\ba_{\hat\alpha} := \hat\bx_{\!,\hat\alpha}$.}
\label{f:ALE}
\end{center}
\end{figure}

Coordinate $\zeta^\alpha$ is generally different from the convective coordinate $\xi^\alpha$, which is used to track material points and which defines the material time derivative
\eqb{l}
\dot{(...)} = \ds\pa{...}{t}\Big|_{\xi^\alpha}\,.
\label{e:dot}\eqe
It is also different from the surface-fixed coordinate $\theta^\alpha$, whose material time derivative defines the tangential fluid velocity, i.e.~$v^\alpha = \dot\theta^\alpha$.
In other words,
\eqb{l}
\bx=\hat\bx(\xi^\alpha,t)\,, 
\label{e:bx_xi}\eqe
is a Lagrangian surface description (that follows the material motion), while
\eqb{l}
\bx = \tilde\bx(\theta^\alpha,t)\,,
\label{e:bx_th}\eqe
is an in-plane Eulerian surface description.\footnote{This notation slightly differs from that of \citet{ALE}, which places no tilde on $\bx$ in \eqref{e:bx_th}, but places a check on $\bx$ in \eqref{e:bx}. This then implies a check will appear on indices that can be written without accent here.}
Eq.~\eqref{e:bx} on the other hand is an arbitrary Lagrangian-Eulerian surface description that contains the two special cases $\zeta^\alpha=\xi^\alpha$ and $\zeta^\alpha=\theta^\alpha$. 
The mappings \eqref{e:bx} and \eqref{e:bx_xi} are illustrated in Fig.~\ref{f:ALE}.
They induce a functional relationship between $\zeta^\alpha$ and $\xi^\alpha$, i.e.~$\zeta^\alpha = \zeta^\alpha(\xi^\beta,t)$ and $\xi^\alpha = \xi^\alpha(\zeta^\beta,t)$.
%-----------------------------------------------------------------
% run ALEplot2
%-----------------------------------------------------------------

Parameterizations \eqref{e:bx}, \eqref{e:bx_xi} and \eqref{e:bx_th} define the three parametric derivatives
\eqb{l}
..._{,\alpha} := \ds\pa{...}{\zeta^\alpha}\,,\quad
..._{,\hat\alpha} := \ds\pa{...}{\xi^\alpha}\,,\quad
..._{,\tilde\alpha} := \ds\pa{...}{\theta^\alpha}\,.
\label{e:comma}\eqe
Applied to $\bx$, this defines the tangent vectors
\eqb{l}
\ba_\alpha := \bx_{\!,\alpha}\,,\quad
\ba_{\hat\alpha} := \hat\bx_{\!,\hat\alpha}\,,\quad
\ba_{\tilde\alpha} := \tilde\bx_{\!,\tilde\alpha}\,.
\label{e:ba}\eqe
Together with the normal vector
\eqb{l}
\bn := \ds\frac{\ba_1\times\ba_2}{\norm{\ba_1\times\ba_2}} = \ds\frac{\ba_{\hat1}\times\ba_{\hat2}}{\norm{\ba_{\hat1}\times\ba_{\hat2}}} 
= \ds\frac{\ba_{\tilde1}\times\ba_{\tilde2}}{\norm{\ba_{\tilde1}\times\ba_{\tilde2}}}\,,
\eqe
they can be used as a basis to decompose general vectors into tangential and normal components, i.e.
\eqb{l}
\bv = v^\alpha\,\ba_{\alpha} + v\,\bn = v^{\hat\alpha}\,\ba_{\hat\alpha} + v\,\bn = v^{\tilde\alpha}\,\ba_{\tilde\alpha} + v\,\bn\,.
\label{e:vsplit}\eqe
Tangent vectors $\ba_{\alpha}$, $\ba_{\hat\alpha}$ and $\ba_{\tilde\alpha}$ define the surface metrics
\eqb{l}
a_{\alpha\gamma} := \ba_{\alpha}\cdot\ba_{\gamma}\,,\quad
a_{\hat\alpha\hat\gamma} := \ba_{\hat\alpha}\cdot\ba_{\hat\gamma}\,,\quad
a_{\tilde\alpha\tilde\gamma} := \ba_{\tilde\alpha}\cdot\ba_{\tilde\gamma}\,,
\eqe
and the curvature tensor components
\eqb{l}
b_{\alpha\gamma} := \ba_{\alpha,\gamma}\cdot\bn\,,\quad
b_{\hat\alpha\hat\gamma} := \ba_{\hat\alpha,\hat\gamma}\cdot\bn\,,\quad
b_{\tilde\alpha\tilde\gamma} := \ba_{\tilde\alpha,\tilde\gamma}\cdot\bn\,.
\label{e:bab}\eqe
Through the inverse metrics
\eqb{l}
[a^{\alpha\gamma}] := [a_{\alpha\gamma}]^{-1} \,,\quad
[a^{\hat\alpha\hat\gamma}] := [a_{\hat\alpha\hat\gamma}]^{-1} \,,\quad
[a^{\tilde\alpha\tilde\gamma}] := [a_{\tilde\alpha\tilde\gamma}]^{-1}\,,
\eqe
the dual tangent vectors
\eqb{l}
\ba^\alpha := a^{\alpha\gamma}\ba_\gamma\,,\quad
\ba^{\hat\alpha} := a^{\hat\alpha\hat\gamma}\ba_{\hat\gamma}\,,\quad
\ba^{\tilde\alpha} := a^{\tilde\alpha\tilde\gamma}\ba_{\tilde\gamma}\,,
\eqe
are defined.\footnote{Following index notation, summation is implied on repeated indices within terms.}
They in turn define the Christoffel symbols
\eqb{l}
\Gamma^\mu_{\alpha\gamma} := \ba^\mu\cdot\ba_{\alpha,\gamma}\,,\quad
\Gamma^{\hat\mu}_{\hat\alpha\hat\gamma} := \ba^{\hat\mu}\cdot\ba_{\hat\alpha,\hat\gamma}\,,\quad
\Gamma^{\tilde\mu}_{\tilde\alpha\tilde\gamma} := \ba^{\tilde\mu}\cdot\ba_{\tilde\alpha,\tilde\gamma}\,.
\label{e:Chris}\eqe

Based on the preceding quantities, further quantities in surface differential geometry, e.g.~see \citet{cism}, 
can be formulated in each basis.
In particular, the surface gradient and surface divergence of a general scalar $\phi$ and vector $\bv$ become
\eqb{l}
\nabla_{\!\mrs\,}\phi = \phi_{,\alpha}\,\ba^\alpha= \phi_{,\hat\alpha}\,\ba^{\hat\alpha} = \phi_{,\tilde\alpha}\,\ba^{\tilde\alpha}\,,
\label{e:sgrads}\eqe
\eqb{l}
\nabla_{\!\mrs\,}\bv = \bv_{\!,\alpha}\otimes\ba^\alpha= \bv_{\!,\hat\alpha}\otimes\ba^{\hat\alpha} = \bv_{\!,\tilde\alpha}\otimes\ba^{\tilde\alpha}
\label{e:sgrad}\eqe
and
\eqb{l}
\divs\bv = \bv_{\!,\alpha}\cdot\ba^\alpha= \bv_{\!,\hat\alpha}\cdot\ba^{\hat\alpha} = \bv_{\!,\tilde\alpha}\cdot\ba^{\tilde\alpha}\,.
\label{e:sdiv}\eqe
The comma here denotes the parametric derivative according to Eq.~\eqref{e:comma}.
In case of $\bv$, it applies to the full vector and not just its components.

The description following in Sec.~\ref{s:bal} exclusively uses basis $\{\ba_1,\,\ba_2,\,\bn\}$ induced by \eqref{e:bx}. 
Basis $\{\hat\ba_1,\,\hat\ba_2,\,\bn\}$ induced by \eqref{e:bx_xi} and basis $\{\tilde\ba_1,\,\tilde\ba_2,\,\bn\}$ induced by \eqref{e:bx_th} are not used further, apart from some derivations.
But it is important to note that they exist and provide alternative descriptions.
Description \eqref{e:bx} is the most general one, that contains the other two as special cases.
Appendix \ref{s:trans} contains the transformation rules that can be used to adapt an expression from one parameterization to another. 
They can be used to show that tensor invariants, such as the curvature tensor invariants $H := a^{\alpha\beta}b_{\alpha\beta}/2$ and $\kappa := \det[b_{\alpha\beta}]/\det[a_{\alpha\beta}]$ are invariant w.r.t.~the choice of basis.
Essentially, all quantities without free index are invariant w.r.t.~the parametrization, which includes its change -- a property often denoted \textit{reparametrization invariance}, e.g.~see \citet{guven18}. 

An important aspect to note, is that the material time derivative \eqref{e:dot} only commutes with the parametric derivative w.r.t.~$\xi^\alpha$, but not with the other two parametric derivatives in \eqref{e:comma}.
This is needed in the following.

\subsection{The ALE expressions for various material time derivatives}

The material velocity is defined as the material time derivative of $\bx$, i.e.
\eqb{l}
\bv = \dot\bx = \ds\pa{\hat\bx}{t}\Big|_{\xi^\alpha}\,.
\label{e:matvel}\eqe
Changing to $\zeta^\alpha$, this expands into
\eqb{l}
\bv = \ds\pa{\bx}{t}\Big|_{\zeta^\alpha} + \pa{\bx}{\zeta^\alpha} \pa{\zeta^\alpha}{t}\Big|_{\xi^\beta}\,.
\label{e:bvz}\eqe
Introducing the velocity of the ALE frame,
\eqb{l}
\bv_\mrm : = \ds\pa{\bx}{t}\Big|_{\zeta^\alpha}\,,
\label{e:vm}\eqe
which in computational methods is also referred to as the mesh velocity, 
and using Eq.~(\ref{e:ba}.1), Eq.~\eqref{e:bvz} can be rewritten into
\eqb{l}
\bv = \bv_\mrm + \dot\zeta^\alpha\, \ba_\alpha\,.
\label{e:bv}\eqe
This expression admits the two special cases:
\begin{enumerate}
\item In-plane Lagrangian description: $\zeta^\alpha = \xi^\alpha$, for which $\dot\zeta^\alpha=0$ and thus $\bv_\mrm=\bv$, and \\[-7mm]
\item In-plane Eulerian description: $\zeta^\alpha = \theta^\alpha$, for which $\dot\zeta^\alpha=v^\alpha$ and thus $\bv_\mrm=v\,\bn$ due to \eqref{e:vsplit}.  \\[-7mm] 
\end{enumerate}
Due to these properties, $\xi^\alpha$ is also referred to as the convected coordinate, while $\theta^\alpha$ is referred to as the surface-fixed coordinate.

In general, $\dot\zeta^\alpha$ characterizes the tangential velocity difference
\eqb{l} 
\dot\zeta^\alpha = \ba^\alpha\cdot(\bv-\bv_\mrm)\,,
\label{e:zetadot}\eqe
according to Eq.~\eqref{e:bv}.
This leads to the following interpretation of Eq.~\eqref{e:bv}: 
The (absolute) material velocity $\bv$ is composed of the (absolute) mesh velocity $\bv_\mrm$ plus the relative material velocity w.r.t.~the mesh, $\dot\zeta^\alpha\, \ba_\alpha$.
A graphical representation of this is shown in the example of Sec.~\ref{s:film}, see Fig.~\ref{f:pflow}.

Likewise to \eqref{e:matvel}, the material acceleration is defined as
\eqb{l}
\dot\bv = \ds\pa{\bv}{t}\Big|_{\xi^\alpha}\,.
\eqe
Changing to $\zeta^\alpha$, this expands into
\eqb{l}
\dot\bv = \ds\pa{\bv}{t}\Big|_{\zeta^\alpha} + \pa{\bv}{\zeta^\alpha} \pa{\zeta^\alpha}{t}\Big|_{\xi^\beta}\,.
\label{e:baz}\eqe
Introducing
\eqb{l}
(...)' := \ds\pa{...}{t}\Big|_{\zeta^\alpha}\,,
\label{e:bvp}\eqe
and using Eqs.~\eqref{e:zetadot} and \eqref{e:sgrad}, Eq.~\eqref{e:baz} can be rewritten into
\eqb{l}
\dot\bv = \bv' + \nabla_{\!\mrs\,}\bv\,\big(\bv-\bv_\mrm\big)\,,
\label{e:vALE}\eqe
which is the analogous surface version of the well-known 3D ALE equation \citep{donea}.
Using Eq.~\eqref{e:sgrad}, Eq.~\eqref{e:vALE} can also be written as
\eqb{l}
\dot\bv = \bv' + \bv_{\!,\alpha}\,\big(v^\alpha-v^\alpha_\mrm\big)\,,
\label{e:vALEa}\eqe
where $v^\alpha := \ba^\alpha\cdot\bv$ and $v^\alpha_\mrm := \ba^\alpha\cdot\bv_\mrm$.
It is emphasized that the temporal and spatial differentiation in $\bv_{\!,\alpha}$ is generally \underline{not} exchangeable, i.e.~$\bv_{\!,\alpha} \neq \dot\ba_\alpha$.\footnote{The identity $\bv_{,\alpha} = \dot\ba_\alpha$, used in \citet{rangamani13} and \citet{sahu17,ALE} for the Eulerian frame, is not general as it relies on the special case $\dot\zeta^\gamma_{,\alpha}=0$. 
This is only the case for particular ALE descriptions such as the Lagrangian description $\zeta^\gamma = \xi^\gamma$, or when $\dot\zeta^\gamma_{,\alpha}$ is negligibly small.} 
Instead
\eqb{l}
\bv_{\!,\alpha} = \dot\ba_\alpha + \dot\zeta^\gamma_{,\alpha}\,\ba_\gamma\,,
\label{e:badot}\eqe
see Appendix \ref{s:va}.
Here,
\eqb{l}
\dot\zeta^\gamma_{,\alpha} := \ds\pa{\dot\zeta^\gamma}{\zeta^\alpha}
= \ds\pa{}{\zeta^\alpha}\bigg(\pa{\zeta^\gamma}{t}\bigg)_{\!\!\xi^\beta}\,.
\label{e:ddzeta}\eqe
Also here the order of differentiation cannot be exchanged (i.e.~$\dot\zeta^\gamma_{,\alpha}$ is generally not equal to $(\zeta^\gamma_{,\alpha}\dot)=0$).
Eq.~\eqref{e:badot} admits the two special cases: 1.~$\zeta^\alpha=\xi^\alpha$, for which $\dot\zeta^\gamma_{,\alpha}=0$, and 2.~$\zeta^\alpha=\theta^\alpha$, for which $\dot\zeta^\gamma_{,\alpha} = v^\gamma_{,\alpha}$.
The former case implies that
\eqb{l}
\bv_{\!,\hat\alpha} = \dot\ba_{\hat\alpha}\,. 
\label{e:bahdot}\eqe

The expansion used in \eqref{e:bvz} and \eqref{e:baz} is a fundamental relation.
It generalizes to
\eqb{l} 
\boxed{\dot{(...)} = (...)' + ..._{,\alpha}\,\dot\zeta^\alpha}\,,
\label{e:ALE}\eqe
and is denoted the \textit{fundamental surface ALE equation} in the following.
As an example, the material time derivative of a scalar field $\phi$, such as the density $\rho$, then follows as
\eqb{l}
\dot\phi := \ds\pa{\phi}{t}\Big|_{\xi^\alpha}  = \phi' + \phi_{,\alpha}\,\dot\zeta^\alpha\,,
\eqe
which, in view of \eqref{e:zetadot}, leads to
\eqb{l}
\dot\phi = \phi' + \phi_{,\alpha}\,\big(v^\alpha-v_\mrm^\alpha\big)\,,
\label{e:rALEa}\eqe
or equivalently
\eqb{l}
\dot\phi = \phi' + \nabla_{\!\mrs\,}\phi\cdot\big(\bv-\bv_\mrm\big)\,.
\label{e:rALE}\eqe

\subsection{Velocity gradient}\label{s:gradv}

An important object for characterizing surface flows is the symmetric surface velocity gradient (also known as \textit{rate-of-deformation tensor})
$\bd := (\nabla_{\!\mrs\,}\bv + \nabla_{\!\mrs\,}\bv^\mathrm{T})/2$, 
which in view of \eqref{e:sgrad} and \eqref{e:badot} becomes
\eqb{l}
2\bd = \bv_{\!,\alpha} \otimes \ba^\alpha + \ba^\alpha \otimes \bv_{\!,\alpha}
\label{e:bd}\eqe
and
\eqb{l}
2\bd = \dot\ba_\alpha \otimes \ba^\alpha + \ba^\alpha \otimes \dot\ba_\alpha 
+ \dot\zeta^\gamma_{,\alpha}\big( \ba_\gamma \otimes \ba^\alpha + \ba^\alpha \otimes \ba_\gamma \big)\,.
\eqe
From this one finds the in-plane components
\eqb{l}
2d_{\alpha\beta} = \dot a_{\alpha\beta} + \dot\zeta^\gamma_{,\alpha}\,a_{\gamma\beta} + a_{\alpha\gamma}\,\dot\zeta^\gamma_{,\beta} 
\label{e:dab}\eqe
and
\eqb{l}
2d^{\alpha\beta} = -\dot a^{\alpha\beta} + \dot\zeta^\alpha_{,\gamma}\,a^{\gamma\beta} + a^{\alpha\gamma}\,\dot\zeta^\beta_{,\gamma}\,,
\eqe
that are connected through $\dot a^{\alpha\beta} = -a^{\alpha\gamma} \dot a_{\gamma\delta}\, a^{\delta\beta}$.
In general, $\bd$ also has out-of-plane components, but those are not relevant to the constitutive models considered in Sec.~\ref{s:consti}.
The in-plane components can be arranged in the in-plane tensor
\eqb{l}
\bd_\mrs := d_{\alpha\beta}\,\ba^\alpha\otimes\ba^\beta = d^{\alpha\beta}\ba_\alpha\otimes\ba_\beta\,.
\label{e:bds}\eqe

\subsection{Vorticity}

Another important object for characterizing flows is the vorticity. 
It derives from the curl of the velocity.
To characterize surface flows we thus introduce the \textit{surface curl} of the velocity,
\eqb{l}
\mathrm{curl}_\mrs\bv := \ba^\alpha\times\bv_{\!,\alpha}\,,
\label{e:scurl_bv}\eqe
analogous to the definitions of surface gradient and surface divergence in Eqs.~\eqref{e:sgrad} and \eqref{e:sdiv}.
The surface vorticity $\omega$ is then defined as the normal component of $\mathrm{curl}_\mrs\bv$, i.e.
\eqb{l}
\omega := \mathrm{curl}_\mrn\bv := (\ba^\alpha\times\bv_{\!,\alpha})\cdot\bn\,.
\label{e:vorticity}\eqe
It can be shown that
\eqb{l}
\mathrm{curl}_\mrn\bv = \divs(\bv\times\bn)
\eqe
and
\eqb{l}
\mathrm{curl}_\mrn\bv = (\bn\times\ba^\alpha)\cdot\bv_{\!,\alpha}\,.
\eqe
The latter identity motivates the definition of the surface curl of a scalar $\phi$ by
\eqb{l}
\mathrm{curl}_\mrs\phi := (\bn\times\ba^\alpha)\,\phi_{,\alpha} = \bn\times\nabla_{\!\mrs\,}\phi \,,
\label{e:scurl_phi}\eqe
which is a vector like $\mathrm{curl}_\mrs\bv$.
The surface curl of $\phi$ can also be written as
\eqb{l}
\mathrm{curl}_\mrs\phi = \phi_{,\alpha}\,\epsilon^{\alpha\beta}\,\ba_\beta\,,
\label{e:scurl_phi2}\eqe
where $[\epsilon^{\alpha\beta}] = [0~1;-1~0]/\sqrt{\det[a_{\gamma\delta}]}$ is the scaled permutation (or Levi-Civita) symbol.

\subsection{Surface stretch}\label{s:J}

The surface stretch $J$ measures the local area change w.r.t.~to the initial configuration of the surface, denoted $\sS_0$.
The current surface point $\bx\in\sS$ has the initial location $\bX=\bx|_{t=0}\in\sS_0$.
Based on the three parameterizations \eqref{e:bx}, \eqref{e:bx_xi} and \eqref{e:bx_th}, this leads to the basis vectors $\bA_\alpha=\ba_\alpha|_{t=0}$, $\bA_{\hat\alpha}=\ba_{\hat\alpha}|_{t=0}$ and $\bA_{\tilde\alpha}=\ba_{\tilde\alpha}|_{t=0}$, and the corresponding surface metrics 
$A_{\alpha\gamma}=a_{\alpha\gamma}|_{t=0}$, $A_{\hat\alpha\hat\gamma}=a_{\hat\alpha\hat\gamma}|_{t=0}$ and $A_{\tilde\alpha\tilde\gamma}= a_{\tilde\alpha\tilde\gamma}|_{t=0}$.
The surface stretch is given by 
\eqb{l}
J = \sqrt{\det[a_{\hat\alpha\hat\gamma}]}\Big/\sqrt{\det[A_{\hat\alpha\hat\gamma}]}\,.
\label{e:J}\eqe
It is obtained in the Lagrangian frame, since only this frame is tracking the physical material motion.
The quantity 
\eqb{l}
J_\mrm := \sqrt{\det[a_{\alpha\gamma}]}\Big/\sqrt{\det[A_{\alpha\gamma}]}\,,
\label{e:Jm}\eqe
on the other hand, tracks (non-physical) area-changes of the ALE frame.
Generally $J_\mrm\neq J$. 

As was already mentioned, the following presentation exclusively uses basis $\{\ba_1,\,\ba_2,\,\bn\}$.
Expression \eqref{e:J} therefore becomes impractical.
Instead, $J$ can be determined from the surface velocity $\bv$ through the evolution law
\eqb{l}
\ds\frac{\dot J}{J} = \divs\bv\,, 
\label{e:dJJ}\eqe
following from \eqref{e:sdiv} and $\dot\ba_{\hat\alpha}\cdot\ba^{\hat\alpha} = \dot J/J$, e.g.~see \citet{cism}.
Since $ \divs\bv = \tr\bd$, one can also write
\eqb{l}
\ds\frac{\dot J}{J} = d^{\alpha\beta} a_{\alpha\beta}\,.
\label{e:dJJ2}\eqe
Using \eqref{e:sdiv}, \eqref{e:badot} and $\dot\ba_\alpha\cdot\ba^\alpha = \dot J_\mrm/J_\mrm$, one can further write
\eqb{l}
\divs\bv = \ds\frac{\dot J_\mrm}{J_\mrm} + \dot\zeta^\alpha_{,\alpha}\,.
\eqe
For area-incompressible flows, where $\dot J=0$, Eq.~\eqref{e:dJJ} leads to the velocity constraint
\eqb{l}
\divs\bv = 0\,.
\label{e:divv}\eqe
Associated with this constraint is an unknown surface tension, the Lagrange multiplier $q$.
It can be different to the physical surface tension $\gamma$, as is seen in Sec.~\ref{s:bending}.

\section{Field equations}\label{s:bal}

In the ALE description, a field equation is required for each of the unknown fields $\rho(\zeta^\alpha,t)$, $\bv(\zeta^\alpha,t)$ and $\bv_\mrm(\zeta^\alpha,t)$.
They follow from mass, momentum and ``mesh" balance as is discussed in the following.
While the former two are well known, the latter is treated in a new manner here.

\subsection{Mass balance}

The surface fluid density $\rho=\rho(\zeta^\alpha,t)$ is governed by the the continuity equation
\eqb{l}
\dot\rho + \rho\,\divs\bv = 0\,,
\label{e:rho}\eqe
which follows from surface mass balance \citep{marsden,sahu17}.\footnote{Both references write the surface divergence of $\bv = v^\alpha\ba_\alpha + v\,\bn$ in the form $\divs\bv = v^\alpha_{;\alpha} - 2Hv$.}
In the Lagrangian frame this is a first order ordinary differential equation (ODE) that only requires the initial condition $\rho(\zeta^\alpha,0) = \rho_0(\zeta^\alpha)$, where $\rho_0$ is the given initial density distribution.
In the ALE frame, where $\dot\rho$ can be expanded according to \eqref{e:rALEa}, this is a first order PDE that additionally requires a boundary condition to capture the mass influx 
\eqb{l}
j^\alpha := \rho\,\big(\bv_\mrm-\bv\big)\cdot\ba^\alpha
\eqe
on all boundaries where $\bv_\mrm\neq\bv$.

\begin{remark} 
Inserting \eqref{e:dJJ} into \eqref{e:rho} shows that ODE \eqref{e:rho} is solved by 
\eqb{l}
\rho = \rho_0/J\,. 
\label{e:rhosol}\eqe
This is convenient in a Lagrangian description, where $J$ is easily obtained from \eqref{e:J}.
In an ALE description, \eqref{e:J} requires reconstructing basis $\ba_{\hat\alpha}$.
To avoid this, one can either solve \eqref{e:dJJ} for $J$ and then get $\rho$ from \eqref{e:rhosol}, or -- equivalently -- solve \eqref{e:rho} for $\rho$ and then get $J$ from \eqref{e:rhosol}.
\end{remark}

\begin{remark} 
If area-incompressibility is assumed, as in the examples in Sec.~\ref{s:anex}, Eqs.~\eqref{e:divv} and \eqref{e:rho} imply $\dot\rho=0$.
Thus the surface density remains constant in time ($\rho(t) = \rho_0$) and is thus known.
The remaining field equation to satisfy is \eqref{e:divv}.
It is now the field equation for the unknown Lagrange multiplier $q(\zeta^\alpha,t)$.
\end{remark}

\subsection{Momentum balance}

The surface fluid velocity $\bv=\bv(\zeta^\alpha,t)$ is governed by the equation of motion, 
\eqb{l}
\rho\,\dot\bv = \bT^\alpha_{;\alpha} + \bff\,,
\label{e:Taa}\eqe
which follows from linear surface momentum balance, see e.g.~\citet{shelltheo}.
Eq.~\eqref{e:Taa} is a very general expression, that includes surface flow on fixed and evolving surfaces, as well as flows without and with bending resistance (e.g.~Wilmore flows).
Here, $\bff = p\,\bn + f_\alpha\,\ba^\alpha$ is a surface load, while
\eqb{l}
\bT^\alpha = \bsig^\mrT \ba^\alpha
\label{e:T}\eqe
is the traction vector on a cut through $\sS$ that is perpendicular to $\ba^\alpha$.
It depends on the Cauchy stress
\eqb{l}
\bsig = N^{\alpha\beta} \ba_\alpha\otimes\ba_\beta + S^\alpha\,\ba_\alpha\otimes\bn\,,
\label{e:bsig0}\eqe
that has the components
\eqb{lll}
N^{\alpha\beta} \is \sigma^{\alpha\beta} + b^\beta_\gamma\,M^{\gamma\alpha}\,, \\[1mm]
S^\alpha \is - M^{\beta\alpha}_{~;\beta}\,,
\label{e:NS}\eqe
for Kirchhoff-Love shells \citep{naghdi72,steigmann99b}.
Here the in-plane membrane stresses $\sigma^{\alpha\beta}$ and the bending stress couples $M^{\alpha\beta}$ are defined by constitution, see Sec.~\ref{s:consti}.
The out-of-plane shear stress $S^\alpha$ on the other hand, follows from $M^{\alpha\beta}$, which is a consequence of assuming Kirchhoff-Love kinematics (i.e.~neglecting out-of-plane shear deformations).
The stress $\sigma^{\alpha\beta}$ is also referred to as the \textit{effective stress} \citep{simo89}. 
$N^{\alpha\beta}$, on the other hand corresponds to the physical stress according to Cauchy's theorem \citep{sahu17}.

Using Eq.~\eqref{e:vALE} or \eqref{e:vALEa}, PDE \eqref{e:Taa} can be expressed in the ALE frame.
In order to solve PDE \eqref{e:Taa}, an initial condition and boundary conditions on $\bv=\bv(\zeta^\alpha,t)$ are needed.
In PDE  \eqref{e:Taa}, one can also replace $\bT^\alpha_{;\alpha}$ by  
\eqb{l}
\bT^\alpha_{;\alpha} = \bsig_{\!;\alpha}^\mrT\,\ba^\alpha =: \divs\bsig^\mrT \,,
\label{e:Ta}\eqe
which follows from \eqref{e:T} and $\bsig^\mrT\ba^\alpha_{;\alpha}=2H\bsig^\mrT\bn=\mathbf{0}$, which in turn follows from $\ba^\alpha_{;\beta} = b^\alpha_\beta\,\bn$, $b^\alpha_\alpha=2H$ and \eqref{e:bsig0}.

In the expressions above, $(...)_{;\alpha}$ denotes the so-called covariant derivative.
It is equal to the parametric derivative $(...)_{,\alpha}$ for objects without free index, such as $\bv$ and $\bsig$, e.g.~see \citet{cism}.

\subsection{Mesh ``balance"}

The mesh velocity $\bv_\mrm=\bv_\mrm(\zeta^\alpha,t)$ is governed by the following set of equations.
Firstly, the condition
\eqb{l}
\bv_\mrm\cdot\bn = \bv\cdot\bn\,,
\label{e:vmn}\eqe
which states that $\bv_\mrm$ must have the same normal component as $\bv$, and which essentially corresponds to an Lagrangian out-of-plane fluid description.
Secondly, an equation for $v_\mrm^\alpha(\zeta^\alpha,t)$, the in-plane component of $\bv_\mrm$, is needed.
The following options will be considered here: \\[-8mm]
\begin{itemize}
\item [I.] \textit{Prescribed mesh motion}. 
In this case $v_\mrm^\alpha$ is prescribed either directly, or indirectly by choosing $\zeta^\alpha=\zeta^\alpha(\xi^\beta,t))$ and then calculating $v_\mrm^\alpha$ from \eqref{e:zetadot}. 
Examples for the latter choice are given in Sec.~\ref{s:anex}.
A special case is prescribing the following suboption: \\[-7mm]
\item [Ia.] \textit{Zero in-plane mesh velocity}. 
In this case
\eqb{l}
\bv_\mrm\cdot\ba_\alpha = 0\,.
\label{e:vm1}\eqe
This corresponds to an Eulerian in-plane fluid description. 
If desired, Eqs.~\eqref{e:vmn} and \eqref{e:vm1} can be combined into
\eqb{l}
\bv_\mrm = (\bn\otimes\bn)\,\bv\,.
\label{e:vm1c}\eqe
It is noted that for evolving surfaces, an Eulerian description, just like a Lagrangian description, can lead to large mesh distortion.
An example is shown in Sec.~\ref{s:bubble2}, see Fig.~\ref{f2:flow}. \\[-7mm]
\item [II.] \textit{Mesh motion defined by membrane elasticity}.
In this case $v_\mrm^\alpha$ is characterized by the membrane PDE
\eqb{l}
\sigma^{\alpha\beta}_{\mrm\,;\beta} = 0\,,
\label{e:vm2}\eqe
that can be derived from \eqref{e:Taa} for the special choice $\rho=0$, $\sig^{\alpha\beta}=\sig^{\alpha\beta}_\mrm$, $M^{\alpha\beta}=0$ and $\bff=\mathbf{0}$.\footnote{Contracting \eqref{e:Taa} by $\ba^\alpha$ and using \eqref{e:bsig0}-\eqref{e:Ta} immediately leads to \eqref{e:vm2}.} 
A possible elasticity model for the mesh is
\eqb{l}
\sigma^{\alpha\beta}_\mrm = \ds\frac{\mu_\mrm}{J_\mrm}\Big(A^{\alpha\beta}-a^{\alpha\beta}\Big)
\label{e:sigm}\eqe
\citep{shelltheo}.
The parameter $\mu_\mrm$ is physically irrelevant, but it can be used to improve the numerical behavior.
Eq.~\eqref{e:vm2} requires boundary conditions for the mesh position.
An example is taking $v^\alpha_\mrm = 0$ on the boundary.
Eqs.~\eqref{e:vm2}-\eqref{e:sigm} can then be solved at every time step for the current in-plane mesh position, see Sec.~\ref{s:wf}.
\end{itemize}
The choice of option depends on the application at hand. 
If the surface does not deform much, or is even fixed, the in-plane Eulerian description (option Ia) is best, as it is simplest. 
Examples are shown in Secs.~\ref{s:shear} and \ref{s:4tex}.
On the other hand, if large surface deformations occur, the Eulerian description can become very inaccurate due to large local mesh distortions, 
and so the elastic mesh description (option II) is best. 
This is demonstrated in the examples of Secs.~\ref{s:bubble2} and \ref{s:bud}.

\begin{remark} 
If membrane elasticity is used in conjunction with a Lagrangian description ($\bv=\bv_\mrm$), the resulting formulation corresponds to the stabilization scheme `A' proposed in \citet{droplet} for liquid menisci.
\end{remark}

\section{Constitutive models for area-incompressible fluid films}\label{s:consti}

This section introduces two known area-incompressible constitutive models for fluid films that either neglect or account for bending elasticity. 
They both are based on the additive membrane stress decomposition
\eqb{l}
\sigma^{\alpha\beta} = \sigma^{\alpha\beta}_\mathrm{el} + \sigma^{\alpha\beta}_\mathrm{inel}\,,
\label{e:sigK}\eqe
of elastic and inelastic stress components, which corresponds to a Kelvin model.
For bending, only elastic behavior is assumed. 
The stresses and bending moments are generally defined through the relations
\eqb{l}
\sig_\mathrm{el}^{\alpha\beta} = \ds\frac{2}{J}\pa{\Psi_0}{a_{\alpha\beta}}\,,
\label{e:2lawcrit_el}\eqe
\eqb{l}
M^{\alpha\beta} = \ds\frac{1}{J}\pa{\Psi_0}{b_{\alpha\beta}}
\label{e:2lawcrit_M}\eqe
and
\eqb{l}
\sig_\mathrm{inel}^{\alpha\beta}\,d_{\alpha\beta}\geq 0\,,
\label{e:2lawcrit_visc}\eqe
which follow from the second law of thermodynamics, see Appendix \ref{s:2law}.
Here $\Psi_0$ is the Helmholtz free energy per reference area.

\subsection{Pure in-plane flow}

The first case considers in-plane flow without (out-of-plane) bending resistance ($M^{\alpha\beta}=0$).
From $\Psi_0 = q\,g$, where $g := J-1$ is the area-incompressibility constraint and $q$ its Lagrange multiplier, and the Newtonian surface fluid\footnote{sometimes also denoted a \textit{Boussinesq-Scriven surface fluid}} model
\eqb{l}
\sigma_\mathrm{inel}^{\alpha\beta} = 2\eta\,d^{\alpha\beta}
\label{e:sigvisc}\eqe 
follows
\eqb{l}
\sig^{\alpha\beta} = q\,a^{\alpha\beta} + 2\eta\,d^{\alpha\beta}\,,
\label{e:sig}\eqe
see \citet{memtheo}.
Since $M^{\alpha\beta}=0$, we find $N^{\alpha\beta}=\sigma^{\alpha\beta}$ and $S^\alpha=0$ from Eq.~\eqref{e:NS}.
Hence the surface tension,
\eqb{l}
\gamma := \ds\frac{1}{2}\,N^{\alpha\beta} a_{\alpha\beta} \,,
\label{e:gamma}\eqe 
simply is $\gamma = q$ due to \eqref{e:dJJ2} and $\dot J = 0$.
With $\bsig = \sig^{\alpha\beta}\,\ba_\alpha\otimes\ba_\beta$ and \eqref{e:bds}, the tensorial form of constitutive model \eqref{e:sig} becomes
\eqb{l}
\bsig = \gamma\,\bi + 2\eta\,\bd_\mrs\,.
\label{e:bsig}\eqe
Here $\bi:=\ba_\alpha\otimes\ba^\alpha$ is the surface identity on $\sS$.
Since $\bi_{,\alpha}\,\ba^\alpha = 2H\bn$, Eq.~\eqref{e:Ta} then yields
\eqb{l}
\bT^\alpha_{;\alpha} = \nablas\gamma + 2H\gamma\,\bn + 2\eta\,\divs\bd_\mrs\,, 
\label{e:Taa1}\eqe
with $\divs\bd_\mrs = \bd_{\mrs,\alpha}\,\ba^\alpha$.
This expression is used in some of the analytical examples of Sec.~\ref{s:anex}.

\begin{remark} 
Even though the stress $\bsig$ only has in-plane components, its divergence contains the out-of-plane component $2H\gamma\,\bn$.
For flows on fixed manifolds, the out-of-plane part of the equation of motion is not needed, and hence \eqref{e:Taa} is projected into the tangent plane, yielding the ``surface Navier Stokes equations" \citep{nitschke12,reuther15,jankuhn18,olshanskii18,lederer20,suchde20}, sometimes also referred to as the ``Navier-Stokes equations on a manifold" \citep{koba17,fries18}.
This projection step is not necessary for the general case of evolving manifolds, as \eqref{e:Taa} naturally contains the out-of-plane equation of motion and its coupling to the in-plane equations of motion. 
Instead of splitting it into in-plane and out-of-plane contributions, which involves nonlinear projection operators and tends to yield lengthy equations, it is simplest to treat \eqref{e:Taa} as a compact, vector-valued PDE for the unknown surface velocity $\bv$.
\end{remark}

\subsection{In-plane flow with bending elasticity}\label{s:bending}

The second case is like the first but with an additional out-of-plane bending energy based on the model of \citet{helfrich73}. 
In this case
\eqb{l}
\Psi_0 = q\,g + J\big(k (H-H_0)^2 + k_\mrg\,\kappa\big)\,,\quad g := J-1\,.
\eqe
Here, $k$ and $k_\mrg$ are bending moduli, and $H_0$ is the so-called \textit{spontaneous curvature} -- the value of the mean curvature $H$ that is energetically favorable. 
From \eqref{e:sigK}--\eqref{e:2lawcrit_M} and \eqref{e:sigvisc} then follow
\eqb{lll}
\sig^{\alpha\beta} \is \big(q + k\,\Delta H^2 - k_\mrg\,\kappa\big)a^{\alpha\beta} - 2k\,\Delta H\,b^{\alpha\beta} + 2\eta\,d^{\alpha\beta} \,,  \\[1mm] 
M^{\alpha\beta} \is \big(k\,\Delta H+2k_\mrg H\big)a^{\alpha\beta} - k_\mrg\,b^{\alpha\beta}\,,
\label{e:sigM}\eqe
\citep{liquidshell} where $\Delta H:=H-H_0$.
Inserting this into (\ref{e:NS}.1) then gives
\eqb{l}
N^{\alpha\beta} = \big(q + k\,\Delta H^2 \big)a^{\alpha\beta} - k\,\Delta H\,b^{\alpha\beta} + 2\eta\,d^{\alpha\beta} \,, 
\eqe
In this case the physical surface tension defined in \eqref{e:gamma} is
\eqb{l}
\gamma = q - k\,H_0\,\Delta H\,. 
\eqe 
The stress tensor can then be determined from \eqref{e:bsig0}.

It can be shown that for a sphere, where $b^{\alpha\beta}=Ha^{\alpha\beta}$,
\eqref{e:bsig} and \eqref{e:Taa1} are still valid (but now need to be used with the new $\gamma$).
From (\ref{e:sigM}.2) follows for a sphere,
\eqb{l}
M^{\alpha\beta} = \big(k\,\Delta H + k_g\,H\big)\,a^{\alpha\beta}\,.
\label{e:Mabs}\eqe

\begin{remark} 
The Helfrich model admits the model of \citet{canham70} as a special case, when $c:=k/2=-k_\mrg$ and $H_0=0$ are chosen.  
This then leads to the simple relations $M^{\alpha\beta}=c\,b^{\alpha\beta}$ and $\gamma=q$.
\end{remark}

\begin{remark} 
In the framework of Kirchhoff-Love kinematics, the stress tensor $\bsig$ can only have the out-of-plane shear component defined by (\ref{e:NS}.2).
A separate constitutive choice for $S^\alpha$ is therefore not possible.
Reissner-Mindlin or Cosserat shell kinematics are needed for choosing $S^\alpha$ independently.
An obvious choice would be to choose $S^\alpha$ proportional to the out-of-plane part of $\bd$.
\end{remark}

\section{Weak form equations}\label{s:wf}

Box~\ref{t:SF} summarizes the governing strong form equations derived in the preceding two sections.
%-------------------------------------------------------------------------------------------------------------------------------
\begin{Box}[h]
\begin{center}
\begin{tabular}{|ll|}
\hline
& \\[-3.5mm]

1. & PDE for the fluid velocity $\bv = \bv(\zeta^\alpha,t)$ from \eqref{e:Taa} and \eqref{e:vALEa}: \\[1mm]
& $\boxed{\rho\,\bv' + \rho\,\bv_{\!,\alpha}\,\big(v^\alpha-v^\alpha_\mrm\big) = \bT^\alpha_{;\alpha} + \bff}\,,$ \\[3mm]
& where the traction $\bT^\alpha$ depends on the stress $\sigma^{\alpha\beta}$ and bending moment $M^{\alpha\beta}$ \\[.5mm]
& through relations \eqref{e:T}-\eqref{e:NS}. Those in turn depend on velocity, Lagrange  \\[.5mm]
& multiplier and position through the constitutive model, e.g.~\eqref{e:sigM}.
\\[3mm]

2. & Either: PDE for the fluid density $\rho = \rho(\zeta^\alpha,t)$ from \eqref{e:rho} and \eqref{e:rALEa}: \\[1mm]
& $\boxed{\rho' + \rho_{,\alpha}\,\big(v^\alpha-v^\alpha_\mrm\big) + \rho\,\divs\bv = 0}$~~for area-compressibility  \\[3mm]
& Or: PDE for the Lagrange multiplier $q = q(\zeta^\alpha,t)$ from \eqref{e:divv} and \eqref{e:sdiv}: \\[1mm]
& $\boxed{\divs\bv = 0}$~~for area-incompressibility
\\[4mm]

3. & Either: Prescribed in-plane mesh velocity $v_\mrm^\alpha$, e.g.~as zero ($ v_\mrm^\alpha = 0$). \\[0.5mm]
& Or: PDE for the in-plane mesh velocity $v_\mrm^\alpha = v_\mrm^\alpha(\zeta^\beta,t)$, e.g.~from \eqref{e:vm2}: \\[1mm]
& $\boxed{\sig_{\mrm\,;\beta}^{\alpha\beta} = 0}\,,$ with a suitable definition of the ``mesh stress" $\sig_\mrm^{\alpha\beta}$, e.g.~\eqref{e:sigm}. \\[3mm]
& In both cases, the normal mesh velocity is constrained by $\boxed{\bv_\mrm\cdot\bn = \bv\cdot\bn}\,.$
\\[3mm]

4. & ODE for the surface position $\bx(\zeta^\alpha,t)$ from \eqref{e:vm}: \\[1mm]
& $\boxed{\bx' = \bv_\mrm}$ 
\\[2mm]

\hline
\end{tabular}
\vspace{-4.5mm}
\end{center}
\caption{Coupled strong form equations for area-compressible or area-incompressible flows on self-evolving manifolds described in the ALE frame. 
Four equations are needed for the four fields $\bx$, $\bv$, $\bv_\mrm$ and $\rho$ or~$q$. \\[-1mm]
}
\label{t:SF}
\end{Box}
%-------------------------------------------------------------------------------------------------------------------------------
%
The formulation consists of four partially coupled, nonlinear PDEs that admit the special cases: \\[-8mm]
\begin{itemize}
\item Prescribed mesh velocity $v_\mrm^\alpha$: 
In this case the third PDE is absent and the problem reduces by one field.
Interesting special cases are $v_\mrm^\alpha=0$ (Eulerian description) and $v_\mrm^\alpha = v^\alpha$ (Lagrangian description).\\[-7mm]
\item Steady-state flow: 
In this case the time derivatives $\bv'$ and $\rho'$ vanish.
The three PDE then only contain spatial derivatives.
However, the surface still evolves in time, as long as $\bv\cdot\bn\neq0$, and hence the problem is still a transient one via the fourth ODE, which needs to be integrated in order to determine the current surface $\sS$. \\[-7mm]
\item Flows on fixed surfaces:
In this case $\bv\cdot\bn=0$. 
Using an Eulerian description, $\bv_\mrm$ becomes zero and the fourth ODE implies $\bx(\zeta^\alpha,t)=\bx(\zeta^\alpha,0) = \bX(\zeta^\alpha)$. \\[-7mm]
\end{itemize}

The equations in Box~\ref{t:SF} are characterized by the following coupling:
All equations depend on the surface deformation $\bx$, either directly or through the appearing differential operators.
The velocity $\bv$ appears in the first and second PDE, and in the normal mesh constraint; but PDE~3 and ODE~4 do not directly dependent on it. 
The mesh velocity $\bv_\mrm$ appears in PDE~1, PDE~2.1, ODE~4 and the normal mesh constraint, but not elsewhere.
Density $\rho$ affects PDE~2.1 directly and PDE~1 through stress $\sigma^{\alpha\beta}$; the other equations are independent from it. 
Lagrange multiplier $q$ only affects PDE~1 through $\sigma^{\alpha\beta}$; all other equations are independent from it.

This work is motivated by providing the foundations for numerical methods, such as the finite element method (FEM).
In order to numerically solve the four coupled equations of Box~\ref{t:SF} with FEM, the weak forms of their PDEs are needed.
A weak form is constructed by multiplying the PDE by suitable test functions, integrating it over the surface and using the surface divergence theorem where advantageous.
The integration is either carried out over the current surface $\sS$, or the reference surface $\sS_0$, depending on what is more convenient.
This results in the following three weak form expressions.
They are restricted to the area-incompressible case.
The novelty here lies in the discussion of different choices of variations, and the weak form for mesh elasticity.

\subsection{Weak form for the fluid velocity}\label{s:wfv}

Multiplying PDE \eqref{e:Taa} with the admissible variation $\delta\bx\in\sV$, integrating over the current surface and using the surface divergence theorem, leads to the weak form 
\eqb{l}
G := G_\mathrm{in} + G_\mathrm{int} - G_\mathrm{ext} = 0 \quad\forall\,\delta\bx\in\sV\,,
\label{e:WFv}\eqe
with
\eqb{lll}
G_\mathrm{in} \is \ds\int_{\sS} \delta\bx\cdot\rho\,\dot\bv\,\dif a\,, \\[4mm]
G_\mathrm{int} \is \ds\int_{\sS} \sig^{\alpha\beta}\,\delta\bx_{,\alpha}\cdot\ba_\beta \, \dif a + \int_{\sS} M^{\alpha\beta}\,\delta\bx_{;\alpha\beta}\cdot\bn \, \dif a\,, \\[4mm]
G_\mathrm{ext} \is \ds\int_{\sS}\delta\bx\cdot\bff\,\dif a
+ \int_{\partial\sS}\delta\bx\cdot\bT\,\dif s
+ \int_{\partial\sS}\delta\bn\cdot\bM\,\dif s\,,
\label{e:Giie}\eqe
and
\eqb{l}
\delta\bx_{;\alpha\beta} = \delta\bx_{,\alpha\beta} - \Gamma^\gamma_{\alpha\beta}\,\delta\bx_{,\gamma}\,,
\label{e:dxab}\eqe
see Appendix \ref{s:wfd}.
Here, the integration is considered over the ALE parameterized surface $\bx=\bx(\zeta^\alpha,t)$ as this is required for a computational formulation in the ALE frame. 
Accordingly the surface variation $\delta\bx$ is considered at fixed $\zeta^\alpha$, such that $\delta\bx_{,\alpha} = \delta\ba_\alpha$, i.e.~variation and parametric differentiation with respect to $\zeta^\alpha$ commute.
Thus, variation $\delta(...)$ is mathematically equivalent to the time derivative $(...)'$. 
An alternative would be to integrate over the Lagrangian surface parameterization $\bx=\hat\bx(\xi^\alpha,t)$ using a surface variation at fixed $\xi^\alpha$. Such a variation would be mathematically equivalent to the material time derivative $\dot{(...)}$, see Appendix \ref{s:wfd}.

The second term in $G_\mathrm{int}$ contains second derivatives (in both $M^{\alpha\beta}$ and $\delta\bx_{;\alpha\beta}$) that can only be properly captured by at least $C^1$-continuous ansatz functions.
Examples in finite element methods for the Helfrich model are subdivision surfaces \citep{feng06} and NURBS -- non-uniform rational B-splines \citep{liquidshell}.
NURBS are used in the example of Sec.~\ref{s:bud}.

Inserting Eq.~\eqref{e:vALEa}, $G_\mathrm{in}$ can be decomposed into the transient and convective parts
\eqb{lll}
G_\mathrm{trans} \is \ds\int_{\sS} \delta\bx\cdot\rho\,\bv'\,\dif a\,, \\[4mm]
G_\mathrm{conv} \is \ds\int_{\sS} \delta\bx\cdot\rho\,\bv_{,\alpha}\,\big(v^\alpha-v^\alpha_\mrm\big)\,\dif a\,,
\eqe
such that $G_\mathrm{in} = G_\mathrm{trans} + G_\mathrm{conv}$.

If desired the integrals can be mapped to the reference configuration using $\dif a = J_\mrm\,\dif A$.
If the fluid film does not resist bending, the internal and external bending moments $M^{\alpha\beta}$ and $\bM$ are excluded from \eqref{e:Giie}.

\subsection{Weak form for the area-incompressibility constraint}\label{s:wfq}

Multiplying the area-incompressibility constraint \eqref{e:divv} with the admissible variation $\delta q\in\sQ$ and integrating over the current surface leads to the weak form
\eqb{l}
\bar G := \ds\int_{\sS_0}\delta q\,\divs\bv\,\dif a = 0\,,\quad \forall\,\delta q \in\sQ\,.
\eqe
%$\sQ$ is now the function space $L^2$.

\subsection{Weak form for the mesh motion}\label{s:wfm}

The two components of the mesh velocity generally need to be treated differently:\\[-8mm] 
\begin{enumerate}
\item \textit{Out-of-plane mesh velocity}: This component is defined by the strong form equation \eqref{e:vmn}, which can either be enforced directly or converted into the weak form
\eqb{l}
\tilde G_\mro := \ds\int_{\sS_0}w\,\bn\cdot\big(\bv_\mrm-\bv\big)\,\dif A = 0 \,,\quad \forall\,w \in\sW_\mrn\,.
\label{e:WFmo}\eqe
Here the integration is written over the reference configuration as this simplifies computations.
Alternatively one can also chose to integrate over the current configuration.
\item \textit{In-plane mesh velocity}: This component can either be prescribed, e.g. as zero, see Eq.~\eqref{e:vm1} or it can be obtained from membrane equilibrium Eq.~\eqref{e:vm2}.
If Eq.~\eqref{e:vm2} is used, it is advantageous to derive the weak form by integration over the current ALE surface using $\dif a = J_\mrm\,\dif A$.
This leads to the weak form
\eqb{l}
\tilde G_\mathrm{iel} := \ds\int_{\sS_0}w_{\alpha;\beta}\,\sig_\mrm^{\alpha\beta}\,J_\mrm\,\dif A = 0 \,,\quad \forall\,w_\alpha \in\sW_\alpha\,.
\label{e:WFmie}\eqe
This follows from (\ref{e:Giie}.2) when $\delta\bx$ is replaced by the test function $\bw=w_\alpha\,\ba^\alpha$, see also Eq.~(187.3) in \citet{shelltheo}.
For the elastic membrane stress in \eqref{e:sigm}, $J_\mrm$ can then be conveniently cancelled.
If Eq.~\eqref{e:vm1} is used, it can be either prescribed in strong form or by the weak form
\eqb{l}
\tilde G_{\mri0} := \ds\int_{\sS_0}w_\alpha\,\ba^\alpha\cdot\bv_\mrm\,\dif A = 0 \,,\quad \forall\,w_\alpha \in\sW_\alpha\,.
\label{e:WFmi0}\eqe
\end{enumerate}
The last statement combines with \eqref{e:WFmo} to
\eqb{l}
\tilde G_0 := \ds\int_{\sS_0}\bw\cdot\big(\bv_\mrm-(\bn\otimes\bn)\,\bv\big)\,\dif A = 0 \,,\quad \forall\,\bw\in\sW\,,
\label{e:WFm}\eqe
which is the weak form of Eq.~\eqref{e:vm1c}.
Combining \eqref{e:WFmo} and \eqref{e:WFmie}, on the other hand, yields
\eqb{l}
\tilde G_\mathrm{el} := \alpha_\mrm \ds\int_{\sS_0}w\,\bn\cdot\big(\bv_\mrm-\bv\big)\,\dif A 
+ \int_{\sS_0}w_{\alpha;\beta}\,\tau_\mrm^{\alpha\beta}\,\dif A = 0 \,,\quad \forall\,\bw\in\sW\,,
\label{e:WFme}\eqe
where $\tau_\mrm^{\alpha\beta} :=J_\mrm\,\sig_\mrm^{\alpha\beta}$.
Here, the factor $\alpha_\mrm$ is included to ensure dimensional consistency between the two terms, which is needed in case $\tau^{\alpha\beta}_\mrm$ (via $\mu_\mrm$) is chosen to have units of membrane stress ([force/length]).
In Eqs.~\eqref{e:WFmo}-\eqref{e:WFme}, $\bw=w_\alpha\,\ba^\alpha + w\,\bn$ is the test function corresponding to mesh velocity $\bv_\mrm$.

\subsection{Summary}

The above weak form statements can be solved for the three fields $\bv(\zeta^\alpha,t)$, $q(\zeta^\alpha,t)$ and $\bv_\mrm(\zeta^\beta,t)$.
In Appendix~\ref{s:FE} a stabilized finite element method is outlined that uses quadratic shape functions for the spatial discretization and the implicit trapezoidal rule for the temporal discretization.
This leads to an algebraic system of three coupled equations that can be solved monolithically using the Newton-Raphson method.
If the mesh velocity is prescribed or eliminated via \eqref{e:vmn} and \eqref{e:vm1}, the algebraic system only consists of two coupled equations.

The fields $\bv$, $q$ and $\bv_\mrm$, along with their variations $\delta\bx$, $\delta q$ and $\bw$, have to be picked such that all integrals above are well defined. 
This is the case for piecewise polynomial functions, as are classically considered in finite element methods.

It is noted that in the above formulation no unknown \textit{mesh pressure} appears.
Such a variable can be used to enforce constraint \eqref{e:vmn} in the framework of Lagrange multiplier methods \citep{sahu24}.
It adds an extra field to the system that needs to be accounted for in the out-of-plane mesh equation.
Here instead, constraint \eqref{e:vmn} is enforced through weak form \eqref{e:WFmo} without introducing such an extra variable. 
Consequently, the out-of-plane mesh equation is not needed either.

\section{Examples}\label{s:anex}

This section presents six analytical and numerical examples for area-incompressible surface flows described in a chosen ALE frame and contrasted with Lagrangian and Eulerian descriptions.
Both fixed and evolving surfaces are considered.
Apart from illustrating the proposed surface ALE theory by providing concrete values to its formulae, they serve as manufactured benchmark solutions for future numerical methods.
For this reason they are reported in detail.

\subsection{1D fluid flow}\label{s:film}

\begin{figure}[h]
\begin{center} \unitlength1cm
\begin{picture}(0,5.6)
\put(-7.95,.0){\includegraphics[height=52mm]{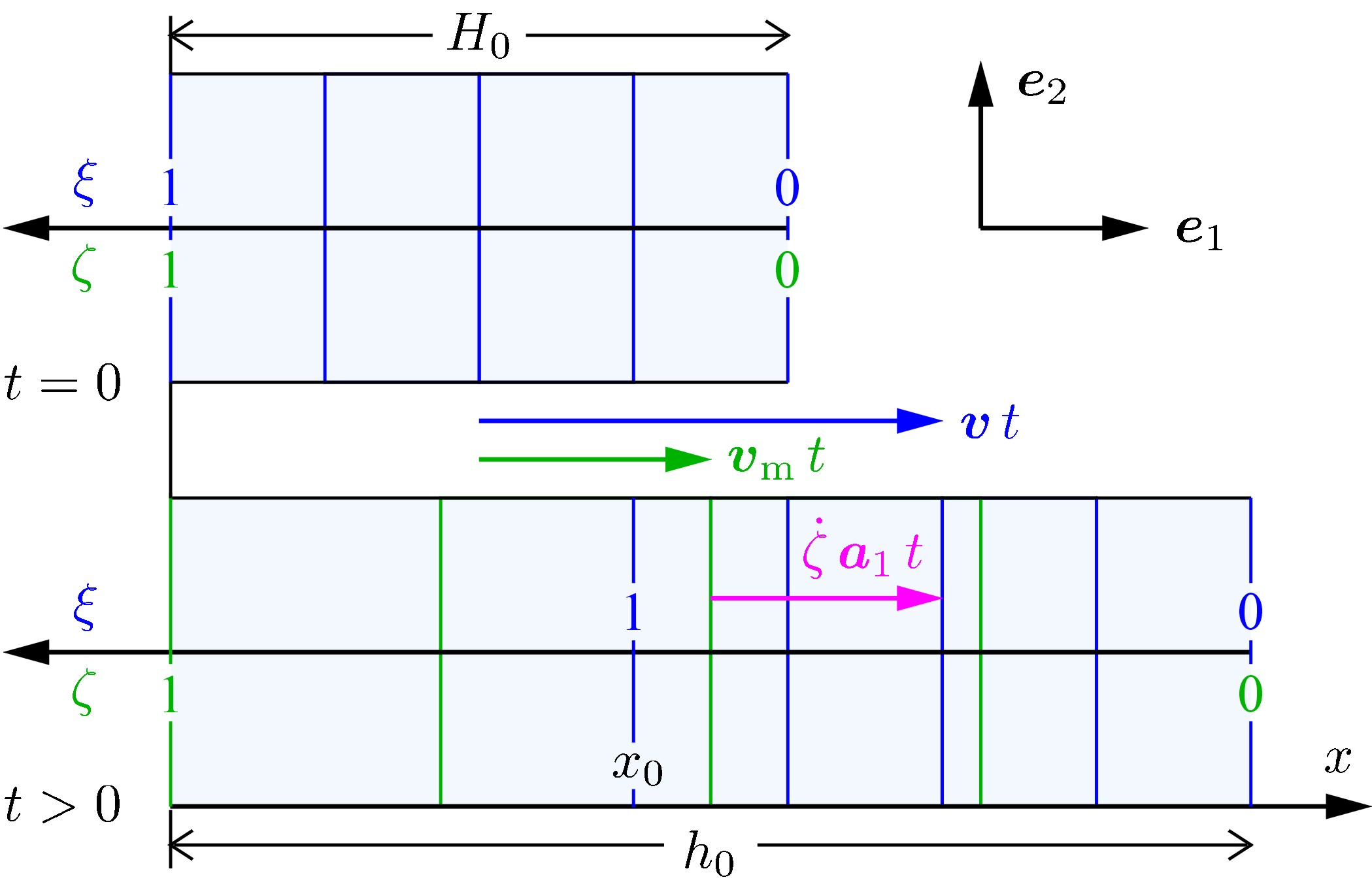}}
\put(0.35,-.25){\includegraphics[height=58mm]{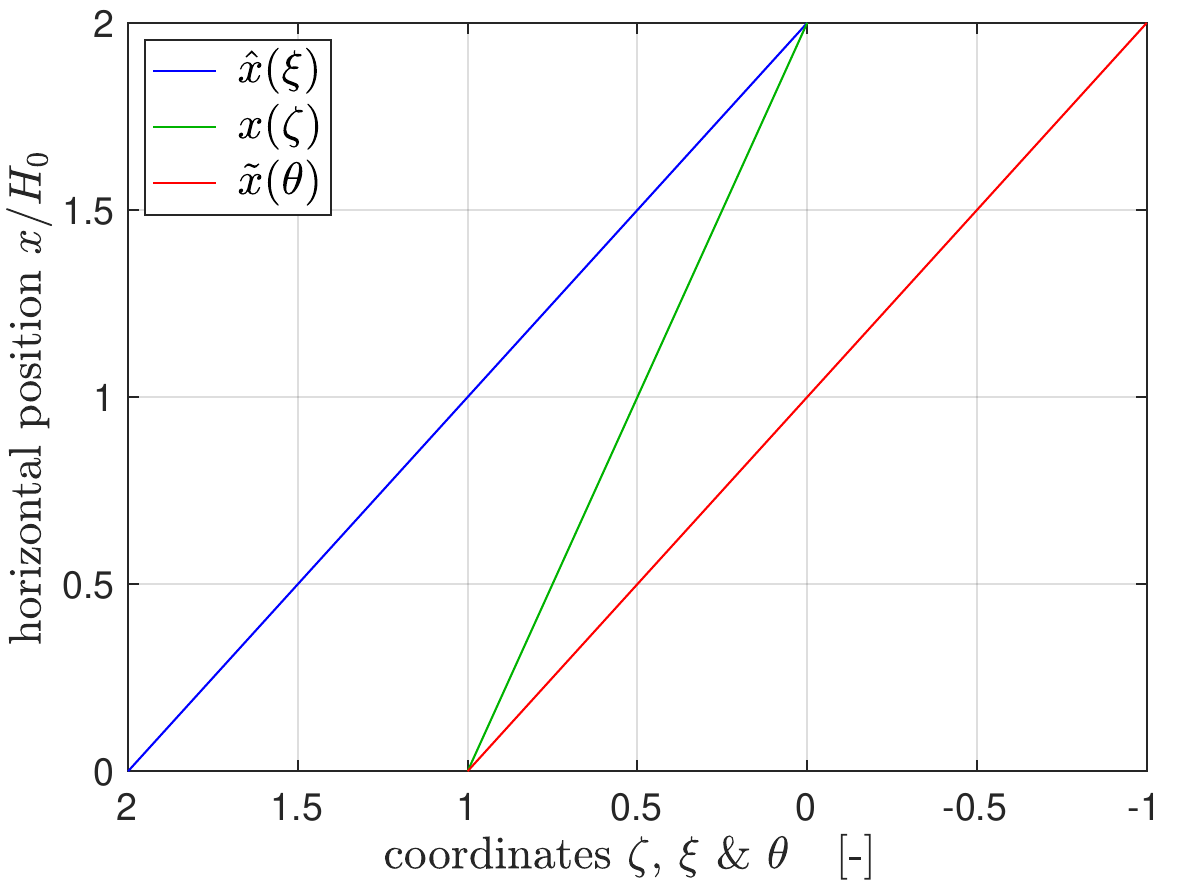}}
\put(-7.95,-.1){\footnotesize (a)}
\put(0.35,-.1){\footnotesize (b)}
\end{picture}
\caption{1D fluid flow: (a) Initial and current configuration, and illustration of Eq.~\eqref{e:bv}.
The Lagrangian and ALE surface parameterizations are shown in blue and green (coordinates $\xi$ and $\zeta$), respectively.  
(b) Position $x = \hat x(\xi)=x(\zeta)=\tilde x(\theta)$ for $x_0 = H_0$ (at $t= H_0/v_\mathrm{in}$). 
Note that coordinates $\zeta$, $\xi$ and $\theta$ run from right to left here.}
\label{f:pflow}
\end{center}
\end{figure}

The first example considers the 1D fluid motion shown in Fig.~\ref{f:pflow}.
It consists of a flat fluid sheet extruded with constant velocity from the left boundary.
Even though the example is very simple, it illustrates the proposed ALE formulation in the curvilinear setting. 
It is also a first step towards the following bubble example.
The motion in Fig.~\ref{f:pflow} can be described by
\eqb{l}
\bx = x\,\be_1 + y\,\be_2\,,
\label{e1:x}\eqe
where position $x = \hat x(\xi^1,t) = x(\zeta^1,t) = \tilde x(\theta^1,t)$ depends on the parameterization, while $y$ is equal for all, i.e. $y=\xi^2 = \zeta^2 = \theta^2$.
To shorten notation we let $\xi:=\xi^1$, $\zeta:=\zeta^1$ and $\theta = \theta^1$. 
The Lagrangian motion is given by
\eqb{l}
x = \hat x(\xi,t) = x_0(t) + (1-\xi)\,H_0\,,
\label{e:x1xi}\eqe
with
\eqb{l}
x_0(t) = v_\mathrm{in}\,t\,,
\eqe
such that
\eqb{l}
X(\xi) = \hat x(\xi,0) = (1-\xi)\,H_0
\eqe
is the initial position. 
Here, length $H_0$, inflow velocity $v_\mathrm{in}$ and directions $\be_\alpha$ are constants. 
%-----------------------------------------------------------------
% run film1D.m
%-----------------------------------------------------------------
Choosing the ALE coordinate
\eqb{l}
\zeta = \ds\frac{H_0}{h_0}\,\xi\,,
\label{e:z1}\eqe
leads to the ALE parametrization
\eqb{l}
x(\zeta,t) = (1-\zeta)\,h_0(t)\,,
\label{e:x1zeta}\eqe
with
\eqb{l}
h_0(t) := H_0 + x_0(t)\,,
\eqe
such that
\eqb{l}
X(\zeta) = x(\zeta,0) = (1-\zeta)\,H_0\,.
\eqe
Since $\dot h_0 = h_0' = x_0' = \dot x_0 = v_\mathrm{in}$, one finds the velocities
\eqb{l}
\bv = \dot\bx = v_\mathrm{in}\,\be_1\,,
\eqe
\eqb{l}
\bv_\mrm = \bx' = (1-\zeta)\,v_\mathrm{in}\,\be_1
\eqe
and
\eqb{l}
\dot\zeta^1 = \dot\zeta = -\zeta\ds\frac{v_\mathrm{in}}{h_0}\,,\quad \dot\zeta^2 = 0\,,
\eqe
from \eqref{e1:x}, \eqref{e:x1xi}, \eqref{e:x1zeta} and \eqref{e:z1}, respectively.
The tangent vectors
\eqb{l}
\ba_1 = \ds\pa{\bx}{\zeta^1} = -h_0\,\be_1\,,\quad
\ba_2 = \ds\pa{\bx}{\zeta^2} = \be_2\,,
\eqe
then give
\eqb{l}
\dot\zeta^\alpha\ba_\alpha = \zeta\,v_\mathrm{in}\,\be_1\,.
\eqe
The velocities $\bv$, $\bv_\mrm$ and $\dot\zeta^\alpha\ba_\alpha$ are visualized in Fig.~\ref{f:pflow} (by showing the displacements $\bv\,t$, $\bv_\mrm\,t$ and $\dot\zeta^\alpha\ba_\alpha\,t$).
It is easy to confirm that they satisfy Eq.~\eqref{e:bv}, which states that the material velocity $\bv$ is composed of the frame velocity $\bv_\mrm$ and the frame-relative material velocity $\dot\zeta^\alpha\ba_\alpha$.
From $\bv_{\!,\alpha}=\mathbf{0}$ and $\dot\ba_1 = -v_\mathrm{in}\,\be_1$ and $\dot\ba_2=\mathbf{0}$ and 
\eqb{l}
\dot\zeta^1_{,1} = -\ds\frac{v_\mathrm{in}}{h_0}\,,
\eqe
one can also confirm that Eq.~\eqref{e:badot} is satisfied.
This underlines that $\bv_\alpha \neq \dot\ba_\alpha$ here.
The equality of $\bv_\alpha$ and $\dot\ba_\alpha$, used in the curvilinear ALE formulation of \citet{ALE}, is therefore not generally satisfied.
Instead they differ here proportional to the inflow velocity $v_\mathrm{in}$.

The Eulerian motion is given by
\eqb{l}
x = \tilde x(\theta) = (1-\theta)\,H_0\,.
\eqe
since it leads to $\bv_\mrm=\mathbf{0}$.
Fig.~\ref{f:pflow}b illustrates the Lagrangian, ALE and Eulerian description of motion.

\subsection{Inflation of a 2D soap bubble}\label{s:bubble2}

\begin{figure}[h]
\begin{center} \unitlength1cm
\begin{picture}(0,4.1)
\put(-7.3,-.2){\includegraphics[height=44mm]{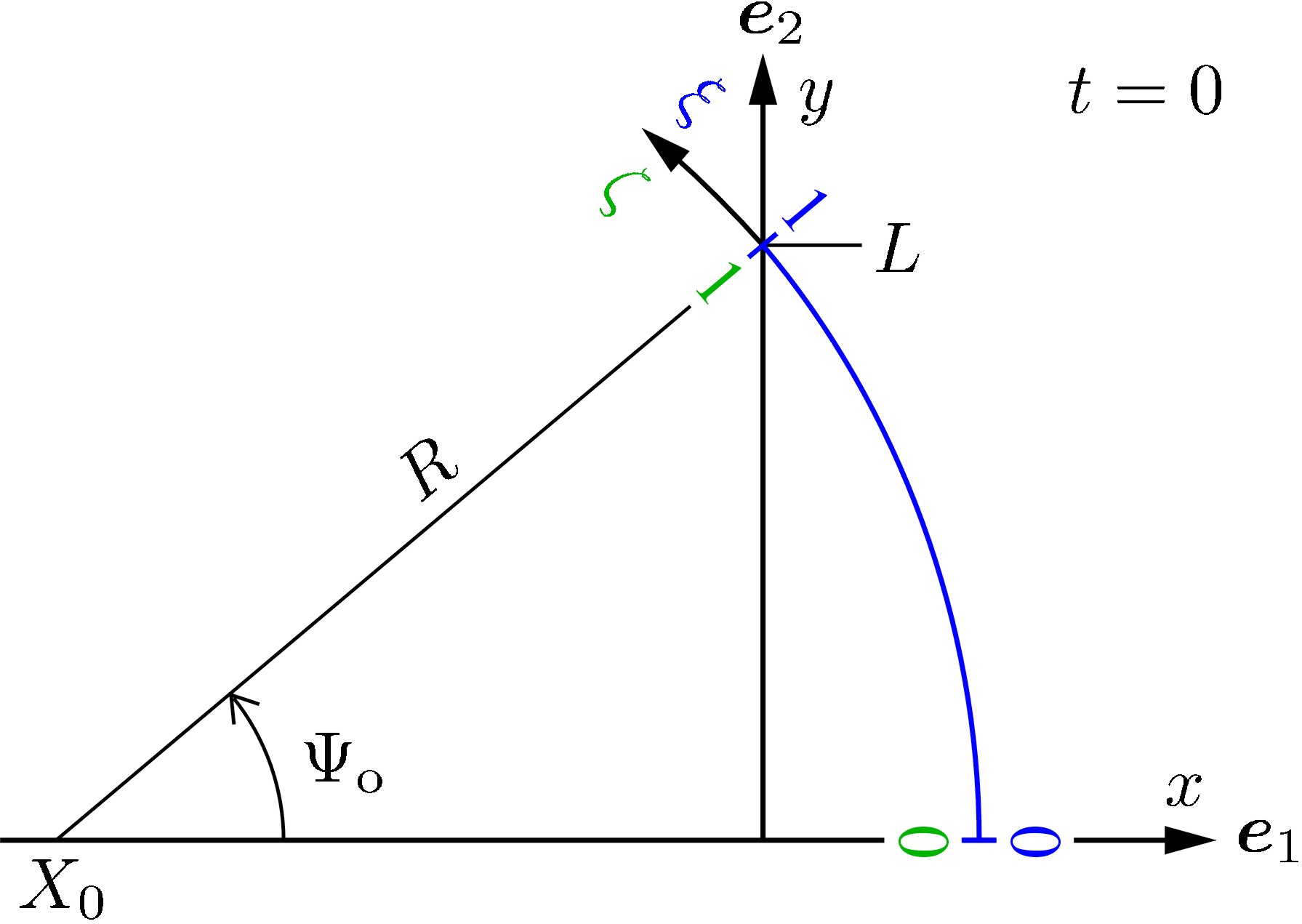}}
\put(0.2,-.2){\includegraphics[height=44mm]{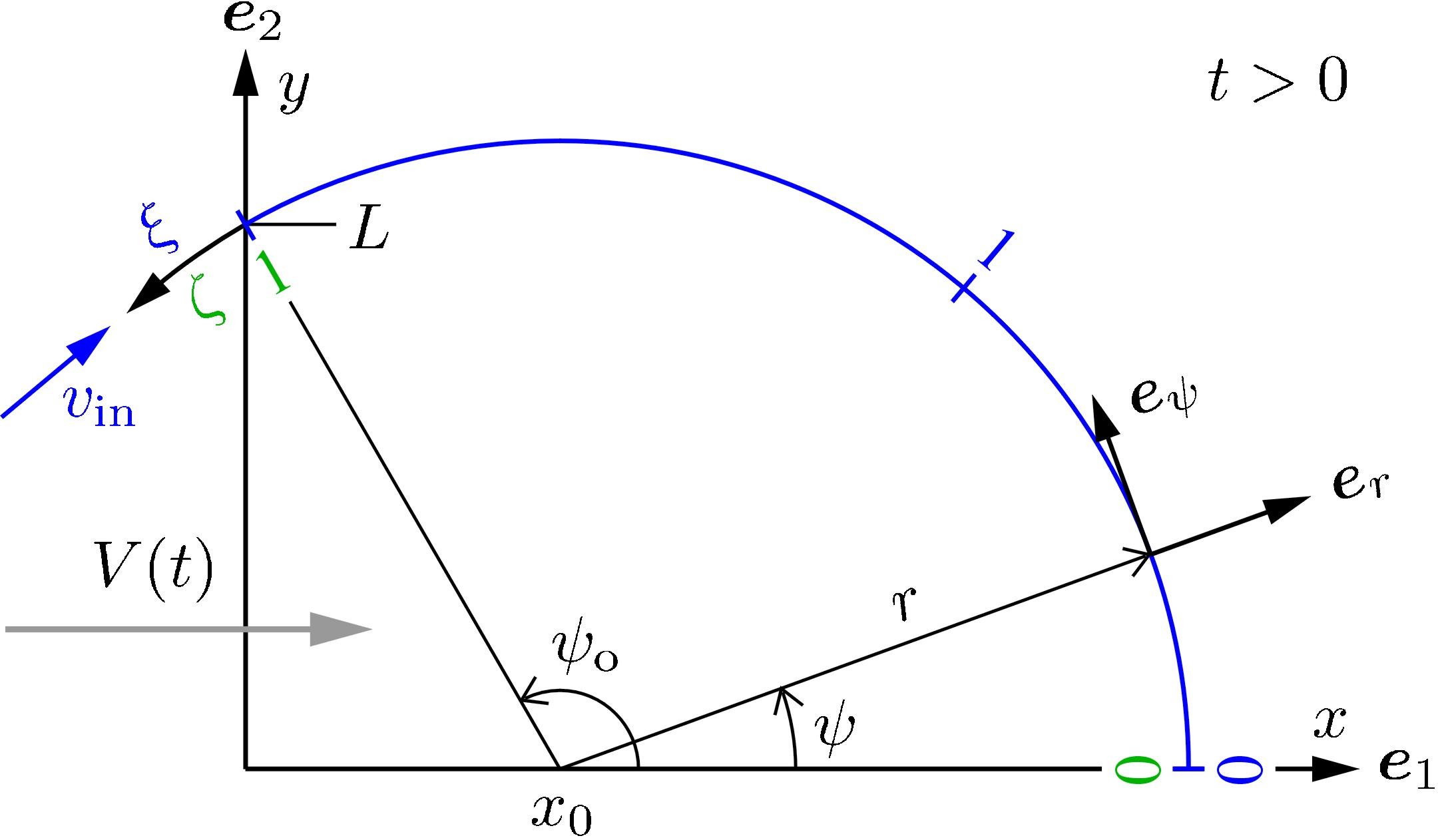}}
\put(-7.95,-.1){\footnotesize (a)}
\put(0.2,-.1){\footnotesize (b)}
\end{picture}
\caption{Inflation of 2D soap bubble: 
(a) Initial and (b) current configuration. 
Fluid inflow is considered at the left boundary such that $J\equiv1$ (i.e. $\divs\bv\equiv0$) over time.
The inflation can be driven by prescribing $v_\mathrm{in}$ or $V(t)$ at the opening.
The Lagrangian and ALE surface parameterizations are shown in blue and green (coordinates $\xi$ and $\zeta$), respectively.}
\label{f2:setup}
\end{center}
\end{figure}

Next, the inflation of a 2D soap bubble as shown in Fig.~\ref{f2:setup} is considered.
%-----------------------------------------------------------------
%-----------------------------------------------------------------
Apart from the kinematics of inflation, the example provides insight to how Lagrangian and Eulerian frame motion leads to severely distorted surface grids, even in simple flow examples, while these distortions can be avoided with the proposed ALE formulation, which then leads to superior numerical methods, as is also shown.

\subsubsection{Surface motion}

As in the example before, no flow in $\be_3$ direction occurs and we again use $\xi=\xi^1$, $\zeta=\zeta^1$ and $\theta=\theta^1$ as shorthand for the main parameter, while $z=\xi^2=\zeta^2=\theta^2$.
The motion of the bubble is described by
\eqb{l}
\bx = x_0\,\be_1 + r\,\be_r + z\,\be_3\,,
\label{e2:x}\eqe
where the center
\eqb{l}
x_0 = -r\,\cos\psi_\mro
\label{e2:x0}\eqe
and radius
\eqb{l}
r = \ds\frac{L}{\sin\psi_\mro}
\label{e2:r}\eqe
depend on the fixed opening height $L$ and the varying opening angle $\psi_\mro=\psi_\mro(t)$, and therefore are only functions of time, i.e.~$x_0=x_0(t)$ and $r=r(t)$.
The only spatially varying objects in \eqref{e2:x} are $z$ and
\eqb{l}
\be_r = \cos\psi\,\be_1 + \sin\psi\,\be_2\,,\quad 0\leq\psi\leq\psi_\mro<\pi\,.
\label{e2:ber}\eqe
The flow is fully described by the Lagrangian parameterization of angle $\psi$,
\eqb{l}
\psi = \hat\psi(\xi,t) = \ds\frac{R}{r}\Psi_\mro\,\xi\,,
\label{e2:hphi}\eqe
where $R = r(0)$ and $\Psi_\mro=\psi_\mro(0)$, see Fig.~\ref{f2:setup}.
A possible ALE parameterization that satisfies $\psi(0,t)=0$ and $\psi(1,t)=\psi_\mro$ is
\eqb{l}
\psi = \psi(\zeta,t) = \psi_\mro\,\zeta\,.
\label{e2:phi}\eqe
From this follows
\eqb{l}
\zeta(\xi,t) = \ds\frac{R\Psi_\mro}{r\psi_\mro}\xi
\label{e2:zeta}\eqe
and
\eqb{l}
\ba_1 = r\,\psi_\mro\,\be_\psi\,,
\label{e2:ba1}\eqe
where
\eqb{l}
\be_\psi = -\sin\psi\,\be_1 + \cos\psi\,\be_2
\label{e2:epsi}\eqe
and Eq.~\eqref{e3:bezeta} from Appendix \ref{s:basis} have been used.
Basis $\ba_2$ on the other hand is constant and equal to $\be_3$.
The function $\zeta(\xi,t)$ defines the in-plane material motion relative to the ALE frame, just like in the example before.

\begin{remark}
Coordinates $\xi$ and $\zeta$ parameterize angle $\psi$ according to \eqref{e2:hphi} and \eqref{e2:phi}. 
Thus, they also parameterize the arc length $r\psi$, but are not necessarily equal to it, i.e.~generally $R\Psi_\mro\neq1$ and $r\Psi_\mro\neq1$. 
\end{remark}

\subsubsection{Surface velocity}

Since $\psi_\mro$, $r$ and $x_0$ are only functions of time, there is no difference between their time derivatives $\dot{(...)}$ and $(...)'$.
Thus, not all time derivatives differ between the Lagrangian and ALE frame.
Given $\dot\psi_\mro$ (that can be prescribed to control the inflation, see below), one finds
\eqb{l}
\dot x_0 = \ds\frac{1}{\sin\psi_\mro}\,r\,\dot\psi_\mro \,,
\label{e2:dx0}\eqe
and
\eqb{l}
\dot r = -\ds\frac{\cos\psi_\mro}{\sin\psi_\mro}\,r\,\dot\psi_\mro\,,
\label{e2:dr}\eqe
from Eqs.~\eqref{e2:x0} and \eqref{e2:r}.
Further, Eqs.~\eqref{e2:hphi} and \eqref{e2:phi} lead to
\eqb{l}
\dot\psi = -\ds\frac{\dot r}{r}\psi\,,
\label{e2:dphi}\eqe
\eqb{l}
\psi' = \dot\psi_\mro\,\zeta
\label{e2:phip}\eqe
and
\eqb{l}
\dot\zeta = \ds\frac{\dot\psi}{\psi_\mro}-\zeta\,\frac{\dot\psi_\mro}{\psi_\mro}\,.
\label{e2:dzeta}\eqe
It can be confirmed that \eqref{e2:dphi}--\eqref{e2:dzeta} satisfy Eq.~\eqref{e:ALE} as $\partial\psi/\partial\zeta=\psi_\mro$.
Inserting \eqref{e2:dphi} and \eqref{e2:phi}, derivative $\dot\zeta$, which describes the in-plane material velocity relative to the ALE frame, can also be written as
\eqb{l}
\dot\zeta = -\ds\frac{\dot{(r\psi_\mro)}}{r\psi_\mro}\,\zeta\,.
\label{e2:dzeta1}\eqe
Using \eqref{e3:bedot}, and writing
\eqb{l}
\be_1 = \cos\psi\,\be_r - \sin\psi\,\be_\psi\,, 
\eqe
the material velocity field follows as
\eqb{l}
\bv = \dot\bx = (\dot x_0\,\cos\psi + \dot r)\,\be_r + (r\,\dot\psi-\dot x_0\,\sin\psi)\,\be_\psi \,,  
\label{e2:v}\eqe
while the ALE frame velocity becomes
\eqb{l}
\bv_\mrm = \bx' = (\dot x_0\,\cos\psi + \dot r)\,\be_r + (r\,\psi'-\dot x_0\,\sin\psi)\,\be_\psi \,.
\label{e2:vm}\eqe
Together with \eqref{e2:ba1}, \eqref{e2:phip}, \eqref{e2:dzeta} and $\dot\zeta^2=0$, they correctly satisfy Eq.~\eqref{e:bv}.
As seen, $\bv$ and $\bv_\mrm$ share the same normal velocity component
\eqb{l}
v_r = (\cos\psi - \cos\psi_\mro)\,\ds\frac{r\,\dot\psi_\mro}{\sin\psi_\mro}
\label{e2:vr}\eqe
along $\be_r$ (as they should), while they differ in their tangential velocity components 
\eqb{l}
v_\psi = r\,\psi_\mro\,\dot\zeta + \bigg(\zeta-\ds\frac{\sin\psi}{\sin\psi_\mro}\bigg)\,r\,\dot\psi_\mro
\label{e2:vp}\eqe
and
\eqb{l}
v_{\mrm\psi} = \bigg(\zeta-\ds\frac{\sin\psi}{\sin\psi_\mro}\bigg)\,r\,\dot\psi_\mro
\label{e2:vmp}\eqe
along $\be_\psi$. 
These expressions follow from inserting \eqref{e2:dx0}--\eqref{e2:dr} and \eqref{e2:phip}--\eqref{e2:dzeta} into \eqref{e2:v}--\eqref{e2:vm}.
At the inflow, where $\zeta=1$ and $\psi=\psi_\mro$, 
the velocity components $v_{\mrm\phi}$ and $v_r$ are zero (the latter since $L$ is fixed),
while the tangential velocity $-v_\psi$ is equal to the inflow velocity $v_\mathrm{in}$ (see Fig.~\ref{f2:setup}), giving
\eqb{l} 
v_\mathrm{in} = -r\,\psi_\mro\,\dot\zeta\big|_{\zeta=1}\,,
\label{e2:vindef}\eqe
which in light of \eqref{e2:dzeta1} becomes
\eqb{l}
v_\mathrm{in} = \dot{(r\psi_\mro)}\,.
\label{e2:vin}\eqe

\begin{figure}[h]
\begin{center} \unitlength1cm
\begin{picture}(0,5)
\put(-7.8,-.2){\includegraphics[height=50mm]{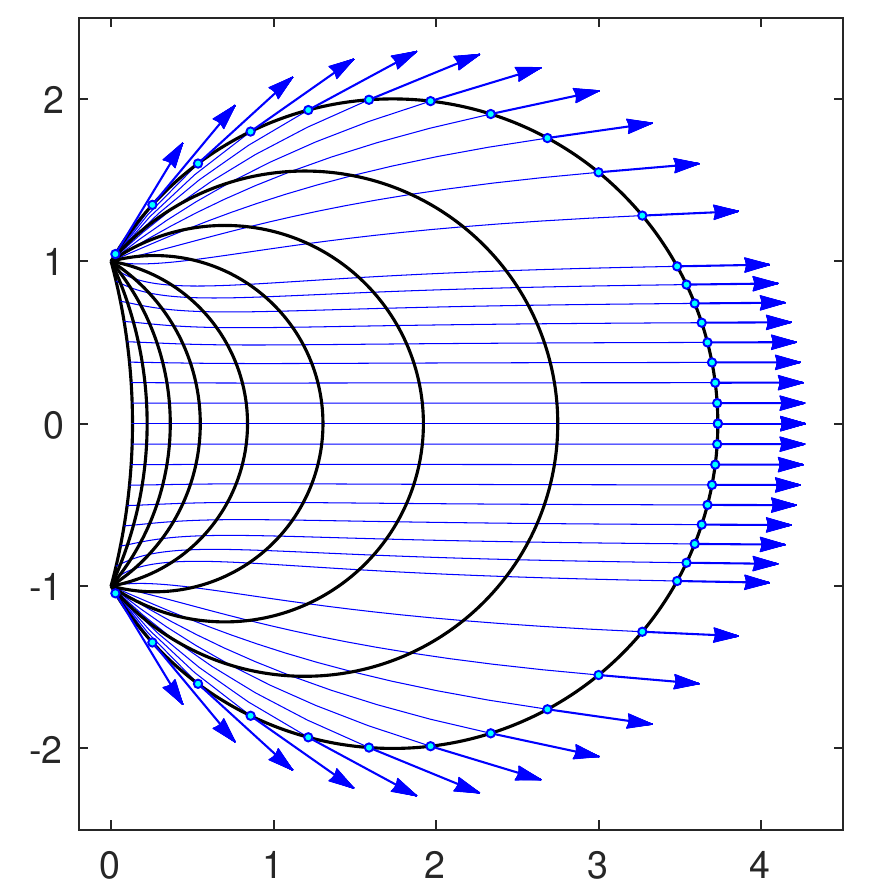}}
\put(-2.4,-.2){\includegraphics[height=50mm]{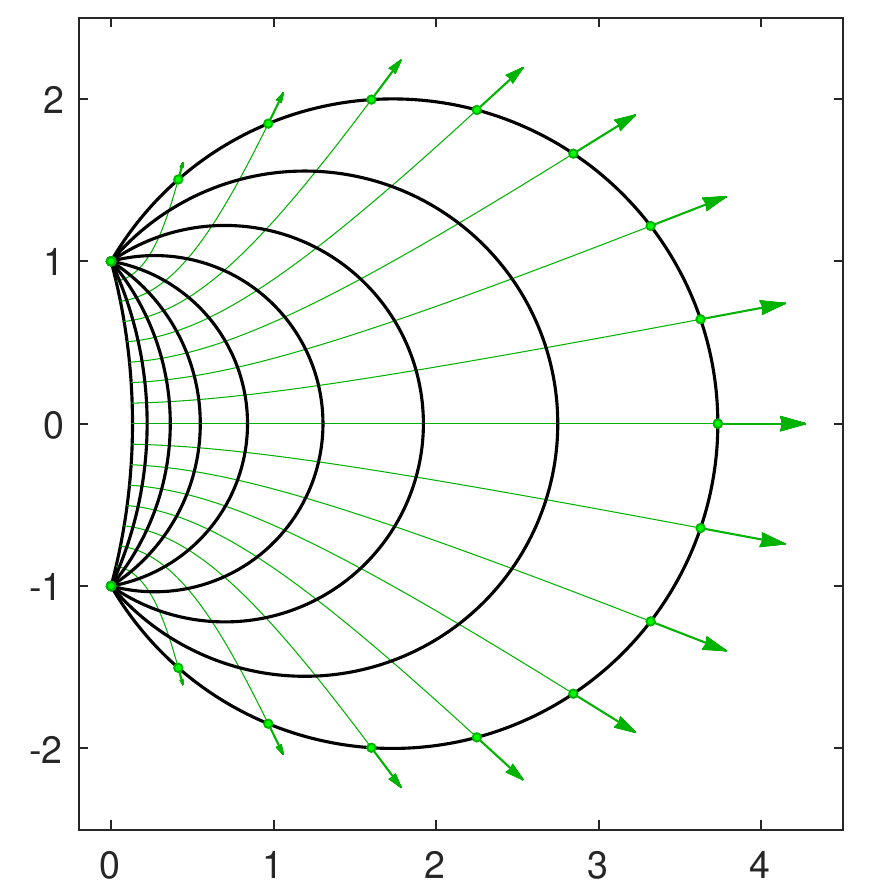}}
\put(3.0,-.2){\includegraphics[height=50mm]{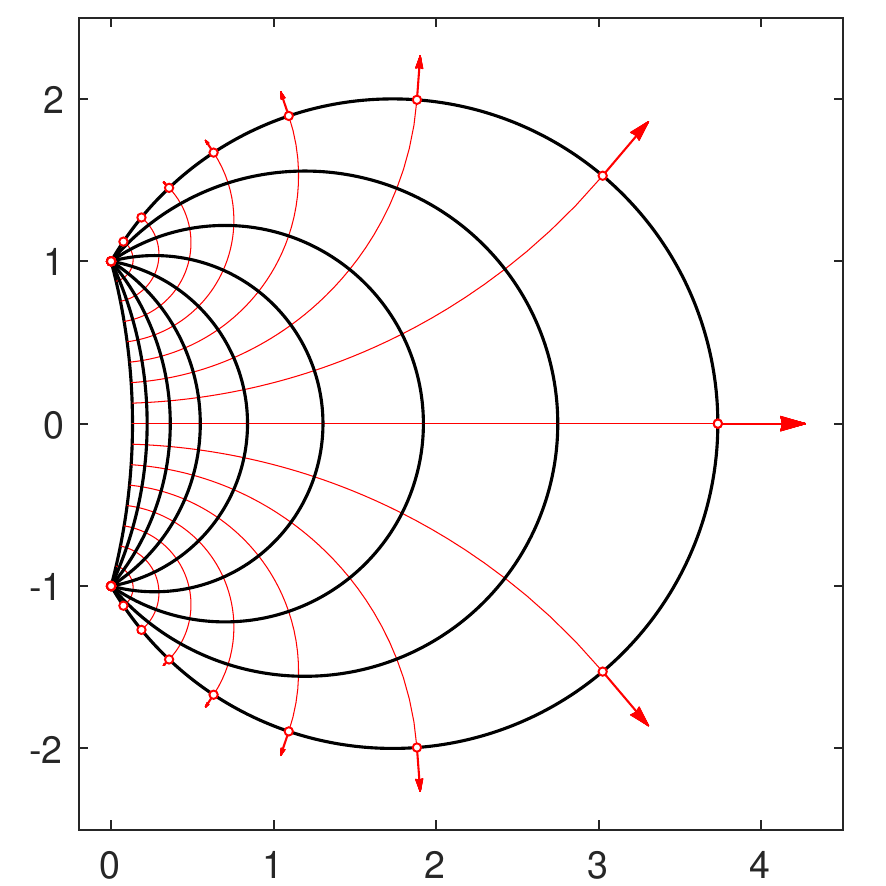}}
\put(-7.95,-.1){\footnotesize (a)}
\put(-2.55,-.1){\footnotesize (b)}
\put(2.85,-.1){\footnotesize (c)}
\end{picture}
\caption{Inflation of a 2D soap bubble: (a) Lagrangian, (b) ALE and (c) Eulerian surface motion and corresponding surface velocity fields $\bv$, $\bv_\mrm$ and $\bv_\mre$. 
Shown is the motion of 17 initially equidistant surface points. 
They remain equidistant in the Lagrangian and ALE motion, but only the former maintains their original distance.
In case of the Lagrangian motion, new material points are drawn in at the boundary (shown by path lines three times more spaced).
The Eulerian motion, which is always normal to the surface, leads to unequal distances between points.
The initial opening angle is chosen as $\Psi_\mro = 15^\circ$.
The black lines show the bubble at $\psi_\mro = [\Psi_\mro,\,25^\circ,\,40^\circ,\,57.5^\circ,\,80^\circ,\,105^\circ,\,125^\circ,\,140^\circ,\,150^\circ]$. 
$L$ is used for normalizing the geometry.}
\label{f2:flow}
\end{center}
\end{figure}

\begin{figure}[h]
\begin{center} \unitlength1cm
\begin{picture}(0,5.6)
\put(-7.9,-.25){\includegraphics[height=58mm]{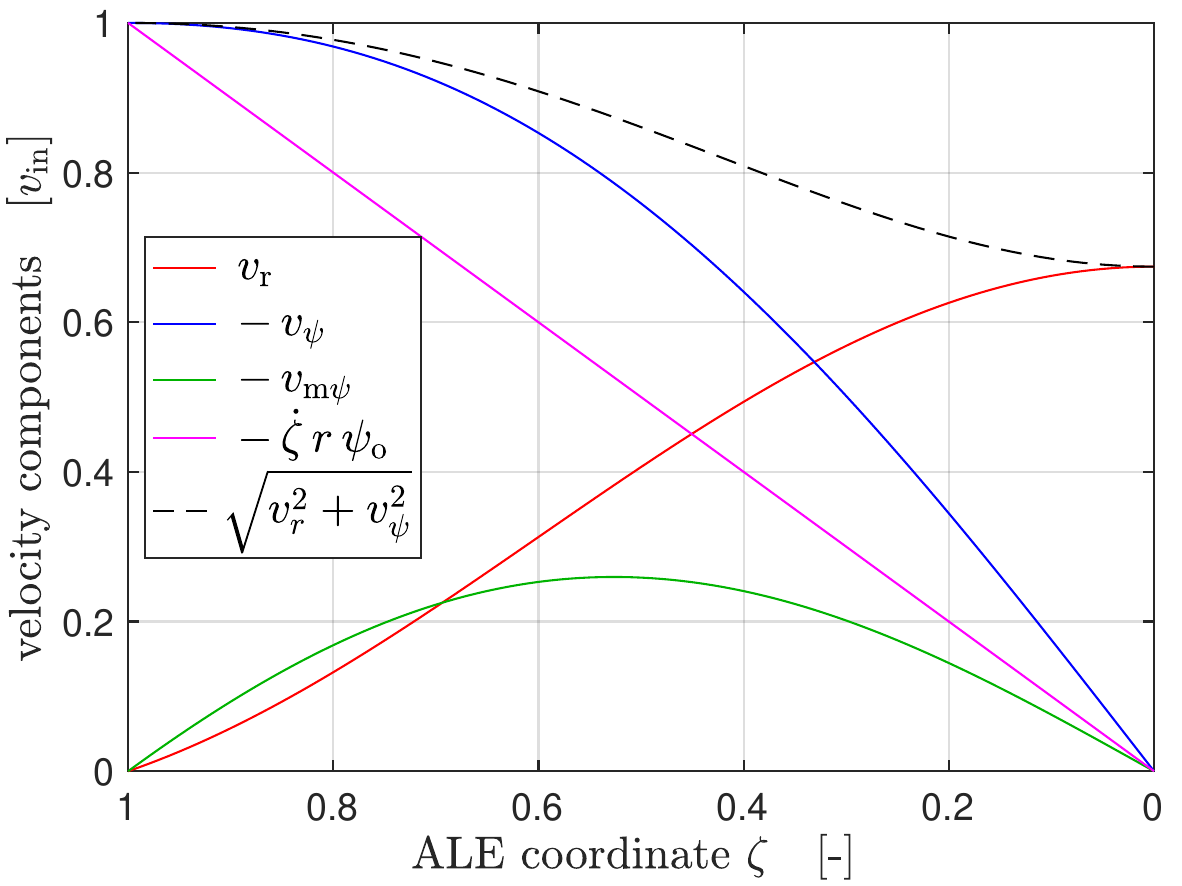}} 
\put(0.3,-.25){\includegraphics[height=58mm]{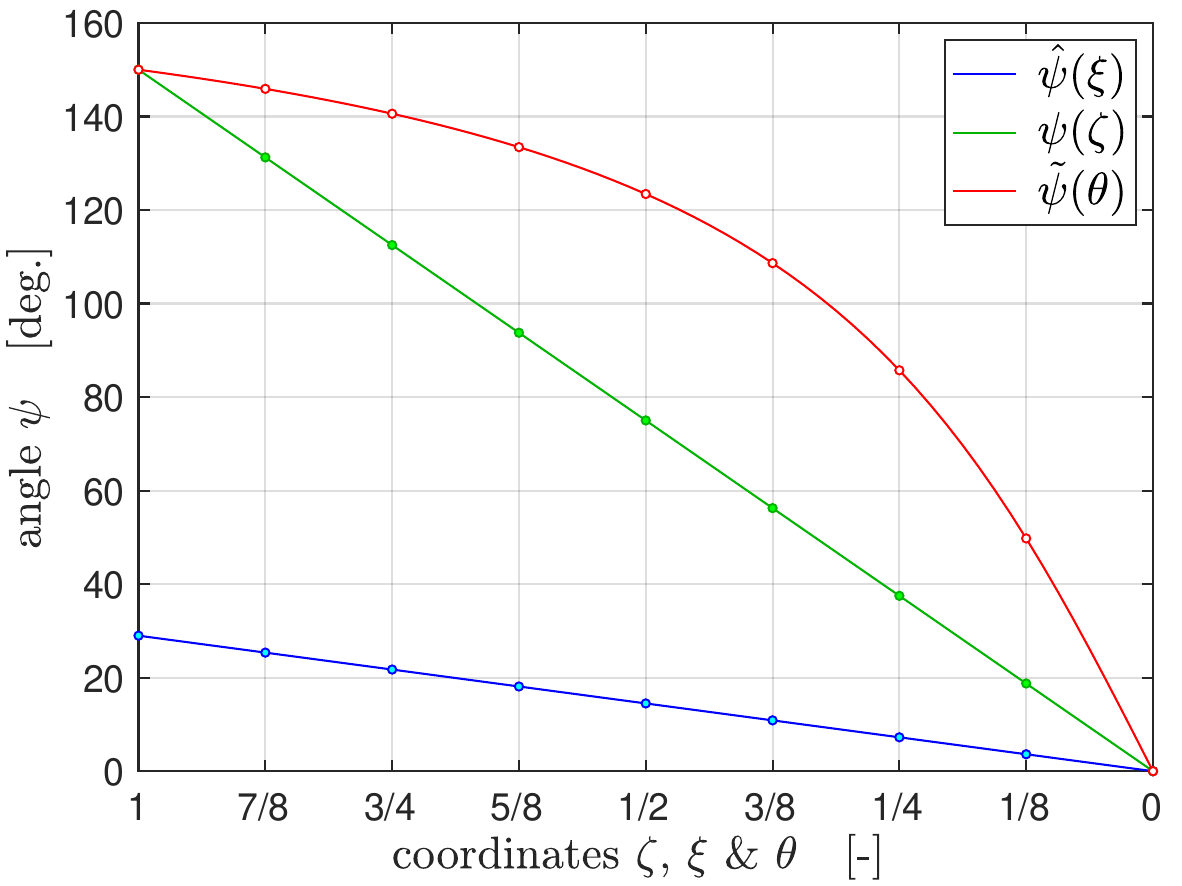}}
\put(-7.95,-.1){\footnotesize (a)}
\put(0.25,-.1){\footnotesize (b)}
\end{picture}
\caption{Inflation of 2D soap bubble: 
(a) Radial and tangential velocity components $v_r(\zeta)$, $v_\psi(\zeta)$ and $v_{\mrm\psi}(\zeta)$, and 
(b) angle~$\psi$ in Lagrangian, $\hat\psi(\xi)$, ALE, $\psi(\zeta)$, and Eulerian, $\tilde\psi(\theta)$, description. 
All for $\psi_\mro = 150^\circ$. 
The coordinates are running from right to left as in Fig.~\ref{f2:setup}. The Lagrangian coordinate continues until $\xi = \psi_\mro\sin\Psi_\mro/(\Psi_\mro\sin\psi_\mro) \approx 5.176$ according to \eqref{e2:zeta} and \eqref{e2:r}.
The left figure shows how the velocity components vary across the surface: $v_r$ is maximum at the tip, and $-v_\psi$ at the inflow. Note $v_\psi = v_{\mrm\psi} + \dot\zeta\,r\,\psi_\mro$.
The circles in (b) mark the positions shown in Fig.~\ref{f2:flow}.}
\label{f2:v}
\end{center}
\end{figure}

The Lagrangian and the ALE motion are visualized in Fig.~\ref{f2:flow}, while Fig.~\ref{f2:v} shows their velocity components.
%-----------------------------------------------------------------
% run soapbubble2D.m
%-----------------------------------------------------------------
%-----------------------------------------------------------------
% run soapbubble2D.m
%-----------------------------------------------------------------
Fig.~\ref{f2:flow} also shows the Eulerian surface motion, which is determined from the condition $v_{\mrm\psi}=0$ for all $t$.
This leads to a nonlinear ODE for $\psi=\tilde\psi(\theta,t)$ that together with \eqref{e2:x} and \eqref{e2:ber} then defines the Eulerian surface motion:
Setting the tangential component of \eqref{e2:vm} to zero and inserting \eqref{e2:dx0} yields
\eqb{l}
\ds\frac{\psi'}{\sin\psi} = \ds\frac{\psi_\mro'}{\sin\psi_\mro}\,,
\eqe
since $\dot\psi_\mro = \psi_\mro'$.
Multiplying by $\dif t$ and considering $\zeta=\theta$ fixed then gives
\eqb{l}
\ds\frac{\dif\psi}{\sin\psi} = \ds\frac{\dif\psi_\mro}{\sin\psi_\mro}\,,
\eqe
which can be integrated on both sides to yield the ODE
\eqb{l}
\ds\frac{1+\cos\psi}{\sin\psi} = c(\theta)\,\ds\frac{1+\cos\psi_\mro}{\sin\psi_\mro}\,. 
\label{e2:thdef}\eqe
The integration constant $c(\theta)$ follows from evaluating \eqref{e2:thdef} at $t=0$, where 
$\psi_\mro = \Psi_\mro$ and $\psi = \Psi_\mro\,\theta:=\Psi(\theta)$ according to \eqref{e2:phi}.
This gives
\eqb{l}
c(\theta) = \ds\frac{1+\cos\Psi(\theta)}{1+\cos\Psi_\mro}\frac{\sin\Psi_\mro}{\sin\Psi(\theta)}\,.
\label{e2:c}\eqe
Picking a $\theta$, one can then evaluate $c(\theta)$ and solve nonlinear equation \eqref{e2:thdef} for $\psi = \tilde\psi(\theta,t)$ at every $t$.
The resulting angle $\tilde\psi(\theta,t)$ and motion $\tilde\bx(\theta,t)$ are shown in Fig.~\ref{f2:v}b and \ref{f2:flow}c.
As the latter figure shows, the Eulerian motion is normal to the surface, and hence it has no tangential component, i.e.~the Eulerian frame velocity is $\bv_\mre = v_r\,\be_r$.
Due to this, grid points become widely separated at the tip of the bubble, as Fig.~\ref{f2:flow}c shows.
This causes problems in numerical methods as is seen in Sec.~\ref{s:numbubble} below.
Also Lagrangian motion is problematic in numerical methods, as new grid points need to be drawn in at the inflow boundary, as Fig.~\ref{f2:flow}a shows.
On the other hand, ALE motion achieves equidistant grid points during inflation, which allows for superior accuracy in numerical methods (see Sec.~\ref{s:numbubble}), which in turn allows for more accurate studies of evolving surface flows.

Next the velocity gradient and its consequences are examined.
From Eqs.~\eqref{e2:v}, \eqref{e2:dphi}, \eqref{e3:bezeta} and $\psi_{,1} = \psi_\mro$ follows
\eqb{l}
\bv_{\!,1} =  \dot r\,\psi\,\psi_\mro\,\be_r\,.
\label{e2:bv1}\eqe
Further,
\eqb{l}
\dot\ba_1 = \dot r\, \psi\,\psi_\mro\,\be_r + v_\mathrm{in}\,\be_\psi
\eqe
and
\eqb{l}
\dot\zeta_{,1} = -\ds\frac{v_\mathrm{in}}{r\psi_\mro}
\eqe
follow from \eqref{e2:ba1}, \eqref{e2:dphi}, \eqref{e2:vin}, \eqref{e3:bedot} and \eqref{e2:dzeta1}.
On the other hand $\bv_{\!,2}=\dot\ba_2=\mathbf{0}$ and $\dot\zeta_{,2}=0$.
Eq.~\eqref{e:badot} is thus satisfied.
With
\eqb{l}
\ba'_1 = - r\,\psi\,\dot\psi_\mro\,\be_r + v_\mathrm{in}\,\be_\psi \,,
\eqe
$\ba_{1,1}=-r\psi^2_\mro\,\be_r$ and $\ba_2'=\ba_{1,2}=\ba_{2,2}=\mathbf{0}$, one can also confirm that \eqref{e:ALE} is satisfied for $\ba_\alpha$.
This follows from \eqref{e2:phi}, \eqref{e2:ba1}, \eqref{e2:phip}, \eqref{e2:dzeta1}, \eqref{e2:vin}, \eqref{e3:bezeta} and \eqref{e3:bedot}. 

With \eqref{e2:bv1}, $\ba^1 = \be_\psi/(r\psi_\mro)$ and \eqref{e:sdiv}, one can then confirm that the flow satisfies \eqref{e:divv}, i.e.~it is area-incompressible, and one can find the rate-of-deformation tensor
\eqb{l}
2\bd = -\dot\psi\,\big( \be_r\otimes\be_\psi + \be_\psi\otimes\be_r \big)
\eqe
according to \eqref{e:bd} and \eqref{e2:dphi},
which implies $\bd_\mrs = \mathbf{0}$.\footnote{Tangential velocity fields that lead to $\bd_\mrs = \mathbf{0}$ are sometimes referred to \textit{Killing vector fields}, e.g.~see 
\citet{jankuhn18}.} 
So the stress according to constitutive model \eqref{e:bsig} will only consist of the surface tension $\gamma$ that equilibrates the internal pressure $p$ through the well known relation $p=\gamma/r$.
Thus, if inertia is negligible and no additional surface forces act, the soap bubble remains perfectly circular during inflation.

\subsubsection{Inflation control}

\begin{figure}[h]
\begin{center} \unitlength1cm
\begin{picture}(0,5.6)
\put(-7.95,-.25){\includegraphics[height=58mm]{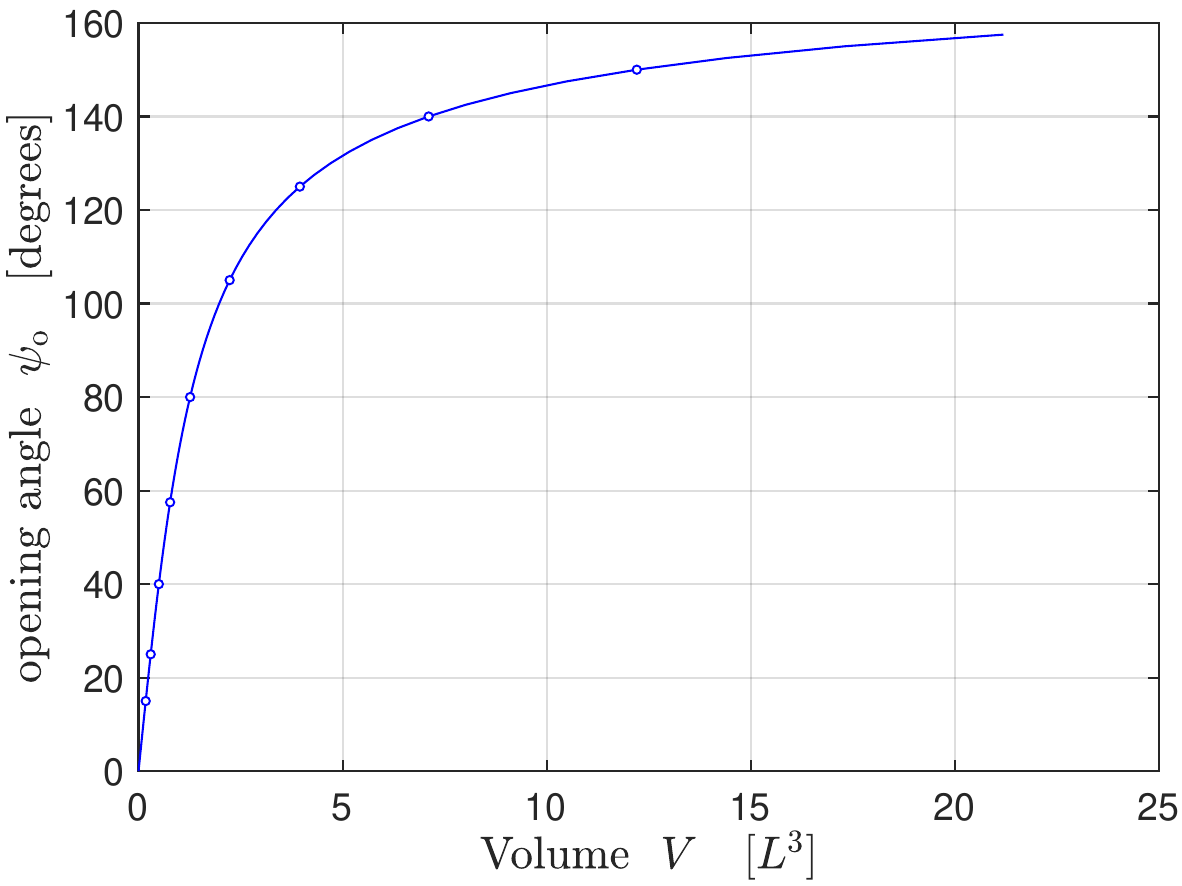}}
\put(0.25,-.25){\includegraphics[height=58mm]{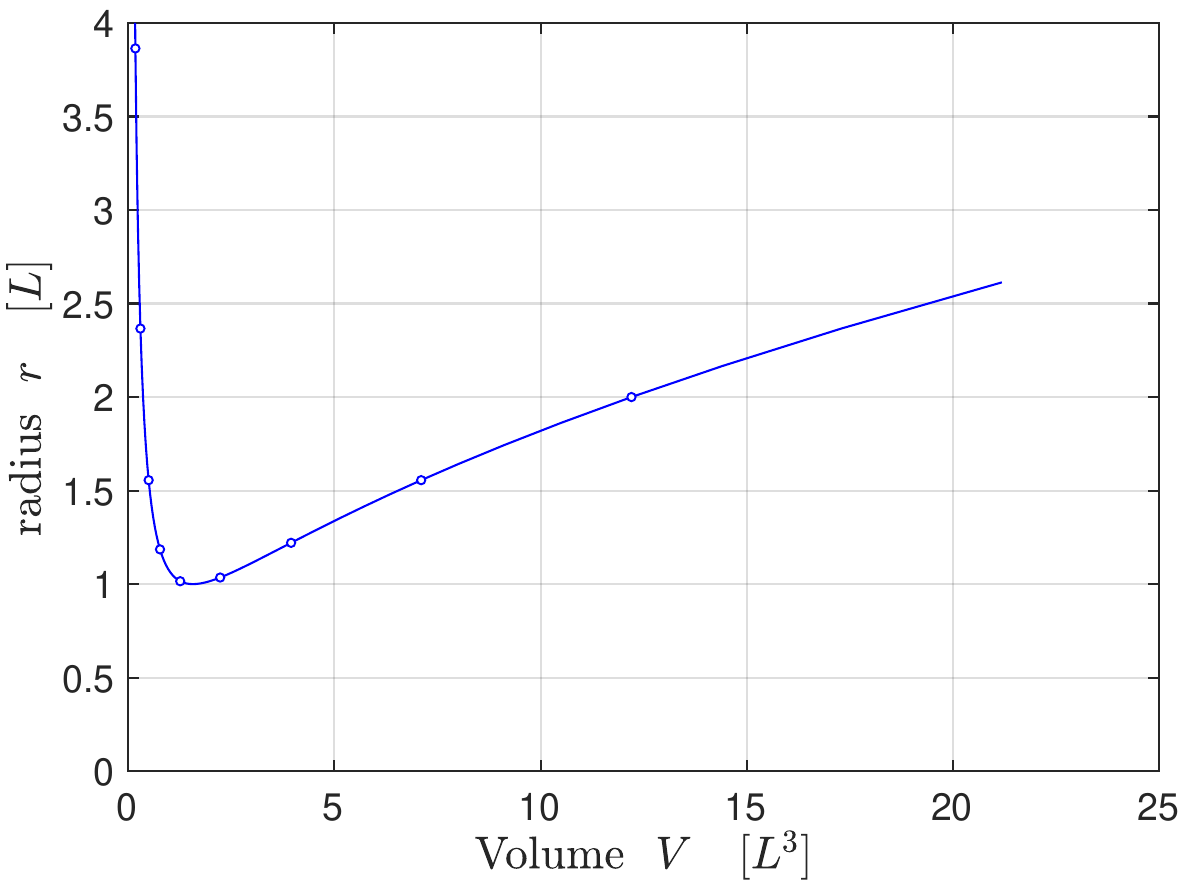}}
\put(-7.95,-.1){\footnotesize (a)}
\put(0.25,-.1){\footnotesize (b)}
\end{picture}
\caption{Inflation of a 2D soap bubble: 
(a) $\psi_\mro(V)$ and (b) $r(\psi_\mro)$ vs.~$V(\psi_\mro)$.
The circles mark the configurations shown in Fig.~\ref{f2:flow}.}
\label{f2:V}
\end{center}
\end{figure}

The inflation can be controlled in various ways.
One can prescribe inflow velocity $v_\mathrm{in}$, giving the opening angle change
\eqb{l}
\dot\psi_\mro = \ds\frac{\sin\psi_\mro}{r\,(\sin\psi_\mro - \psi_\mro\,\cos\psi_\mro)}\,v_\mathrm{in}
\label{e2:dphi0-v}\eqe
from \eqref{e2:vin} and \eqref{e2:dr}.
This then specifies the bubble motion according to the equations above.
Prescribing $v_\mathrm{in}$ corresponds to pushing material into the bubble, which may not be a stable way to inflate the bubble numerically.
A stable alternative is to prescribe the bubble volume\footnote{per unit depth $L$} $V$ since this pulls, instead of pushes, material into the bubble.
V(t) is related to the opening angle $\psi_\mro$ via
\eqb{l}
V = r^2\,(\psi_\mro - \sin\psi_\mro\,\cos\psi_\mro)\,,
\eqe
according to Fig.~\ref{f2:setup}b.
Taking a time derivative of this and using \eqref{e2:dr} then gives
\eqb{l}
\dot\psi_\mro = \ds\frac{\sin\psi_\mro}{2r^2\,(\sin\psi_\mro-\psi_\mro\,\cos\psi_\mro)}\,\dot V\,.
\label{e2:dphi0-V}\eqe
Comparing \eqref{e2:dphi0-v} and \eqref{e2:dphi0-V} shows that volume and inflow velocity are related by $\dot V=2r\,v_\mathrm{in}$.
Fig.~\ref{f2:V} shows the change $\psi_\mro$ and $r$ due to given $V$. 
%-----------------------------------------------------------------
% run soapbubble2D.m OR pSoapStrip
%-----------------------------------------------------------------
\\
Instead of the volume, the internal pressure $p(t)$ can also be prescribed.
However, in contrast to $V(t)$, $p(t)$ generally cannot be chosen as a monotonically increasing function.
This is due to the minimum attained by $r$ during the inflation process (see Fig.~\ref{f2:V}), which leads to a maximum in $p$ for constant surface tension $\gamma$, due to the relation $\gamma = r\,p$.
Prescribing $p$ numerically therefore can also become unstable.

\subsubsection{Numerical solution}\label{s:numbubble}

It is finally shown that the equidistant grid spacing of the ALE description leads to much more accurate numerical results than the unequal grid spacing of the Eulerian description.
Therefore the 2D soap bubble is solved with the finite element (FE) method using either Eulerian mesh motion (described by weak form \eqref{e:WFm}) or elastic mesh motion (described by weak form \eqref{e:WFme}).
The latter leads to an equidistant FE mesh for this problem.
Quadratic Lagrange interpolation is used for all fields as described in Appendix \ref{s:FE}.
The physical soap bubble properties in the simulation are taken as $L = 10$mm, $\gamma = 25\mu$N/mm, $\eta = 0.04\mu$N ms/mm and $\rho = 0.2\mu$Nms$^2$/mm$^3$, while the mesh stiffness is taken as $\mu_\mrm = 0.5$nN/mm$^2$.

\begin{figure}[h]
\begin{center} \unitlength1cm
\begin{picture}(0,5.6)
\put(-7.95,-.25){\includegraphics[height=58mm]{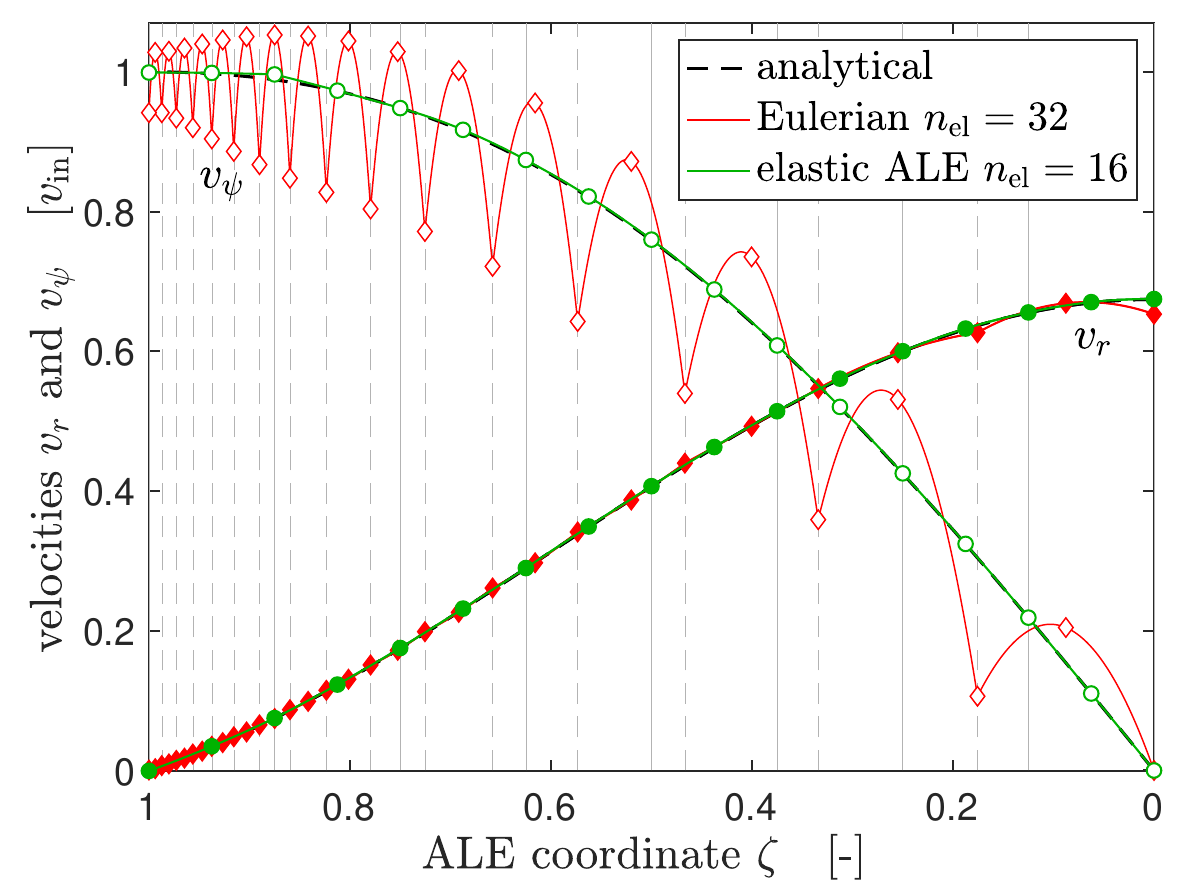}}
\put(0.25,-.25){\includegraphics[height=58mm]{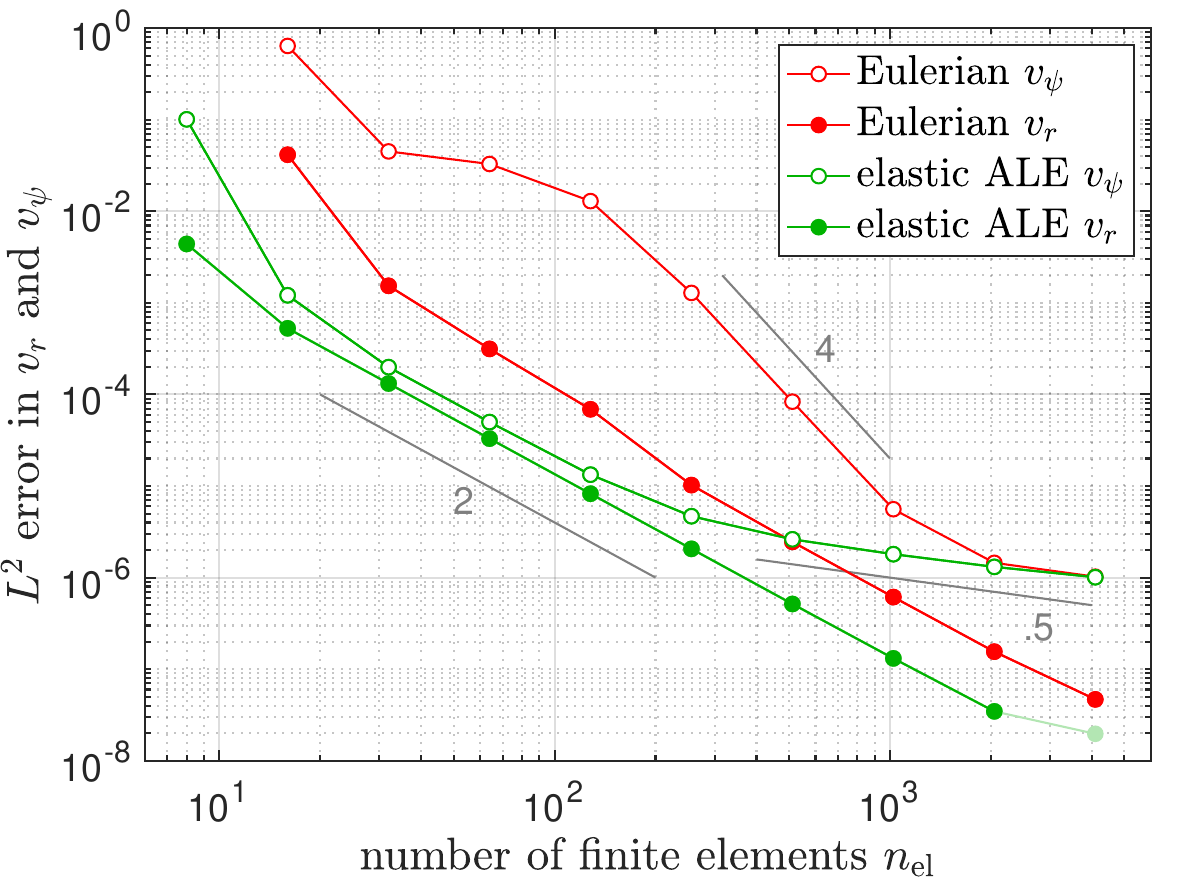}}
\put(-7.95,-.1){\footnotesize (a)}
\put(0.25,-.1){\footnotesize (b)}
\end{picture}
\caption{Inflation of a 2D soap bubble: 
(a) Finite element (FE) result and (b) FE convergence of velocity components $v_r$ and $v_\psi$ for Eulerian and elastic ALE mesh motion. 
The latter is around 10 times more accurate in $v_r$ and over 100 times more accurate in $v_\psi$ for coarse meshes. 
The vertical grey lines in (a) show the FE boundaries: dashed for the Eulerian mesh with 32~FE, solid for the ALE mesh with 16~FE (which is exactly the mesh in Fig.~\ref{f2:flow}b).
The initial FE convergence rate is at least~$O(n_\mathrm{el}^{-2}) = O(h^2)$.}
\label{f2:FE}
\end{center}
\end{figure}

Fig.~\ref{f2:FE}a shows the FE results for velocity components $v_r$ and $v_\psi$ in comparison to the analytical result of Eqs.~\eqref{e2:vr} and \eqref{e2:vp}.  
%-----------------------------------------------------------------
% run pSoapStrip & pSoapStripL2 
%-----------------------------------------------------------------
As seen, the elastic ALE result is much more accurate, even for a coarser mesh.
This can also be seen in the error plot of Fig.~\ref{f2:FE}b:
The ALE result is around 10 times more accurate for $v_r$ and more than 100 times more accurate for $v_\psi$ for coarse meshes.
The difference decreases for finer meshes, but it remains positive.
Interesting, the convergence rate for the tangential velocity $v_\psi$ decreases for finer meshes, which calls for further study.
Also interesting is that the convergence behavior of $v_\psi$ in the Eulerian case is reminiscent of membrane locking in thin shells \citep{B2M1}.

\subsection{Shear flow on a sphere}\label{s:shear}

The third example considers simple shear flow on a sphere.
The example provides insight into the coupling of in-plane surface flow with out-of-plane surface forces and deformations.
The latter remain small here and hence the ALE mesh motion can be set to zero here, i.e.~the ALE frame coincides with an in-plane Eulerian one.
It is further shown that this is a much better way to treat the example than using a Lagrangian frame.
\\[-6mm] %%%

\subsubsection{Surface description}

The spherical surface can be described by
\eqb{l}
\bx = r\,\be_r\,,
\label{e4:bx}\eqe
where radius $r$ is now considered fixed and the direction vectors
\eqb{lllll}
\be_r \dis \cos\zeta^2\,\be_{r_0} \plus \sin\zeta^2\,\be_3\,,\quad -\frac{\pi}{2}\leq\zeta^2\leq\frac{\pi}{2}\,, \\[1mm]
\be_{r_0} \dis \cos\zeta^1\,\be_1 \plus \sin\zeta^1\,\be_2\,,\quad 0\leq\zeta^1\leq 2\pi\,,
\label{e4:bero}\eqe
generally depend on $\zeta^\alpha = \zeta^\alpha(\xi^\beta,t)$, see Fig.~\ref{f:basis3} in Appendix \ref{s:basis}.
Here $\{\be_1,\be_2,\be_3\}$ is the basis of a fixed Cartesian reference coordinate system.
Before specifying $\zeta^\alpha(\xi^\beta,t)$, which defines the material flow relative to the ALE frame (= Eulerian frame here), one can already analyze the geometry.
Introducing the local orthonormal basis $\{\be_r,\be_\phi,\be_\theta\}$ with
\eqb{llllllll}
\be_\phi \dis -\sin\phi\,\be_1 \plus \cos\phi\,\be_2\,, &  \phi \dis \zeta^1\,, \\[1mm]
\be_\theta \dis -\sin\theta\,\be_{r_0} \plus \cos\theta\,\be_3\,, & \theta \dis \zeta^2\,,
\label{e4:bezeta}\eqe
and using Eq.~\eqref{e:bezeta} from Appendix \ref{s:basis}, one finds the tangent vectors
\eqb{l}
\ba_1 = r\,\cos\theta\,\be_\phi\,,\quad
\ba_2 = r\,\be_\theta\,,
\label{e4:baa}\eqe
surface metric
\eqb{l}
\big[a_{\alpha\gamma}\big] = r^2\left[\begin{array}{cc}
\cos^2\theta & 0 \\
0 & 1
\end{array}\right]\,,
\eqe
dual tangent vectors
\eqb{l}
\ba^1 = \ds\frac{1}{r\,\cos\theta}\,\be_\phi\,,\quad
\ba^2 = \frac{1}{r}\,\be_\theta\,,
\label{e4:bad}\eqe
surface normal
\eqb{l}
\bn = \be_r\,,
\eqe
and curvature components
\eqb{l}
\big[b_{\alpha\gamma}\big] = -r\left[\begin{array}{cc}
\cos^2\theta & 0 \\
0 & 1
\end{array}\right]
\label{e4:bab}\eqe
in the ALE frame.
Consequently, $H = -1/r$ and $\kappa = 1/r^2$.
Further, \eqref{e4:bx} implies $\bv_\mrm = \mathbf{0}$.

\subsubsection{Surface flow}

\begin{figure}[h]
\begin{center} \unitlength1cm
\begin{picture}(0,10.45)
\put(0.87,5.3){\includegraphics[height=53mm]{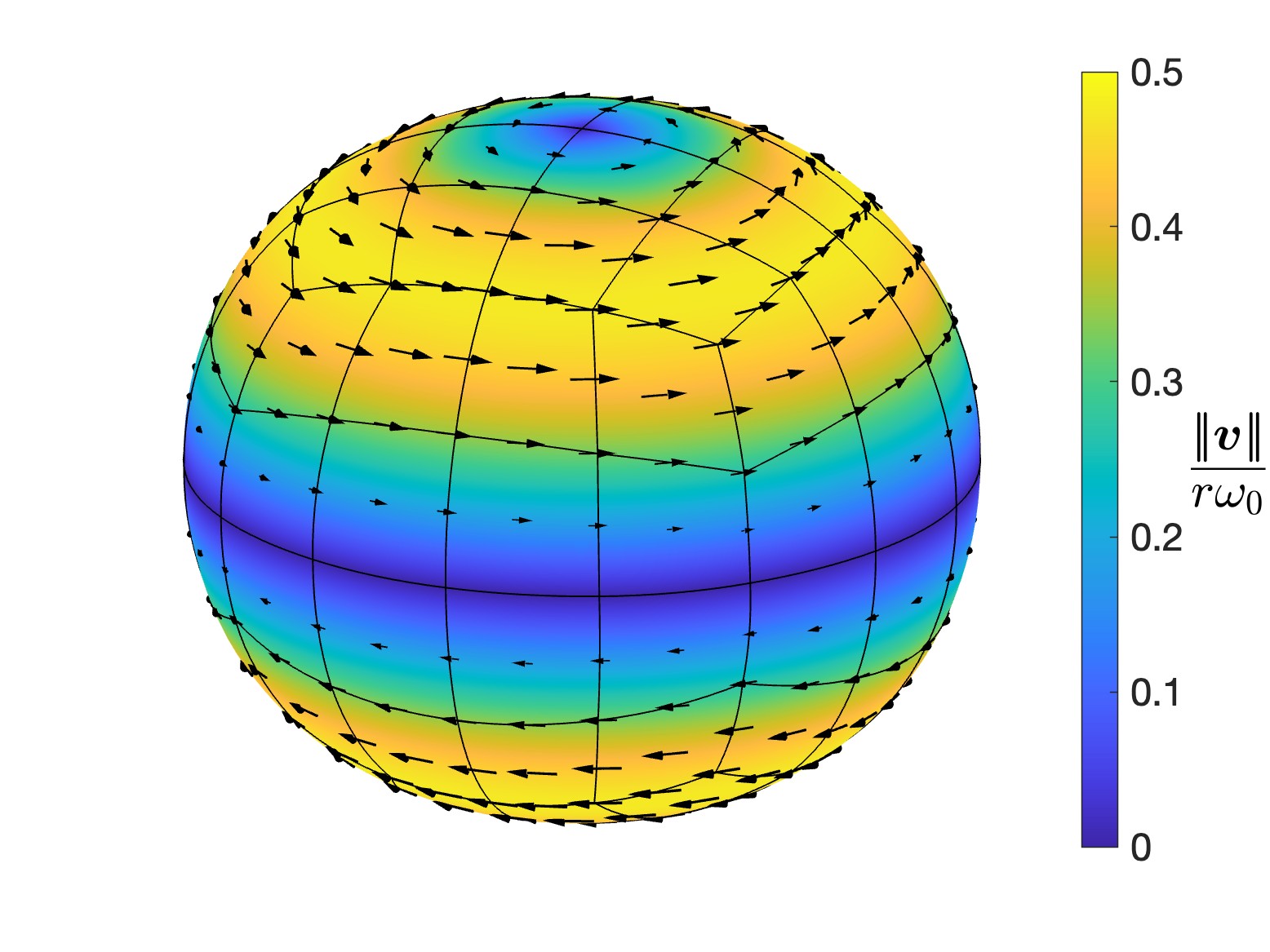}}
\put(-8.16,5.3){\includegraphics[height=53mm]{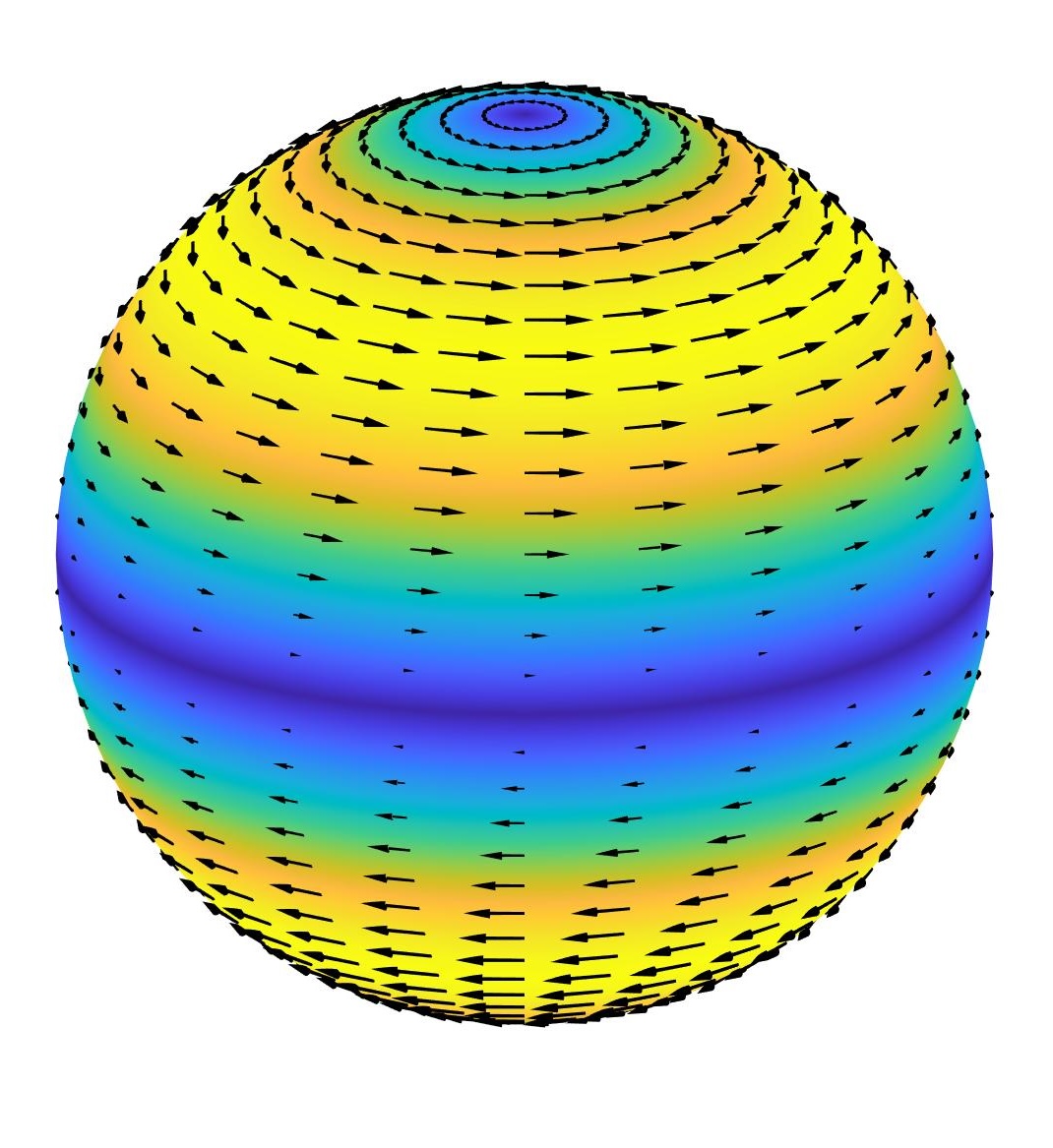}}
\put(-3.26,5.3){\includegraphics[height=53mm]{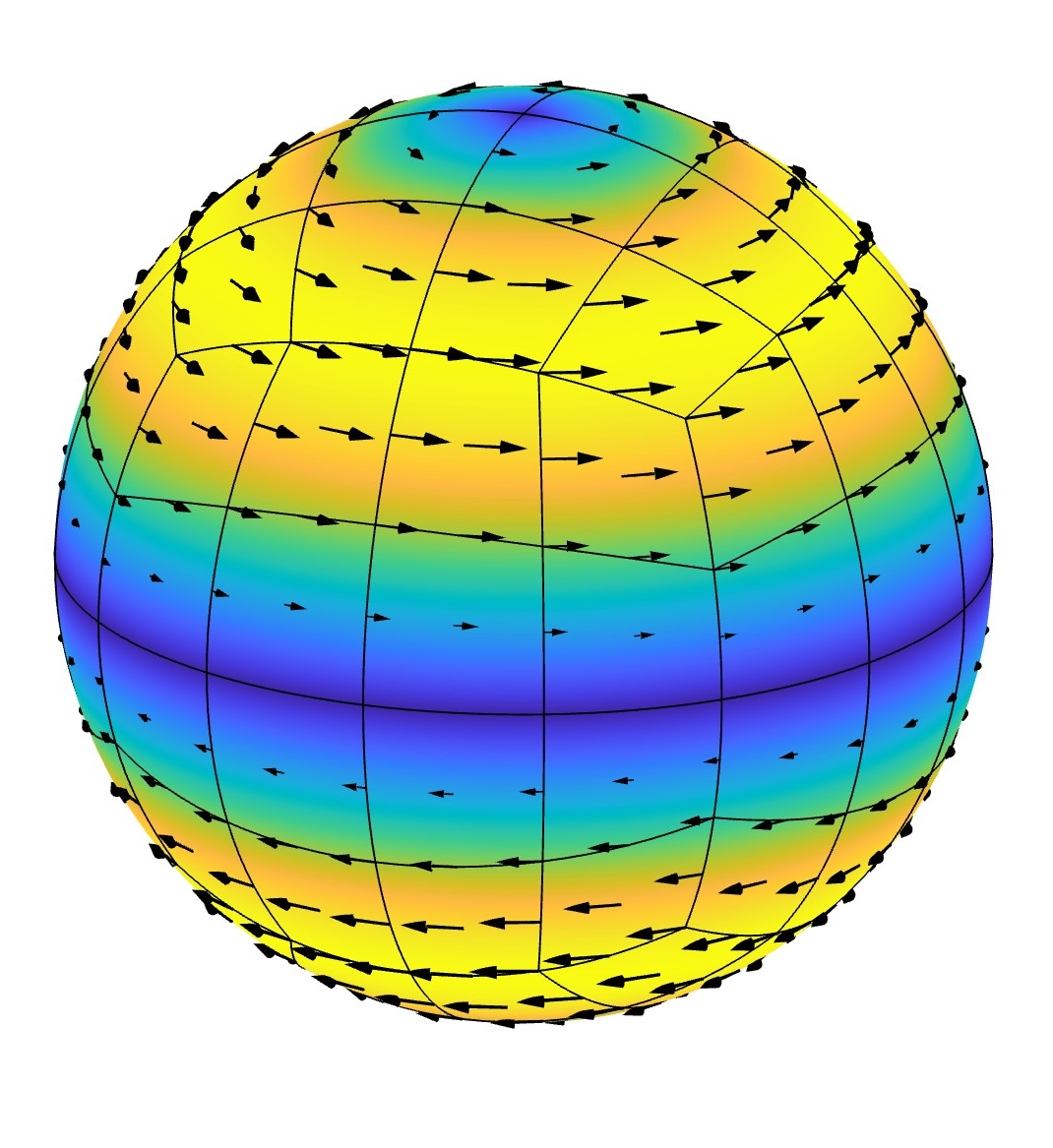}}
\put(-7.9,-.2){\includegraphics[height=57.5mm]{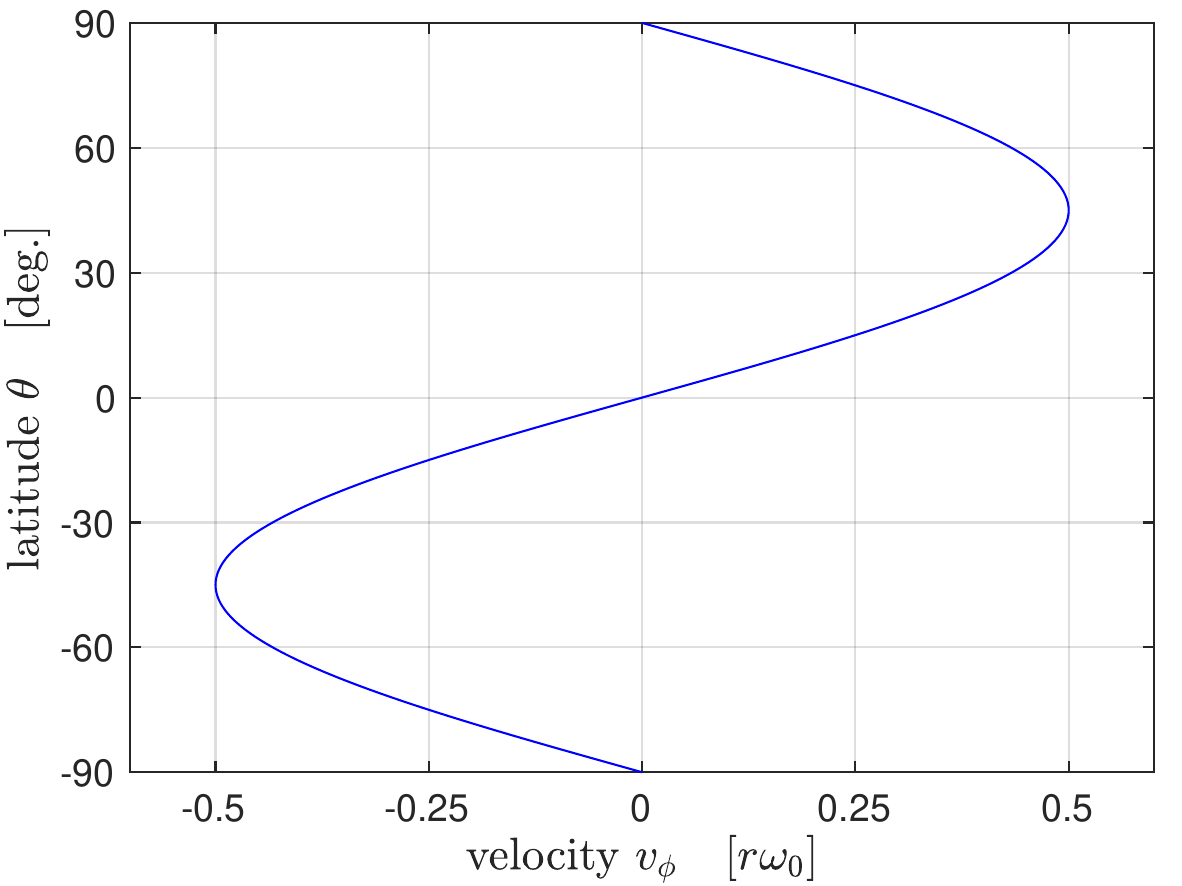}}
\put(0.2,-.2){\includegraphics[height=58mm]{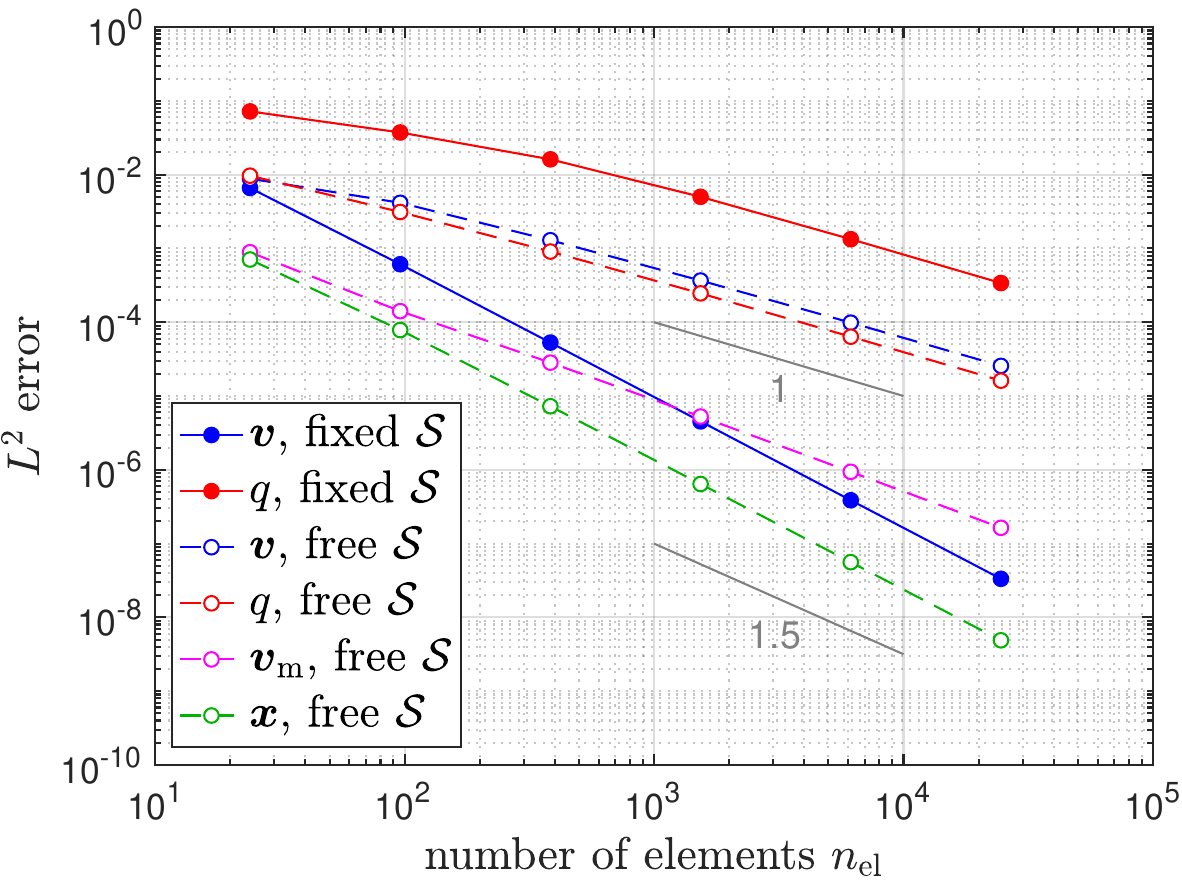}}
\put(-7.95,5.9){\footnotesize (a)}
\put(-3.05,5.9){\footnotesize (b)}
\put(1.85,5.9){\footnotesize (c)}
\put(-7.7,-.1){\footnotesize (d)}
\put(0.25,-.1){\footnotesize (e)}
\end{picture}
\caption{Shear flow on a sphere: 
(a) analytical and (b) numerical flow field $\bv$ of a fixed surface, and (c) numerical flow field of a freely evolving surface;  
(d) analytical velocity profile $v_\phi(\theta)$; (e) finite element error for fluid velocity $\bv$, surface tension $q$, mesh velocity $\bv_\mrm$ and surface position $\bx$, converging at optimal rates:
$O(n^{-1.5}_\mathrm{el})=O(h^3)$ for fixed surface $\bv$ and free surface $\bx$, and $O(n^{-1}_\mathrm{el})=O(h^2)$ otherwise.}
\label{f:shearflow}
\end{center}
\end{figure}

Now consider the shear flow defined by
\eqb{lll}
\zeta^1 \is \theta^1 = \xi^1 + \omega_0 t \sin\xi^2 \,, \\[1mm]
\zeta^2 \is \theta^2 = \xi^2 =: \theta\,,
\label{e4:zeta}\eqe
which leads to 
\eqb{lll}
\dot\zeta^1 \is \omega_0 \sin\theta \,, \\[1mm]
\dot\zeta^2 \is 0\,,
\eqe
\eqb{l}
\bigg[\ds\pa{\zeta^\alpha}{\xi^\gamma}\bigg] = \left[\begin{array}{cc}
1 & \omega t\cos\theta \\
0 & 1
\end{array}\right]
\eqe
and
\eqb{l}
\big[\dot\zeta^\alpha_{,\gamma}\big] = \left[\begin{array}{cc}
0 & \omega_0\cos\theta \\
0 & 0
\end{array}\right],
\eqe
for constant angular velocity $\omega_0$.
Using Eq.~\eqref{e:bedot} from Appendix \ref{s:basis}, then leads to the frame velocities
\eqb{l}
\dot\ba_1 = -r\omega_0\,\sin\theta\cos\theta\,\be_{r_0}\,,\quad
\dot\ba_2 = -r\omega_0\sin^2\theta\,\be_\phi\,,
\label{e4:badot}\eqe
the material velocity
\eqb{l}
\bv = r\omega_0\sin\theta\cos\theta\,\be_\phi
\label{e4:bv}\eqe
and the material acceleration 
\eqb{l}
\dot\bv = -r\omega_0^2\sin^2\theta\cos\theta\,\be_{r_0}\,. 
\eqe
From $\bv$ and \eqref{e:bezeta} follow
\eqb{lll}
\bv_{\!,1} \is -r\omega_0\sin\theta\cos\theta\,\be_{r_0}\,, \\[1mm]
\bv_{\!,2} \is r\omega_0\,(2\cos^2\theta-1)\,\be_\phi\,,
\label{e4:bvzeta}\eqe
which can also be obtained equivalently from \eqref{e:badot}.
From the preceding equations follows, that the ALE equation \eqref{e:vALEa} decomposes into the separate equations
\eqb{l}
\dot\bv = \bv_{\!,\alpha}\,v^\alpha\,,\quad
\bv' = \bv_{\!,\alpha}\,v^\alpha_\mrm
\eqe
for this example.
The velocity field and its nonzero component $v_\phi := \bv\cdot\be_\phi$ are visualized in Fig.~\ref{f:shearflow}
%-----------------------------------------------------------------
% run sphereflow2.m & sphereflow.m
% run sShearFlow/ vShearFlowNS/ mat{1}.case = 1 & pShearFlow.m/ jALE = 0, jcase = 1 
% run sShearEvolve/ vShearEvolveNS2 & pShearEvolve2
% axis(1.01*[-1 1 -1 1 -1 1]); zoom(1.395) ; 1375 - 1070
% run pShearEvolve
%-----------------------------------------------------------------
and compared to the numerical solution following in Sec.~\ref{s:numshear}.
The velocity field leads to the surface vorticity
\eqb{l}
\omega = \omega_0\,(2\sin^2\theta-\cos^2\theta) 
\eqe
according to definition \eqref{e:vorticity} and relation (\ref{e4:be3}.1).

\subsubsection{Surface forces}

\begin{figure}[h]
\begin{center} \unitlength1cm
\begin{picture}(0,5.6)
\put(-7.9,-.2){\includegraphics[height=58mm]{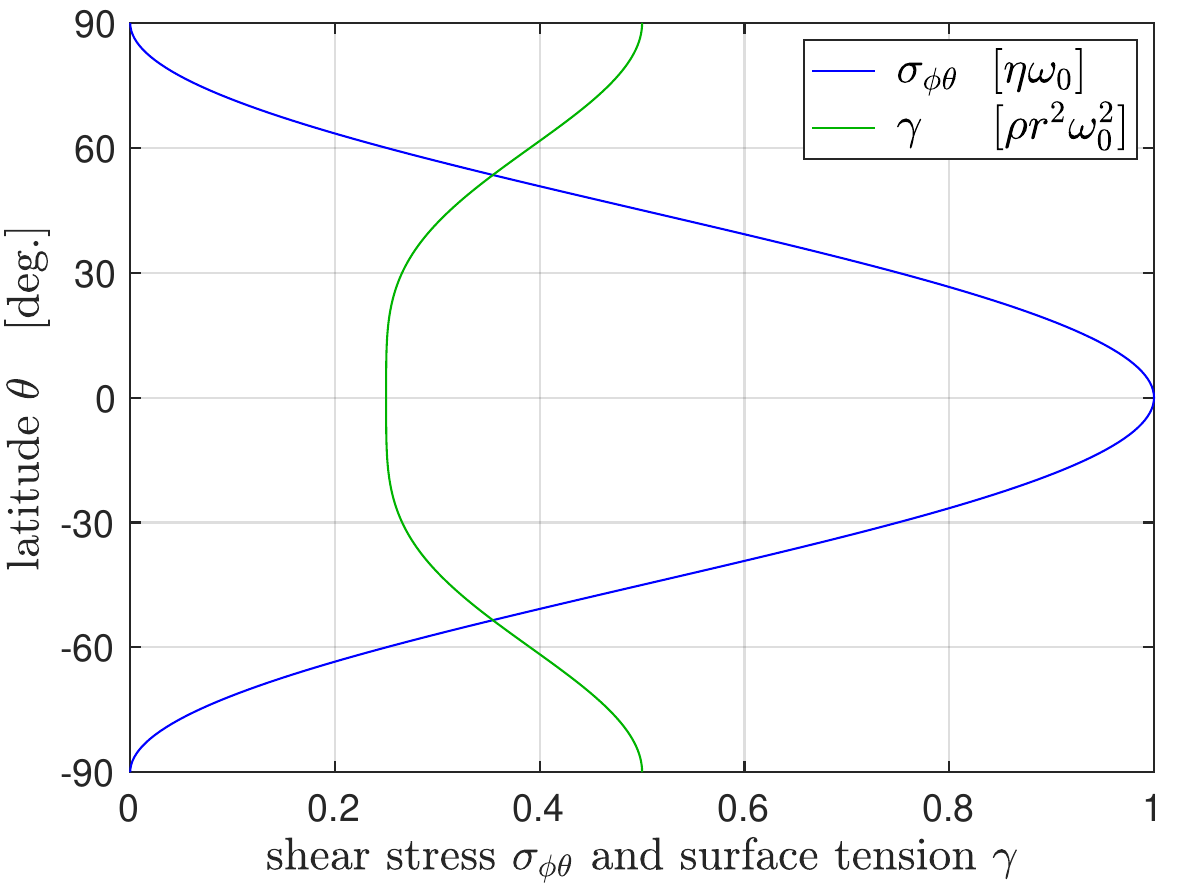}}
\put(0.3,-.2){\includegraphics[height=58mm]{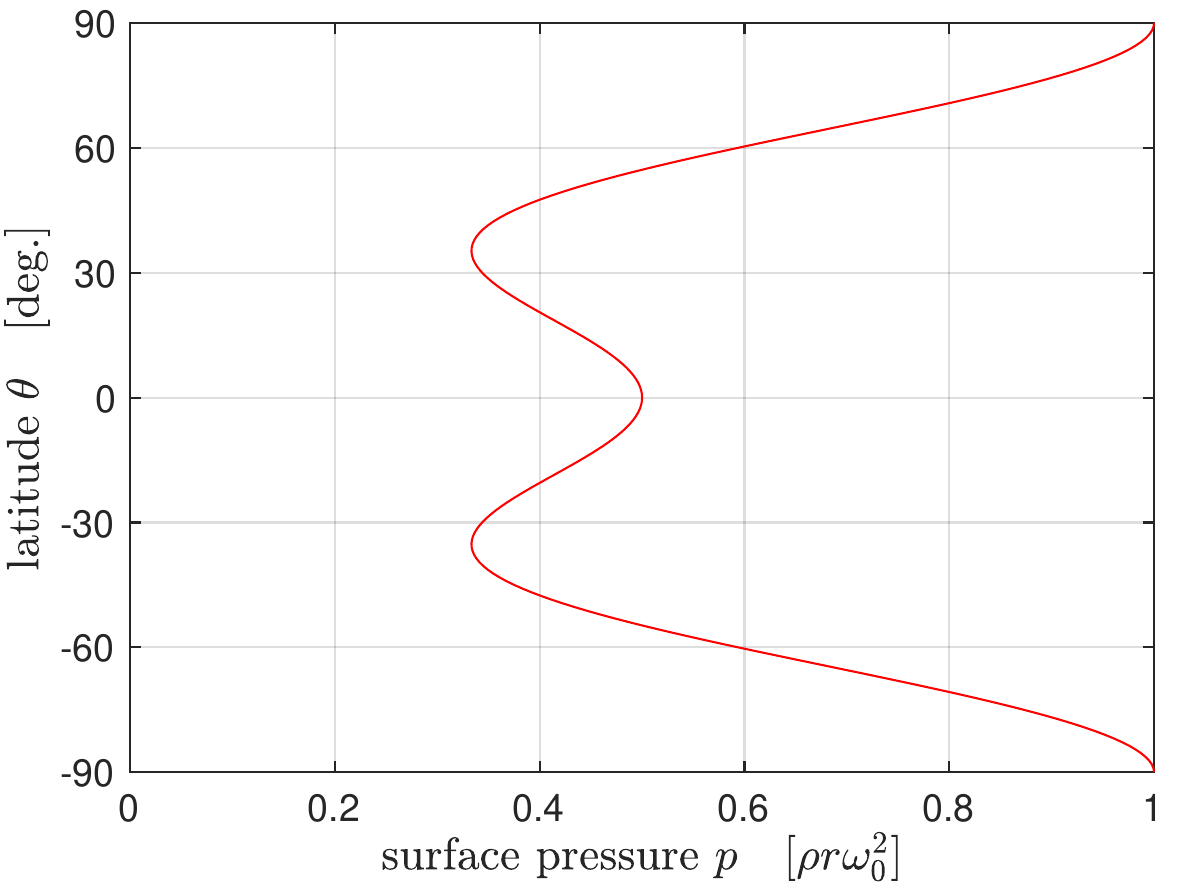}}
\put(-7.9,-.1){\footnotesize (a)}
\put(0.3,-.1){\footnotesize (b)}
\end{picture}
\caption{Shear flow on a rigid sphere: (a) shear stress $\sigma_{\phi\theta}(\theta)$ and surface tension $\gamma(\theta)$, and (b) surface pressure $p(\theta)$ for Case 1 considering $\gamma_0 = \rho r^2\omega_0^2/4$.}
\label{f:shearflow2}
\end{center}
\end{figure}

From \eqref{e:bd}, \eqref{e4:bad}, \eqref{e4:bvzeta} and (\ref{e4:be3}.1) now follows the rate-of-deformation tensor
\eqb{l}
2\bd = \omega_0\cos^2\theta\,\big(\be_\phi\otimes\be_\theta + \be_\theta\otimes\be_\phi\big)
-\omega_0\sin\theta\cos\theta\,\big(\be_\phi\otimes\be_r + \be_r\otimes\be_\phi\big)\,,
\label{e4:bd}\eqe
which has the in-plane part
\eqb{l}
2\bd_\mrs = \omega_0\cos^2\theta\,\big(\be_\phi\otimes\be_\theta + \be_\theta\otimes\be_\phi\big)
\eqe
and results in the stress
\eqb{l}
\bsig = \gamma\,\bi + \eta\,\omega_0\cos^2\theta\,\big(\be_\phi\otimes\be_\theta + \be_\theta\otimes\be_\phi\big)\,,
\eqe
according to constitutive model \eqref{e:bsig} (that is valid for both the material models in Sec.~\ref{s:consti}).
From \eqref{e:Taa1} then follows
\eqb{l}
\bT^\alpha_{;\alpha} = \big(\gamma_{,1} - 4\eta\,\omega_0\sin\theta\cos^2\theta \big)\,\ba^1 + \gamma_{,2}\,\ba^2 - \ds\frac{2\gamma}{r}\,\bn\,,
\eqe
since $\divs\bd_\mrs = \bd_{\mrs,\alpha}\,\ba^\alpha$ with
\eqb{l}
\bd_{\mrs,1}\,\ba^1 = \bd_{\mrs,2}\,\ba^2 = -\omega_0\,\sin\theta\cos^2\theta\,\ba^1
\eqe
in this example.
Due to rotational symmetry $\gamma_{,1} = 0$.
The equation of motion \eqref{e:Taa} is then satisfied for the body force components 
\eqb{lll}
f_1 \is 4\eta\,\omega_0\sin\theta\cos^2\theta\,,
\\[3mm]
f_2 \is \rho\,r^2\,\omega_0^2\,\sin^3\theta\cos\theta - \gamma_{,2}\,, \\[1mm]
p \is \ds\frac{2\gamma}{r} - \rho\,r\,\omega_0^2\,\sin^2\theta\cos^2\theta\,,
\label{e4:manfac}\eqe
that admit %
the two special cases:\\
\underline{Case 1:} $f_2 = 0$. Then (\ref{e4:manfac}.2) can be integrated to yield
\eqb{l}
\gamma = \ds\frac{\rho\,r^2\,\omega_0^2}{4}\sin^4\theta + \gamma_0\,.
\label{e4:manfac1}\eqe
In this case, we can also write $\bff = \eta_\mrs\,\bv + p\,\bn$, where $\eta_\mrs := 4\eta/r^2$ can be interpreted as a viscous surface friction coefficient.
\\
\underline{Case 2:} $\gamma = \gamma_0 = $ const. Thus $\gamma_{,2} = 0$ and we can also write $\bff = \eta_\mrs\,\bv + \rho\,\dot\bv + p_0\,\bn$, where 
$p_0 = 2\gamma_0/r$.
\\
Fig.~\ref{f:shearflow2} shows the stress components $\sigma_{\phi\theta} = \be_\phi\cdot\bsig\be_\theta$ and $\gamma$, as well as the surface pressure $p$ for Case 1 considering $\gamma_0 = \rho r^2\omega_0^2/4$.
%-----------------------------------------------------------------
% run sphereflow.m
%-----------------------------------------------------------------
Eq.~\eqref{e4:manfac} and its two cases provide two analytical solutions for the steady flow in Eq.~\eqref{e4:bv}.
It is emphasized that this solution contains out-of-plane body forces. 
Only the precise pressure distribution of Eq.~(\ref{e4:manfac}.3) keeps the surface in spherical shape. 
Other distributions will lead to shape changes, as the following subsection shows.

\subsubsection{Numerical solution}\label{s:numshear}

The flow example is solved with the finite element formulation based on weak form \eqref{e:WFm} using the discretization described in Appendix~\ref{s:FE}.
In the FE solution the surface can either be fixed, or left free to deform so that $\bv$ and $\bv_\mrm$ are allowed to have out-of-plane components.
The flow field for 96 quadratic FE is shown in Fig.~\ref{f:shearflow} for two cases:
Fig.~\ref{f:shearflow}b shows the FE solution when the varying body force field from \eqref{e4:manfac}-\eqref{e4:manfac1} is prescribed, in which case the surface remains spherical, as this is the equilibrium shape.
Fig.~\ref{f:shearflow}c, on the other hand, shows the FE solution when the interior pressure is set to constant $p = \rho\,r\,\omega_0^2$.
Now the perfect sphere is no longer in equilibrium and deforms into the new equilibrium shape seen in the figure, which has smaller volume but the same surface area -- it shrinks by $9.41\%$ vertically and expands by $2.44$\% around the equator.
This flow-induced deformation is well into the geometrically nonlinear regime.
In the first case the FE convergence behavior can be assessed, as the analytical solution from above is valid.
The convergence rates are $O(n_\mathrm{el}^{-1.5}) = O(h^3)$ for $\bx$ and fixed surface $\bv$, and $O(n_\mathrm{el}^{-1}) = O(h^2)$ for all other $\bv$, $q$ and $\bv_\mrm$, as Fig.~\ref{f:shearflow}c shows.
This is consistent with the optimal rates for nonlinear membranes and incompressible flows \citep{wriggers-fe,dohrmann04}.
For a free surface, $\bv$, $q$ and $\bv_\mrm$ depend on the surface discretization in $\bx$, and hence converge with one order lower.

\subsubsection{Lagrangian parametrization}

Alternatively (but not advantageously), the system can be described in the material frame.
Using \eqref{e:bexi}, one finds
\eqb{l}
\ba_{\hat1} = \ba_1\,,\quad
\ba_{\hat2} = \omega_0 t\cos\theta\,\ba_1 + \ba_2
\label{e4:bahat}\eqe
and
\eqb{l}
\big[a_{\hat\alpha\hat\gamma}\big] = r^2\left[\begin{array}{cc}
\cos^2\theta & \omega_0 t\cos^3\theta \\
\omega_0 t\cos^3\theta & 1 + \omega_0^2 t^2 \cos^4\theta
\end{array}\right]\,.
\eqe
From this follows
\eqb{l}
\big[\dot a_{\hat\alpha\hat\gamma}\big] = r^2\left[\begin{array}{cc}
0 & \omega_0\cos^3\theta \\
\omega_0\cos^3\theta & 2\omega_0^2 t \cos^4\theta
\end{array}\right]
\eqe
and
\eqb{l}
\big[a^{\hat\alpha\hat\gamma}\big] = \ds\frac{1}{r^2}\left[\begin{array}{cc}
\cos^{-2}\theta + \omega_0^2 t^2 \cos^2\theta & -\omega_0 t\cos\theta \\
-\omega_0 t\cos\theta & 1
\end{array}\right]\,,
\eqe
so that
\eqb{l}
\ba^{\hat1} = \ds\frac{1}{r\cos\theta}\be_\phi - \frac{\omega_0 t \cos\theta}{r}\,\be_\theta\,,\quad
\ba^{\hat2} = \ba^2\,.
\label{e4:badhat}\eqe
From \eqref{e4:bahat} and \eqref{e4:badot} then follow
\eqb{lll}
\dot\ba_{\hat1} \is -r\omega_0\sin\theta\cos\theta\,\be_{r_0}\,,\\[1mm]
\dot\ba_{\hat2} \is r\omega_0 \big(\cos^2\theta-\sin^2\theta\big)\,\be_\phi -r\omega_0^2t\sin\theta\cos^2\theta\,\be_{r_0}\,.
\eqe
Together with \eqref{e:sgrad}, \eqref{e:bahdot} and \eqref{e4:badhat} this leads to the same $\bd$ as in \eqref{e4:bd}, as it is supposed to.
This shows that for shear flows the Lagrangian parameterization, although physically equivalent, is mathematically much more complicating than the ALE one.

\subsection{Spinning sphere}\label{s:spin}

\begin{figure}[h]
\begin{center} \unitlength1cm
\begin{picture}(0,5.7)
\put(-8.8,-.12){\includegraphics[height=60mm]{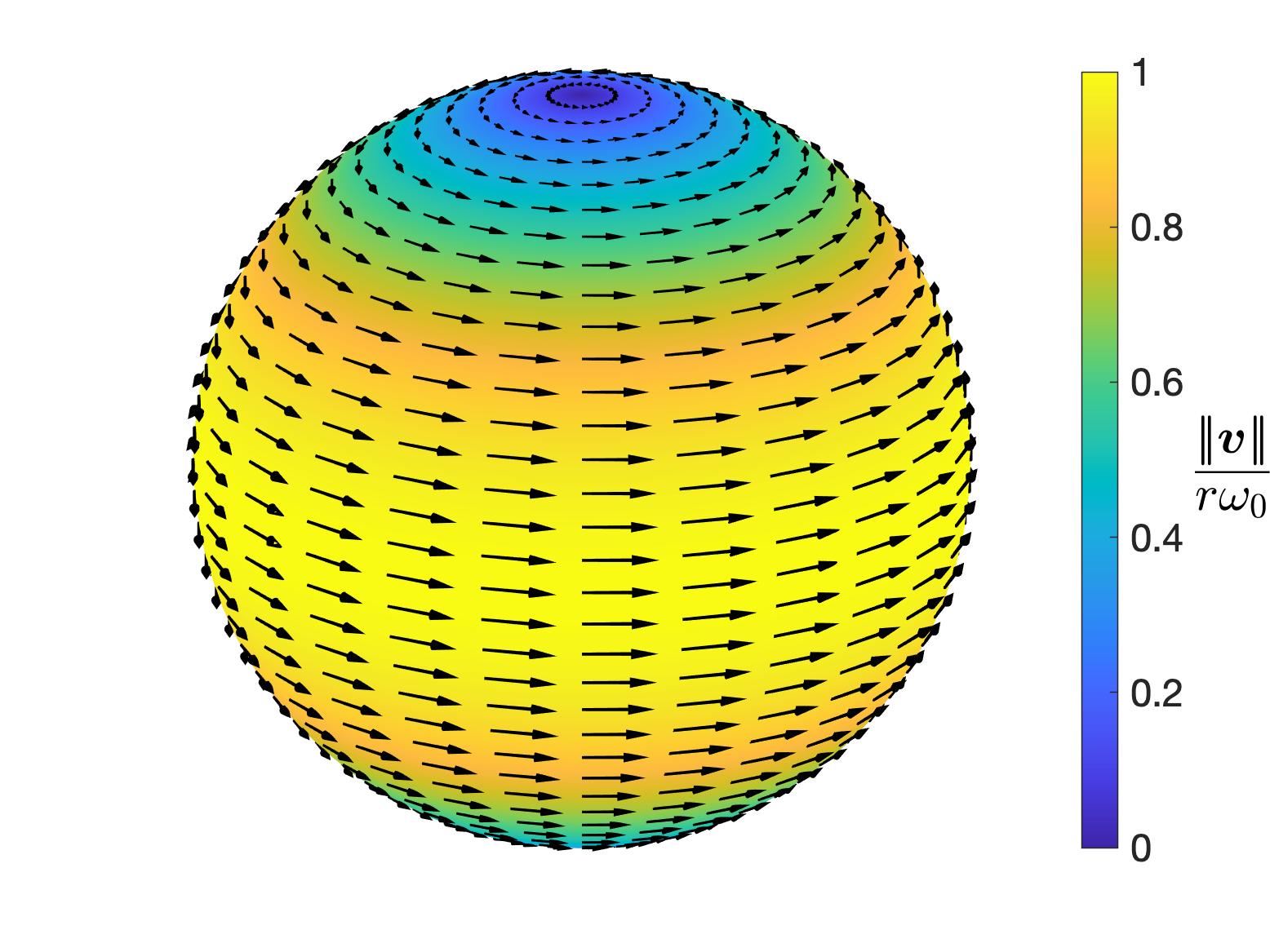}}
\put(0.3,-.2){\includegraphics[height=58mm]{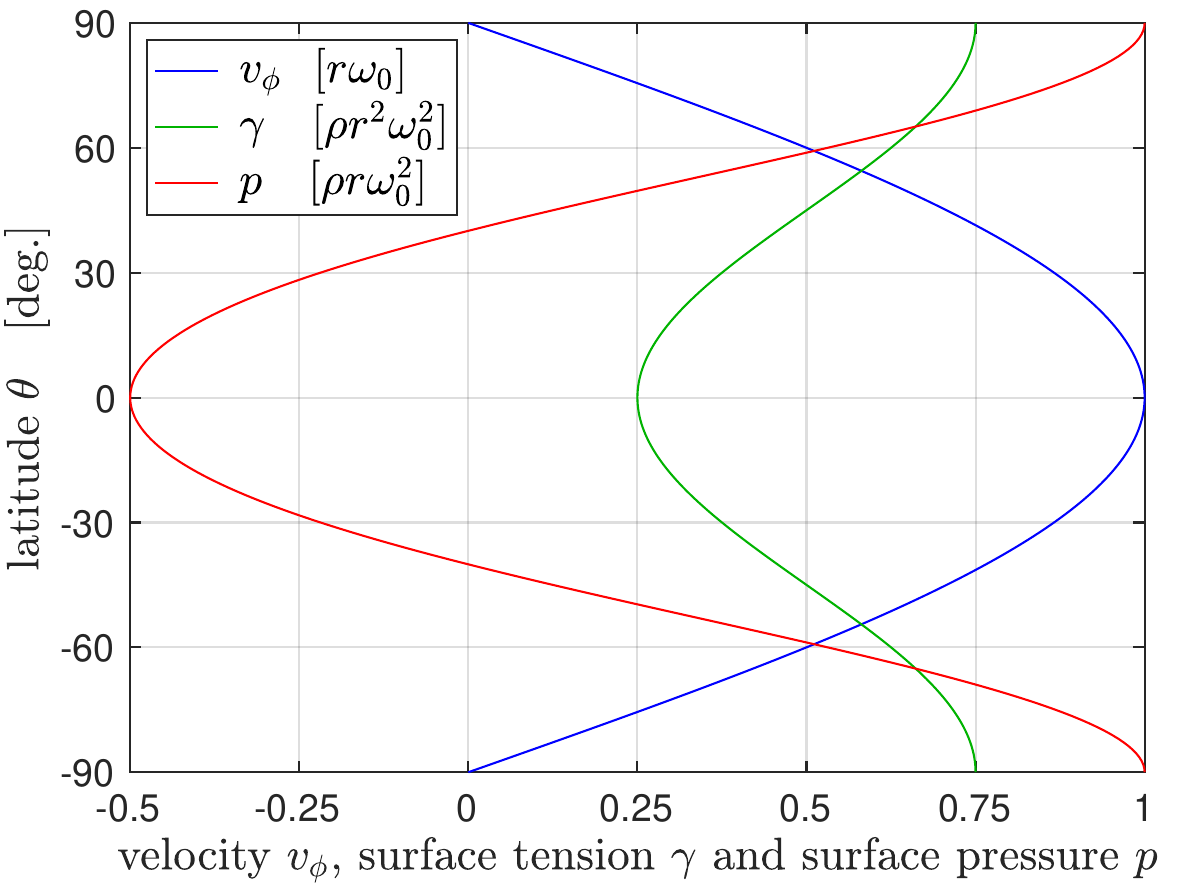}}
\put(-7.7,-.1){\footnotesize (a)}
\put(0.25,-.1){\footnotesize (b)}
\end{picture}
\caption{Spinning sphere: (a) flow field $\bv$ and (b) profiles of surface velocity $v_\phi(\theta)$, surface tension $\gamma(\theta)$ and surface pressure $p(\theta)$ considering $\gamma_0 = \rho r^2\omega_0^2/4$.}
\label{f:spinsphere}
\end{center}
\end{figure}

The previous example can be easily modified to describe the rigid body rotation of a spinning sphere.
In that case, (\ref{e4:zeta}.1) needs to be simply replaced by
\eqb{l}
\zeta^1 = \theta^1 = \xi^1 + \omega_0 t\,,
\eqe
leading to $\dot\zeta^1 = \omega_0$, $\dot\zeta^2 = 0$, $\zeta^\alpha_{,\hat\gamma} = \delta^\alpha_{\hat\gamma}$ and $\dot\zeta^\alpha_{,\gamma}=0$.
The ALE description in \eqref{e4:bx}--\eqref{e4:bab} remains unchanged.
But now velocity and acceleration become
\eqb{lll}
\bv \is r\omega_0\cos\theta\,\be_\phi \\[1mm]
\dot\bv \is -r\omega_0^2\cos\theta\,\be_{r_0}\,,
\eqe
and lead to the gradients
\eqb{lllll}
\bv_{\!,1} \is \dot\ba_1 \is -r\omega\cos\theta\,\be_{r_0}\,, \\[1mm]
\bv_{\!,2} \is \dot\ba_2 \is -r\omega\sin\theta\,\be_\phi\,,
\eqe
As a consequence, the vorticity is
\eqb{l}
\omega = 2\omega_0\sin\theta\,,
\eqe
which is maximum at the poles, while the rate-of-deformation tensor becomes
\eqb{l}
2\bd = -\omega_0\cos\theta\,\big(\be_\phi\otimes\be_r + \be_r\otimes\be_\phi\big)\,,
\eqe
which has zero in-plane  part ($\bd_\mrs=\mathbf{0}$).
Therefore the stress only consists of surface tension, i.e.
\eqb{l}
\bsig = \gamma\,\bi\,.
\eqe
The body force required to equilibrate this stress under dynamic conditions now becomes
\eqb{lll}
f_1 \is 0\,, \\[2.5mm]
f_2 \is \rho\,r^2\,\omega^2_0\,\sin\theta\cos\theta - \gamma_{,2}\,, \\[1mm]
p \is \ds\frac{2\gamma}{r} - \rho\,r\,\omega^2_0\,\cos^2\theta\,,
\label{e:manfac2}\eqe
since still  $\gamma_{,1} = 0$ due to rotational symmetry.
The special case $f_2 = 0$ can then be maintained for the surface tension
\eqb{l}
\gamma =  \ds\frac{\rho\,r^2\,\omega_0^2}{2}\sin^2\theta + \gamma_0\,.
\eqe
The flow field, surface tension and surface pressure are visualized in Fig.~\ref{f:spinsphere} for the choice $\gamma_0 = \rho r^2\omega_0^2/4$.
%-----------------------------------------------------------------
% run sphereflow3.m
% run sphereflow.m
%-----------------------------------------------------------------
This shows that even a simple spinning sphere requires varying body forces and surface tension to stay in spherical shape.
 
\begin{remark}
A general rigid body motion contains a translation $\bc(t)$ apart from a rotation. 
Adding $\bc$ to motion \eqref{e4:bx} results in $\bv_\mrm = \dot\bc$ and $\bv = \dot\bc + r\omega\cos\theta\,\be_\phi$.
Since $\bc$ must be constant in $\zeta^\alpha$, the velocity gradient and stress are the same as before. 
Only the body force needs the extra term $\rho\,\ddot\bc$ to account for acceleration $\ddot\bc$ in case it is non-zero.
\end{remark}

\subsection{Octahedral vortex flow on a sphere}\label{s:4tex}

\begin{figure}[h]
\begin{center} \unitlength1cm
\begin{picture}(0,10.40)
\put(-8.8,4.8){\includegraphics[height=60mm]{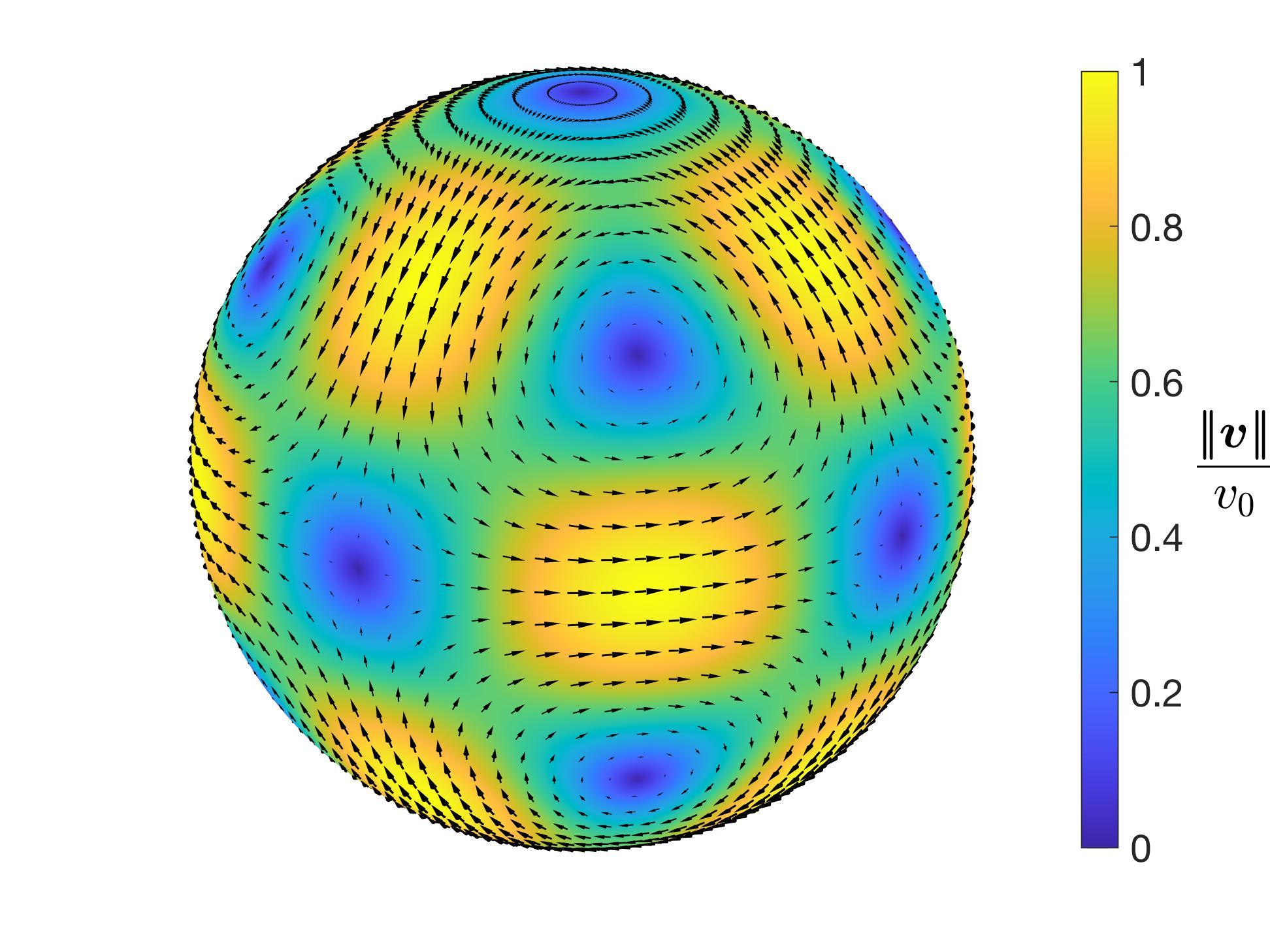}}
\put(-.4,4.8){\includegraphics[height=60mm]{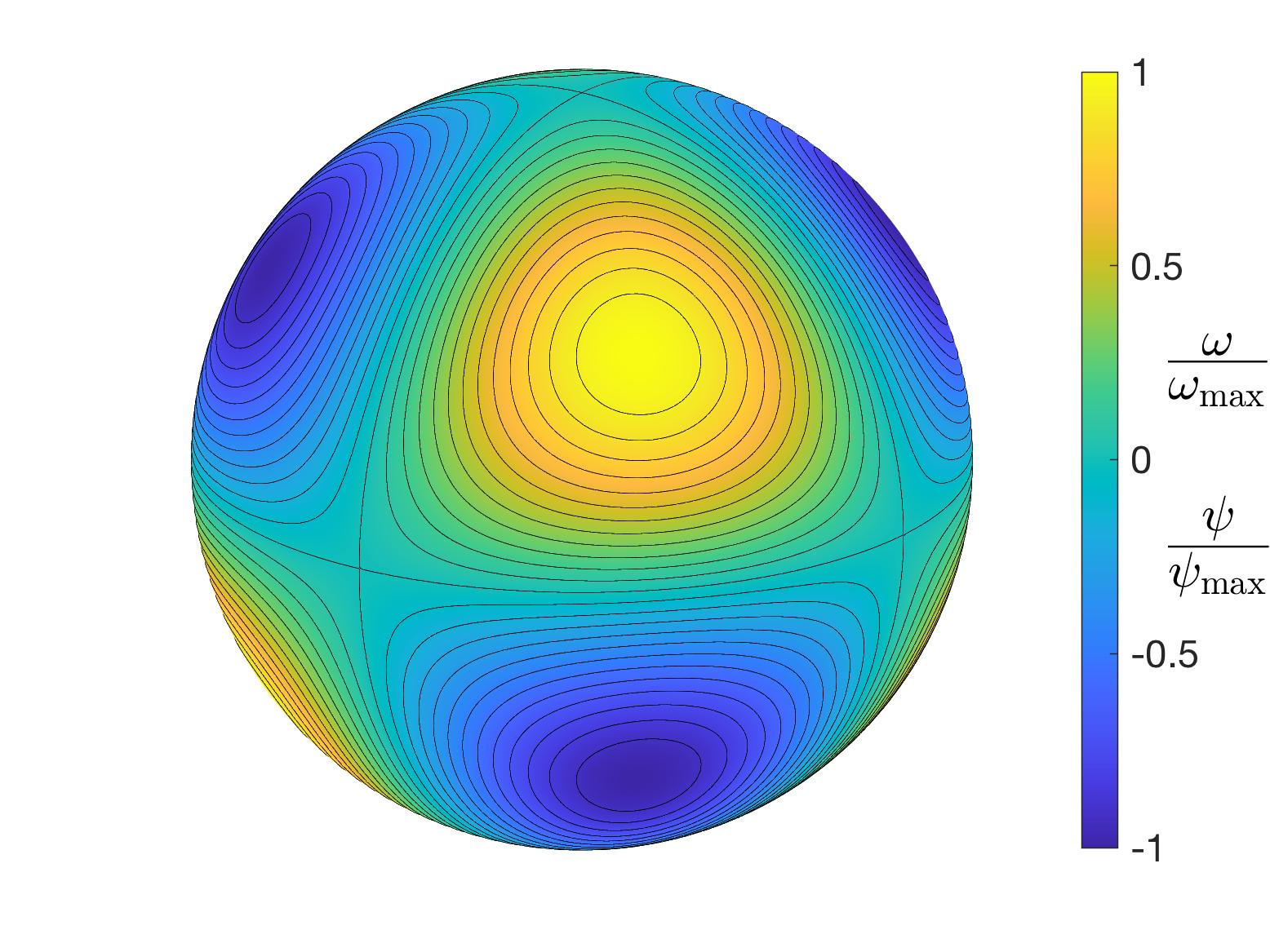}}
\put(-8.8,-.7){\includegraphics[height=60mm]{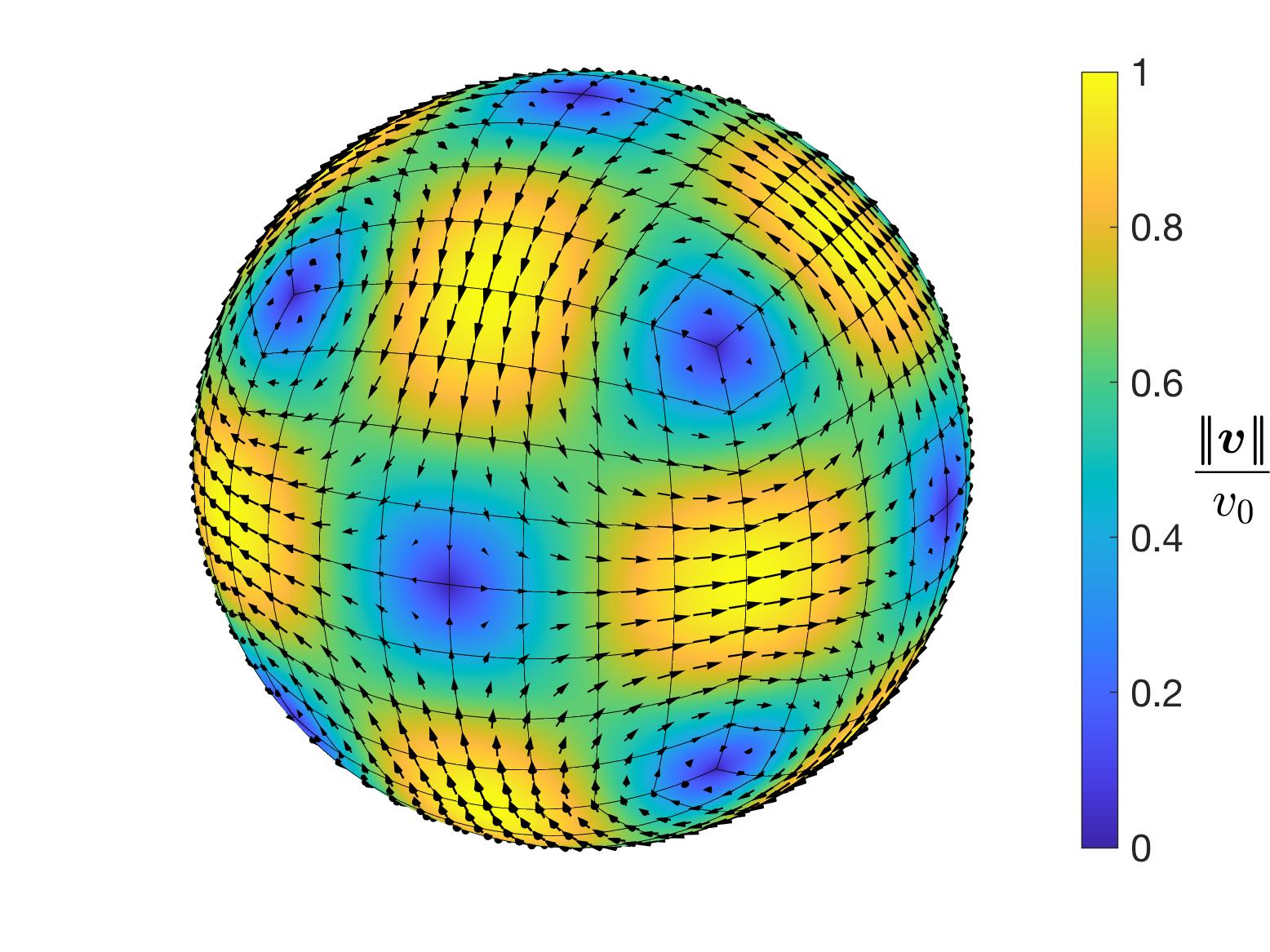}}
\put(-.4,-.7){\includegraphics[height=60mm]{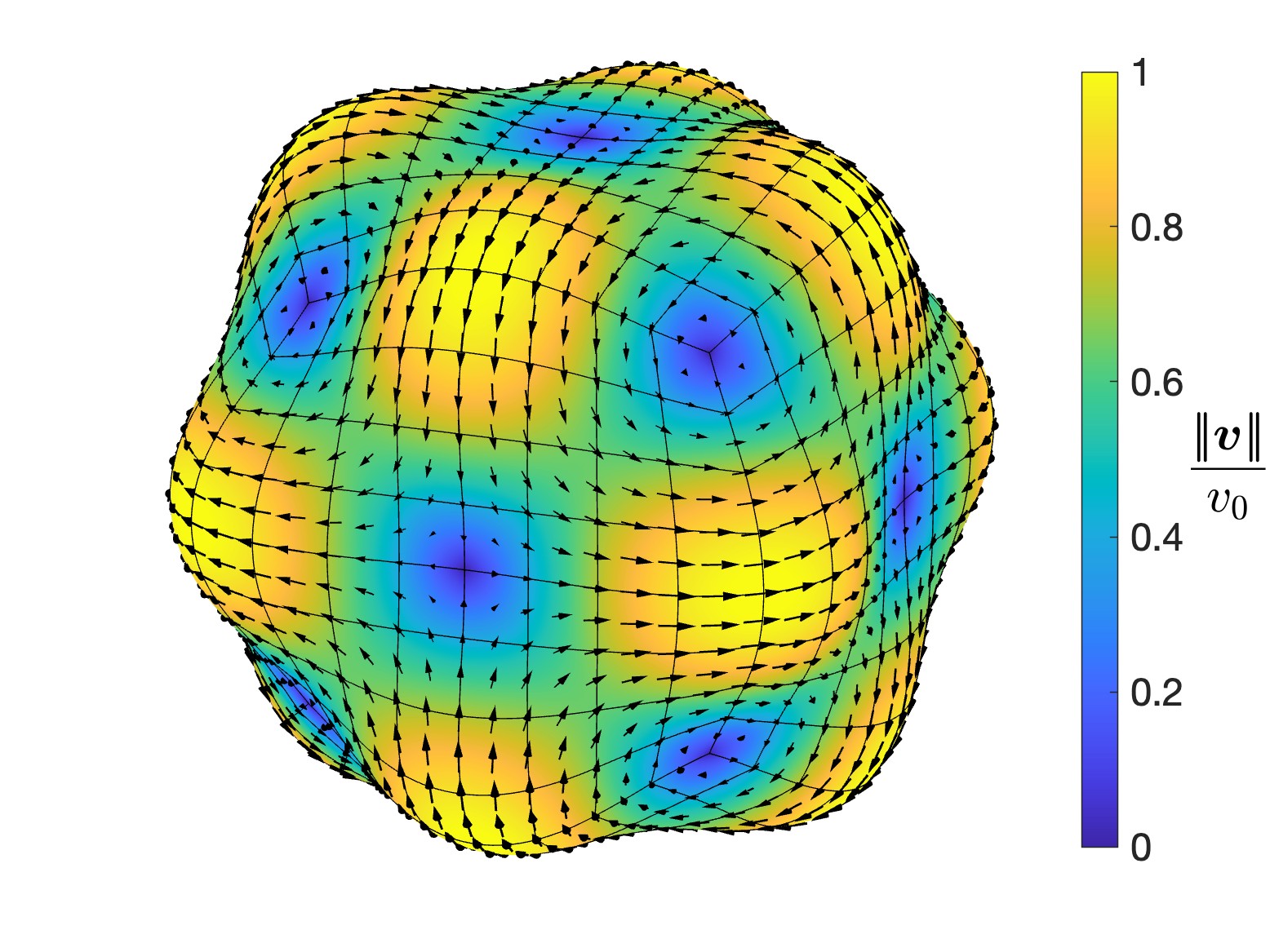}}
\put(-7.7,5.4){\footnotesize (a)}
\put(0.7,5.4){\footnotesize (b)}
\put(-7.7,-.1){\footnotesize (c)}
\put(0.7,-.1){\footnotesize (d)}
\end{picture}
\caption{Octahedral vortex flow on a sphere: 
(a) Analytical flow field $\bv$ and (b) vorticity $\omega$ and stream function $\psi$ normalized by their maximum values $v_0$, $\psi_\mathrm{max}$ and $\omega_\mathrm{max}$. 
The latter two occur at $\phi=\pi/4$ and $\theta=\arctan(\sqrt{2}/2)$ and are equal to $\psi_\mathrm{max} = 2v_0 r/(3\sqrt{3})$ and $\omega_\mathrm{max} = 8v_0/(\sqrt{3}r)$.
The analytical flow field in (a) is compared to the numerical flow field for a fixed (c) and free surface (d).
The radial surface displacement in (d) ranges between -1.18\% and 1.13\%. 
It is scaled-up by a factor of~10 to increase its visibility.\\[-5mm]}
\label{f:4tex-sphere}
\end{center}
\end{figure}

The shear flow solution of Sec.~\ref{s:shear} has no $\phi$-dependency and is steady for fixed surface and mesh.
The following example therefore presents a transient example with $\phi$-dependency.
It considers eight counter-rotating vortices on a sphere arranged in the pattern of an octahedron. 
The example is motivated by the surface flow induced by large scale surface budding on spherical vesicles \citep{liquidshell}.
It is adapted from a comparable example of \citet{nitschke12}.
The spherical surface is described by the same parameterization from before, see Eqs.~\eqref{e4:bx}--\eqref{e4:bab}.
As before, $\phi = \zeta^1$ and $\theta = \zeta^2$ are used.
The flow in this example is derived from a stream function and thus automatically satisfies the area-incompressibility constraint \eqref{e:divv}.
The chosen stream function is
\eqb{l}
\psi = v_0\,r\,\sin2\phi\,\sin\theta\,\cos^2\!\theta\,.
\label{e7:psi}\eqe
The (tangential) surface velocity then follows from the surface curl
\eqb{l}
\bv = \mathrm{curl}_\mrs\psi 
\label{e7:bv}\eqe
defined in Eq.~\eqref{e:scurl_phi}.
This gives the components
\eqb{lllll}
v^1 \is \dot\zeta^1 \is \ds\frac{v_0}{r}\sin2\phi\,\big(2\sin^2\!\theta - \cos^2\!\theta\big)\,, \\[3mm]
v^2 \is \dot\zeta^2 \is \ds\frac{v_0}{r}\cos2\phi\,\sin2\theta
\label{e7:zeta}\eqe
in the $\{\ba_\alpha\}$ basis, i.e.~$\bv = v^\alpha\,\ba_\alpha$.
They can be easily transformed into the components
\eqb{lll}
v_\phi \is v_0\,\sin2\phi\,\big(2\sin^2\!\theta - \cos^2\!\theta\big)\cos\theta\,, \\[1mm]
v_\theta \is v_0\,\cos2\phi\,\sin2\theta
\label{e7:v_pt}\eqe
of the spherical basis $\{\be_\phi,\,\be_\theta\}$ using \eqref{e4:baa}, i.e.~$\bv = v_\phi\,\be_\phi + v_\theta\,\be_\theta$.
Fig.~\ref{f:4tex-sphere} shows the magnitude and components of the flow field as well as the stream function $\psi$ and surface vorticity $\omega$. 
%-----------------------------------------------------------------
% run sphereflow4.m with jpr = 1 & 2
% run sOctoFlow
% run pOctoEvolve2
%-----------------------------------------------------------------
The latter follows from \eqref{e:vorticity}, which in view of \eqref{e7:bv} is equal to the surface Laplacian of $\psi$,
\eqb{l}
\omega = \Delta_\mrs\psi = a^{\alpha\beta}\,\psi_{;\alpha\beta}\,,
\eqe
and turns out to be equal to $\omega = -12\psi/r^2$ in the present example.
It is noted that \eqref{e7:zeta} is a system of ODEs from which one can reconstruct $\zeta^\alpha = \zeta^\alpha(\xi^\beta,t)$ if desired.

\begin{figure}[h]
\begin{center} \unitlength1cm
\begin{picture}(0,5.6)
\put(-8.8,-.22){\includegraphics[height=60mm]{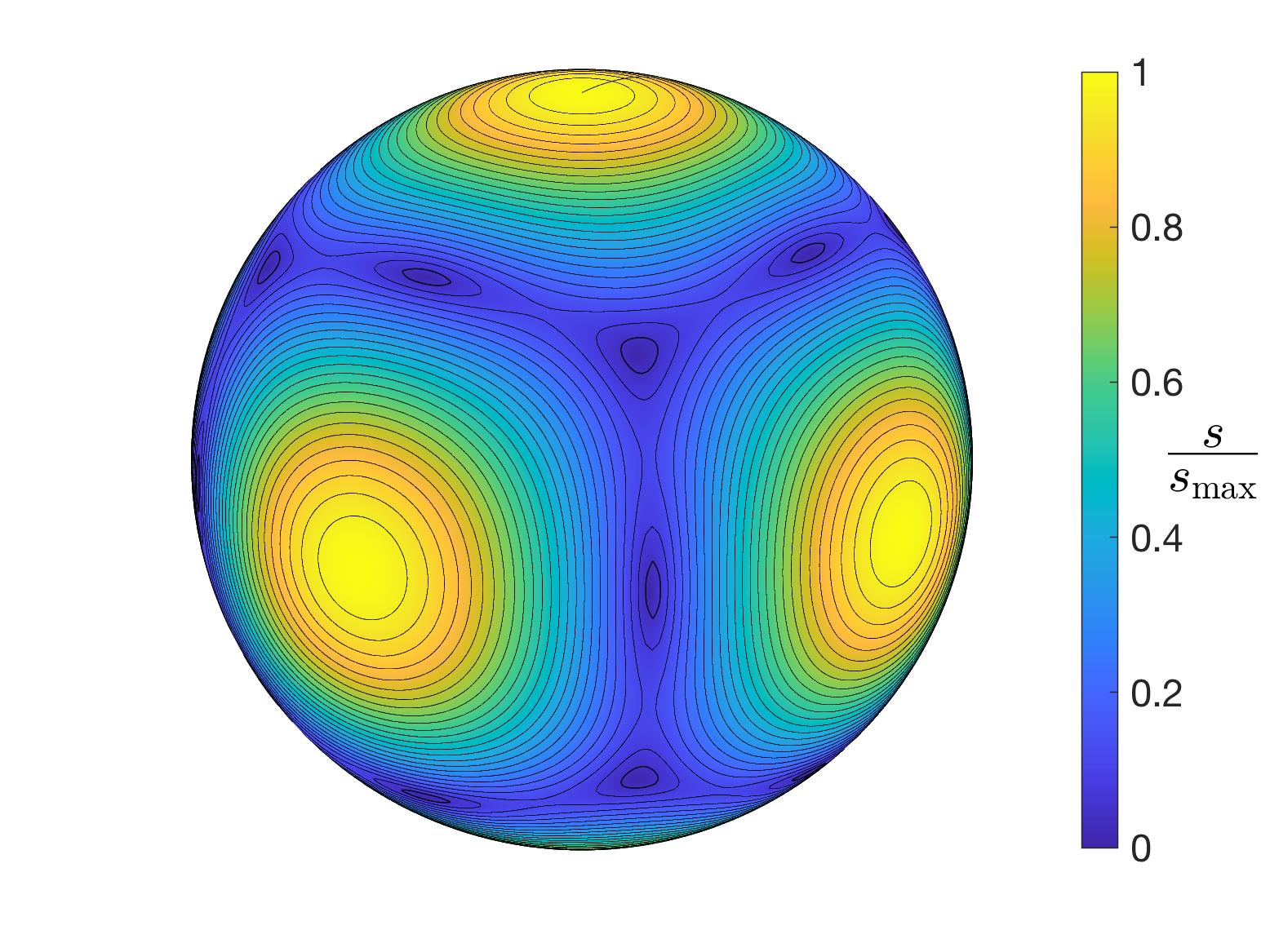}}
\put(0.3,-.3){\includegraphics[height=58mm]{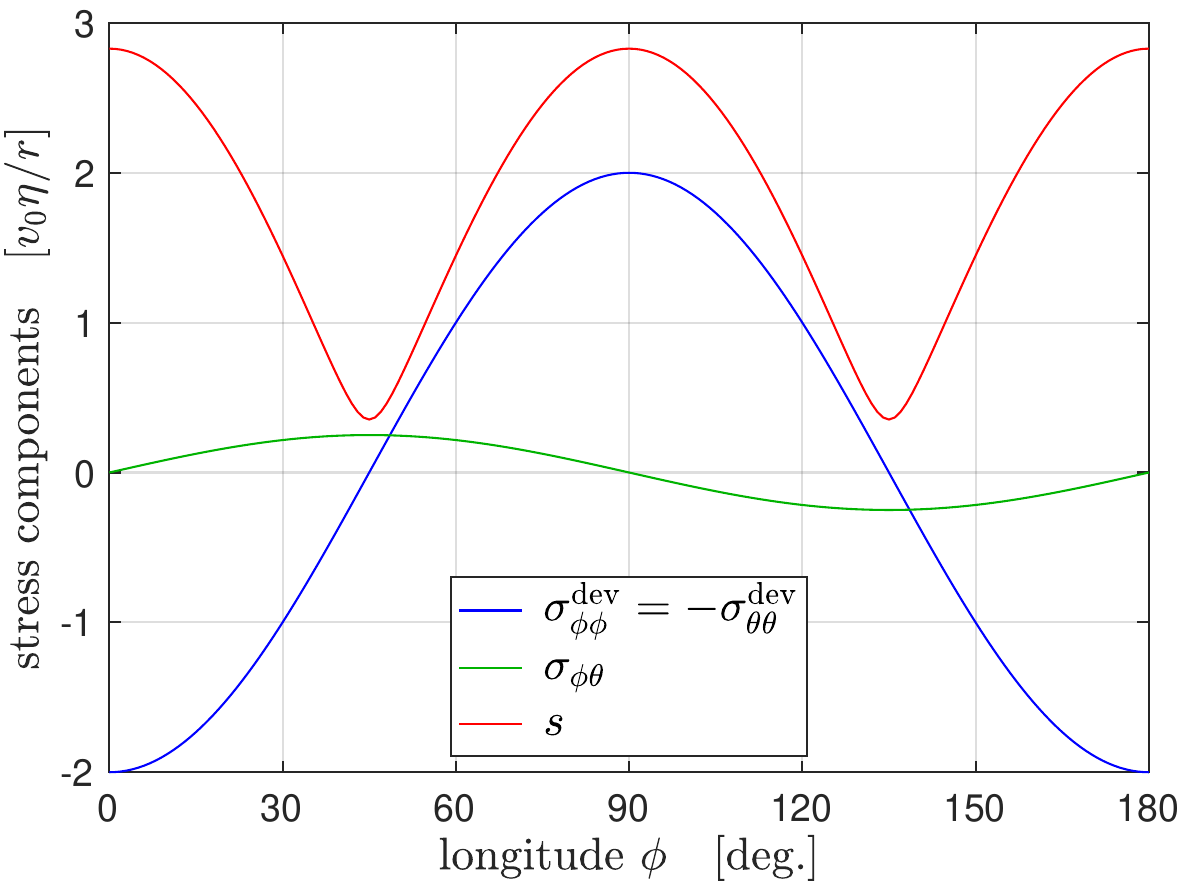}}
\put(-7.7,-.1){\footnotesize (a)}
\put(0.15,-.1){\footnotesize (b)}
\end{picture}
\caption{Octahedral vortex flow on a sphere: (a) shear stress field $s$ and (b) stress components $\sig_{\phi\phi}^\mathrm{dev}$, $\sig_{\theta\theta}^\mathrm{dev}$, $\sig_{\phi\theta}$ and $s$ vs.~$\phi$ for $\theta=30^\circ$. 
The maximum of $s$ is $s_\mathrm{max} = \sqrt{32}v_0\eta/r$.}
\label{f:4tex-sphere2}
\end{center}
\end{figure}

From
\eqb{lll}
\bv_{\!,1} \is -c_\theta\big[2v_0\,c_{2\phi}\,c_{2\theta}\,\be_\phi + v_0\,s_{2\phi}\big(5c^2_\theta+2s^2_\theta\big)s_\theta\,\be_\theta + v_\phi\,\be_r \big]\,, \\[1mm]
\bv_{\!,2} \is v_0\,s_{2\phi}\big(7c^2_\theta-2s^2_\theta\big)s_\theta\,\be_\phi + 2v_0\,c_{2\phi}\,c_{2\theta}\,\be_\theta - v_\theta\,\be_r
\label{e7:bva}\eqe
one can find the in-plane rate-of-deformation tensor
\eqb{l}
2\bd_\mrs = \ds\frac{v_0}{r}\Big[4c_{2\phi}\,c_{2\theta}\,\big(\be_\theta\otimes\be_\theta - \be_\phi\otimes\be_\phi\big) 
+ s_{2\phi}\,(3c_{2\theta}-1)\,s_\theta\,\big(\be_\phi\otimes\be_\theta+\be_\theta\otimes\be_\phi\big) \Big]\,.
\eqe
Here the abbreviations $s_{...} := \sin...$ and $c_{...} := \cos{...}$ have been used.
This leads to the stress components
\eqb{lll}
\sig_{\phi\phi} \is \gamma - 4\ds\frac{v_0\eta}{r}\,\cos2\phi\,\cos2\theta\,, \\[3mm]
\sig_{\theta\theta} \is \gamma + 4\ds\frac{v_0\eta}{r}\,\cos2\phi\,\cos2\theta\,, \\[3mm]
\sig_{\phi\theta} \is \ds\frac{v_0\eta}{r}\,\sin2\phi\,\big(3\cos2\theta-1)\sin\theta\,,
\eqe
according to \eqref{e:bsig}.
A meaningful quantity to examine is the surface shear stress \citep{phaseshell}
\eqb{l}
s = \sqrt{\sig^{\alpha\beta}_\mathrm{dev}\,\sig_{\alpha\beta}^\mathrm{dev}}\,,\quad \sig^{\alpha\beta}_\mathrm{dev} := \sig^{\alpha\beta} - \gamma\,a^{\alpha\beta}\,,
\eqe
which becomes
\eqb{l}
s = \sqrt{\sig_{\phi\phi}^\mathrm{dev}\sig_{\phi\phi}^\mathrm{dev} + \sig_{\theta\theta}^\mathrm{dev}\sig_{\theta\theta}^\mathrm{dev} 
+ 2\sig_{\phi\theta}^\mathrm{dev}\sig_{\phi\theta}^\mathrm{dev}}
\eqe
in spherical coordinates and is shown in Fig.~\ref{f:4tex-sphere2}.
%-----------------------------------------------------------------
% run sphereflow4.m with jpr = 5
%-----------------------------------------------------------------
It confirms the octahedral symmetry of the present flow example.

\begin{figure}[h]
\begin{center} \unitlength1cm
\begin{picture}(0,5.1)
\put(-8.8,-.7){\includegraphics[height=60mm]{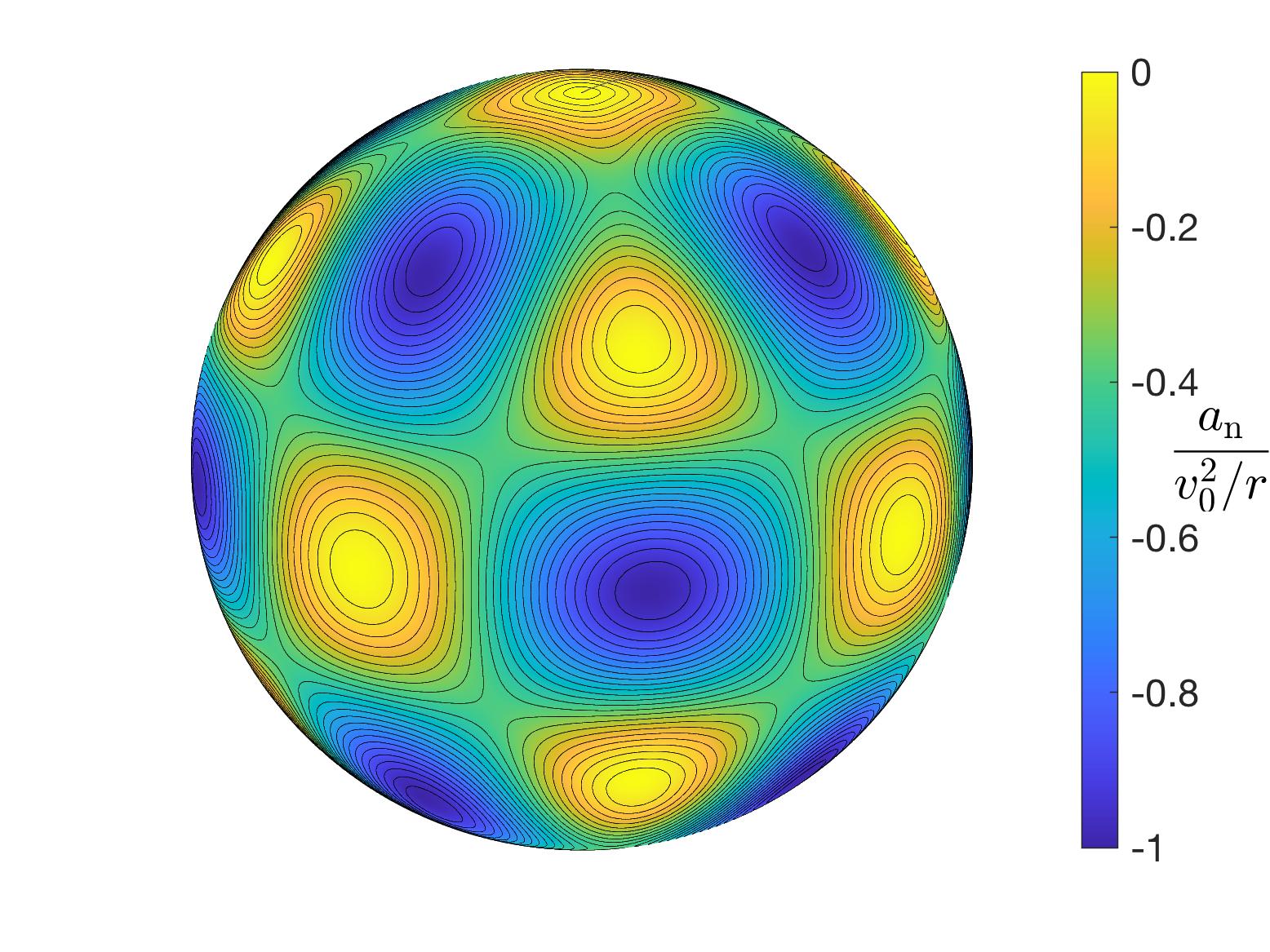}}
\put(-.4,-.7){\includegraphics[height=60mm]{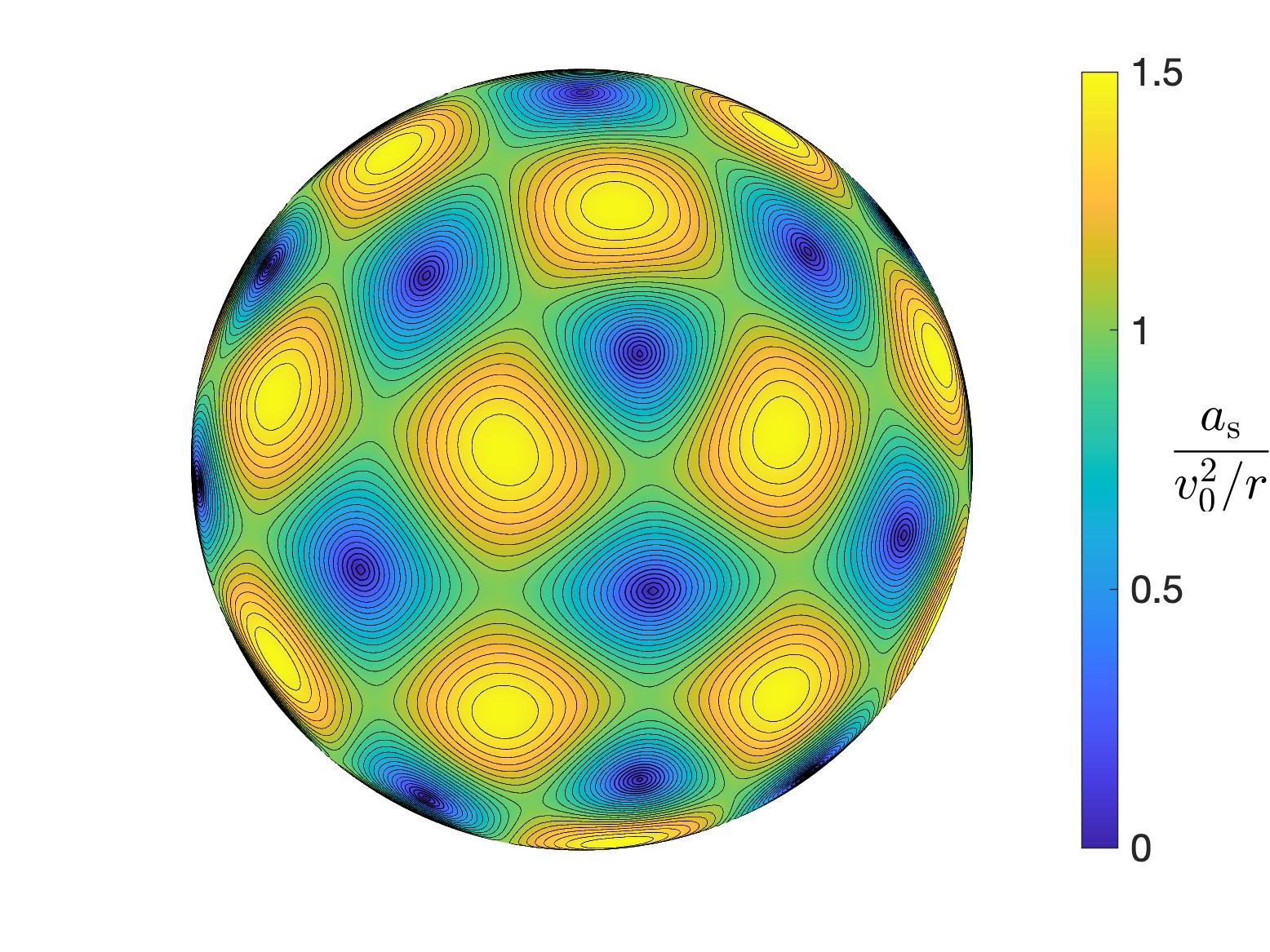}}
\put(-7.7,-.1){\footnotesize (a)}
\put(0.7,-.1){\footnotesize (b)}
\end{picture}
\caption{Octahedral vortex flow on a sphere: (a) normal component $a_\mrn$ and (b) magnitude of the tangential component $a_\mrs:=\norm{\ba_\mrs}$ of the steady-state part of the material acceleration.}
\label{f:4tex-sphere3}
\end{center}
\end{figure}

From a tedious but straightforward calculation one can find the simple relation
\eqb{l}
\bd_{\mrs,\alpha}\,\ba^\alpha = - \ds\frac{5}{v_0}\bv\,,
\eqe
such that
\eqb{l}
\bT^\alpha_{;\alpha} = \gamma_{,\alpha}\,\ba^\alpha + 2H\gamma\,\bn - \ds\frac{10\eta}{v_0}\bv\,,
\eqe
according to \eqref{e:Taa1}.
Choosing $\gamma=$ const., this satisfies PDE \eqref{e:Taa} for the manufactured body force
\eqb{l}
\bff = \rho\,\dot\bv + \eta_\mrs\,\bv + p_0\,\bn \,,
\eqe
with the viscous friction coefficient $\eta_\mrs :=10\eta/r^2$ and static surface pressure $p_0 := 2\gamma/r$.
Taking the material time derivative of $\bv = v_\phi\,\be_\phi + v_\theta\,\be_\theta$, and providing $\dot v_\phi$, $\dot v_\theta$, $\dot\be_\phi$ and $\dot\be_\theta$ through Eqs.~\eqref{e7:v_pt} and \eqref{e:bedot}, leads to the material acceleration
\eqb{l}
\dot\bv = \ba_\mrs + a_\mrn\,\bn\,,
\label{e:octodotv}\eqe
with the components
\eqb{lll}
\ba_\mrs \is a_\phi\,\be_\phi + a_\theta\,\be_\phi\,,\quad a_\mrn = -\ds\frac{\norm{\bv}^2}{r}\,, \\[3.5mm]
a_\phi \is \ds\frac{\dot v_0}{v_0}v_\phi - 2v_0\,c_{2\phi}\,c_{2\theta}\,c_\theta\,\dot\phi 
+ v_0\,s_{2\phi}\,\big(7c_\theta^2-2s_\theta^2\big)\,s_\theta\,\dot\theta\,, \\[3.5mm]
a_\theta \is \ds\frac{\dot v_0}{v_0}v_\theta + 2v_0\,c_{2\phi}\,c_{2\theta}\,\dot\theta 
- v_0\,s_{2\phi}\,\big(5c_\theta^2+2s_\theta^2\big)\,s_\theta\,c_\theta\,\dot\phi\,. 
\eqe
Comparing \eqref{e:octodotv} with \eqref{e:vALEa} and using \eqref{e7:zeta} and \eqref{e7:bva}, reveals
the transient part
\eqb{l}
\bv' = \ds\frac{\dot v_0}{v_0}\bv + \bv_{,\alpha}\,v^\alpha_\mrm \,. 
\eqe
The acceleration $\dot\bv$ thus contains a transient part that is proportional to $\dot v_0$ and is only in-plane, while all the remaining terms are steady-state terms and are proportional to $v_0^2$.
The acceleration becomes negligible for small $\dot v_0$ and small $v_0$.
Fig.~\ref{f:4tex-sphere3} shows the steady-state part of the in-plane and out-of-plane acceleration components.
%-----------------------------------------------------------------
% run sphereflow4.m with jpr = 6 & 7
%-----------------------------------------------------------------
Note that $a_\mrn$ is negative, i.e.~the normal acceleration $a_\mrn\,\bn$ points inward.
It has to be equilibrated by the varying surface pressure $p = p_0 + \rho\,a_\mrn$ for the surface to remain spherical.
If the pressure is taken constant, the shape changes as the numerical solution in Fig.~\ref{f:4tex-sphere}d shows for $p = 3p_0$.
As seen, the surface bulges out, where the velocity is high, and in, where the velocity is small.
Further, the outward bulges are shifted in flow direction, which agrees with intuition.
This solution has been obtained by a FE computation using the discretization described in Appendix~\ref{s:FE}.
It has been confirmed that the FE convergence rates of Fig.~\ref{f:shearflow}e are maintained here, both for the fixed surface solution in Fig.~\ref{f:4tex-sphere}c and the free surface solution in Fig.~\ref{f:4tex-sphere}d.

It is finally noted that the FE code used here has been debugged and verified using the analytical examples proposed above.
They are characterized by increasing complexity and thus suitable for code development.
They test all flow components, in-plane as well as out-of-plane.
The last three examples are derived with zero mesh motion, but a mesh motion can be easily superimposed for further testing.
%This will be discussed in future work.
This is discussed in \citet{ALEcomp}.

\subsection{Vesicle budding}\label{s:bud}

\begin{figure}[H]
\begin{center} \unitlength1cm
\begin{picture}(0,5.55)
\put(-8.25,2.5){\includegraphics[height=31mm]{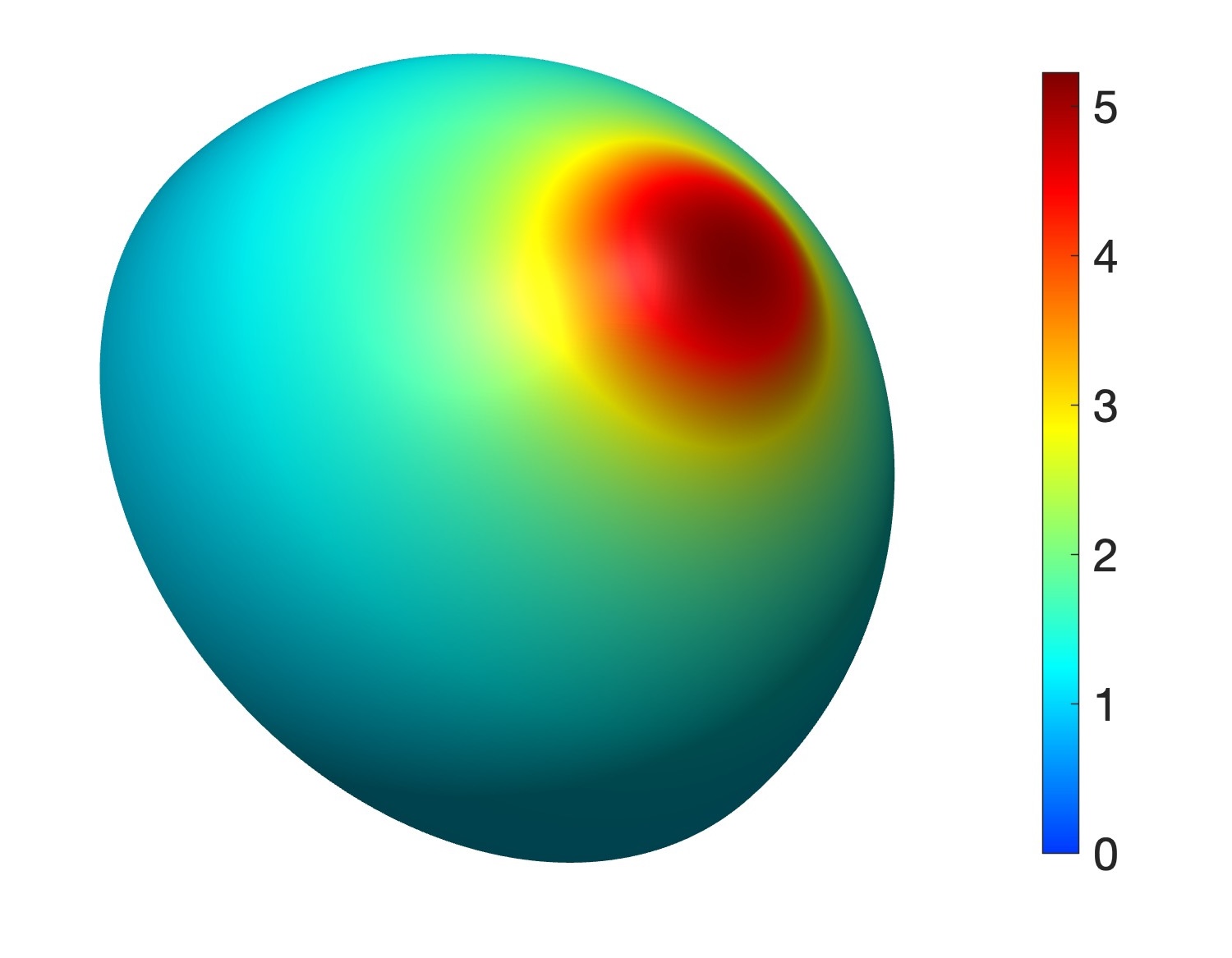}}
\put(-5.2,2.5){\includegraphics[height=31mm]{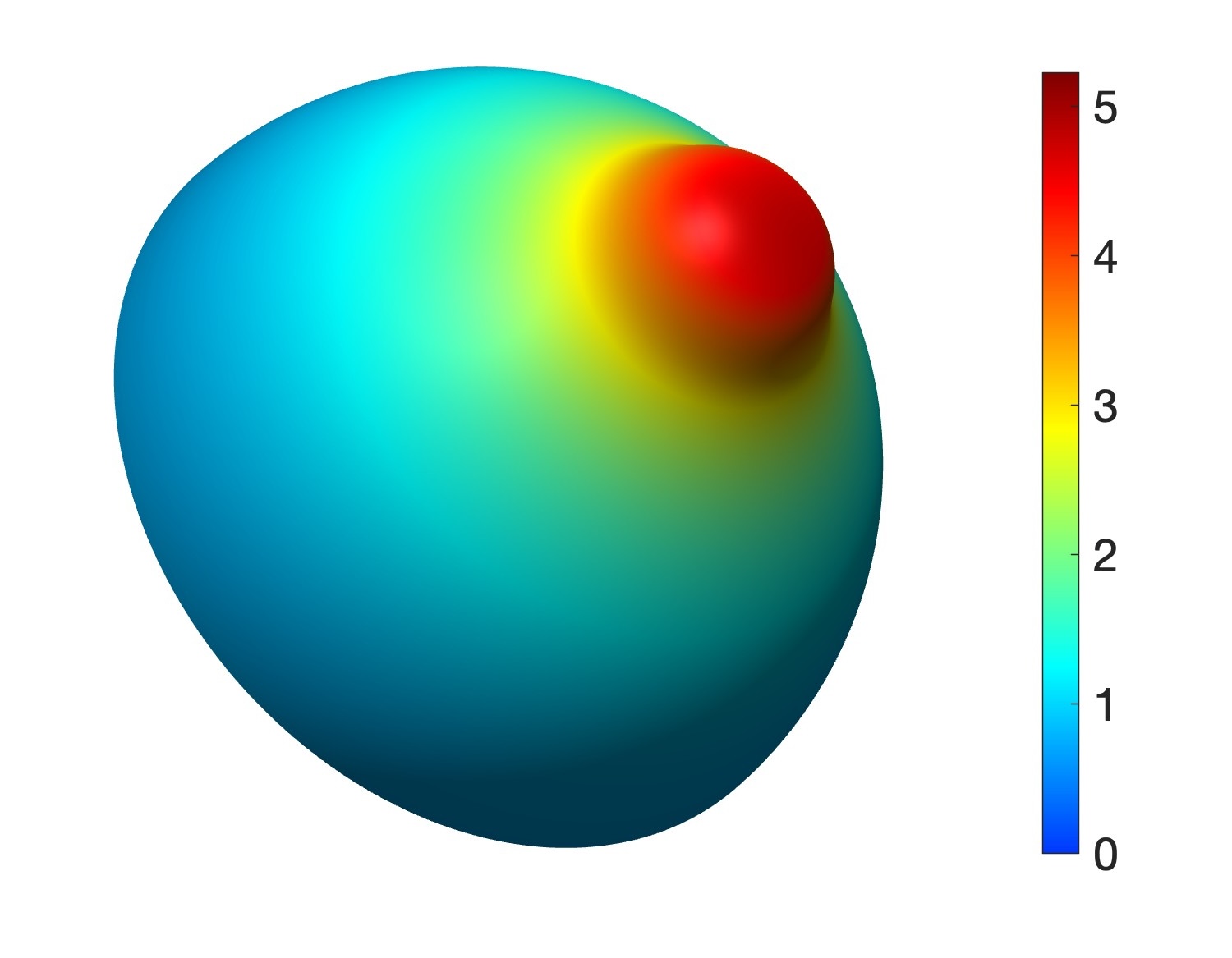}}
\put(-2.22,2.5){\includegraphics[height=31mm]{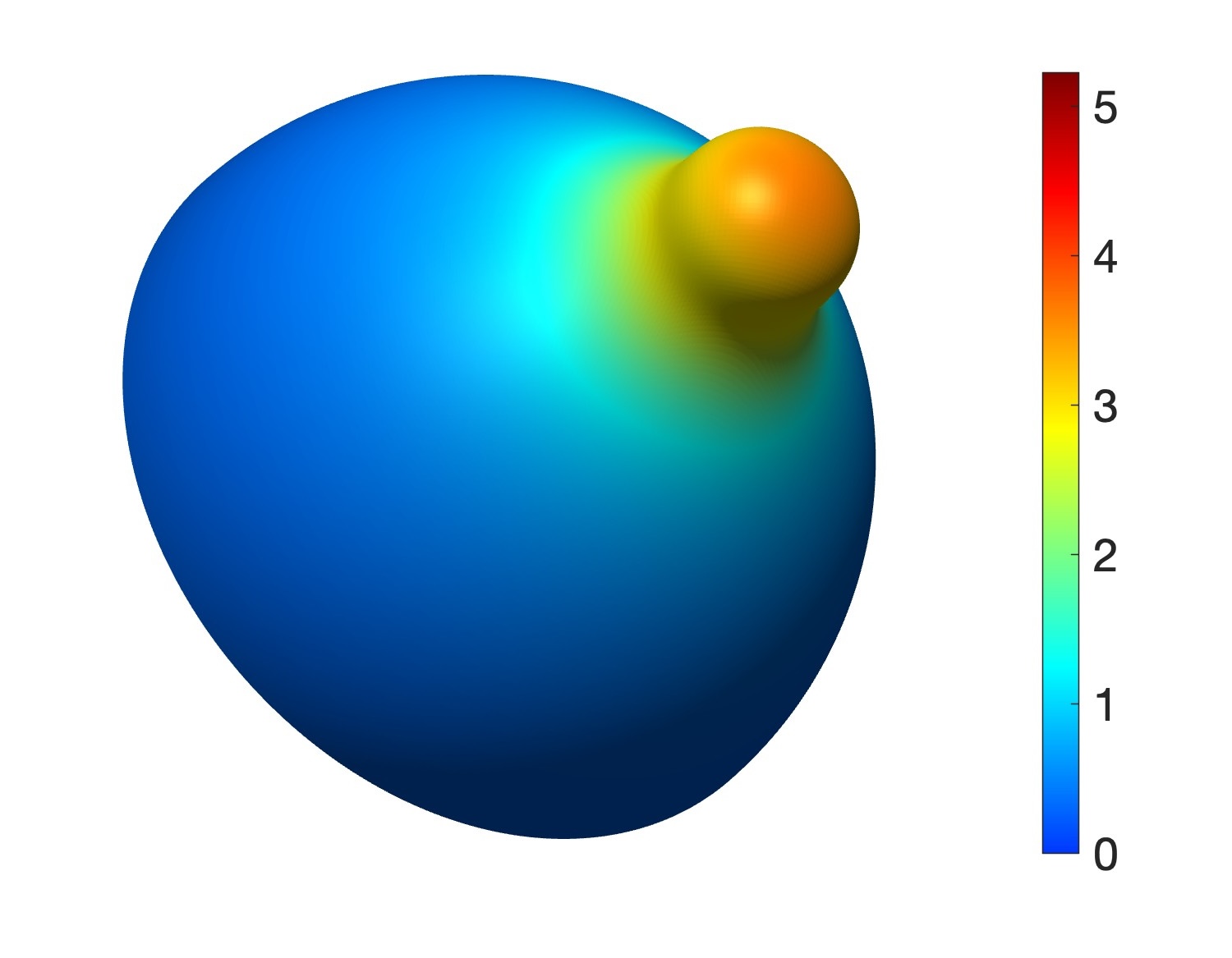}}
\put(0.76,2.5){\includegraphics[height=31mm]{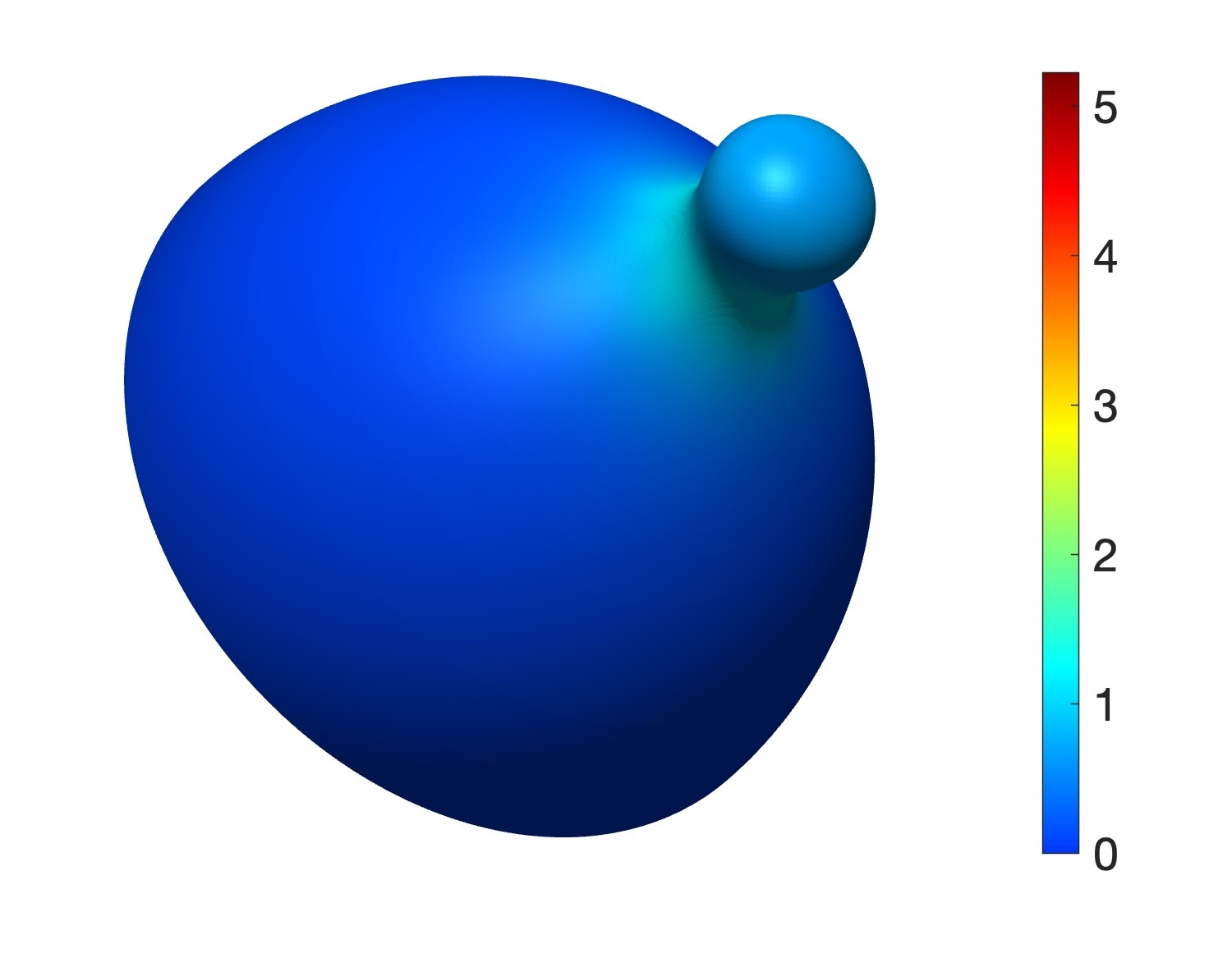}}
\put(3.74,2.5){\includegraphics[height=31mm]{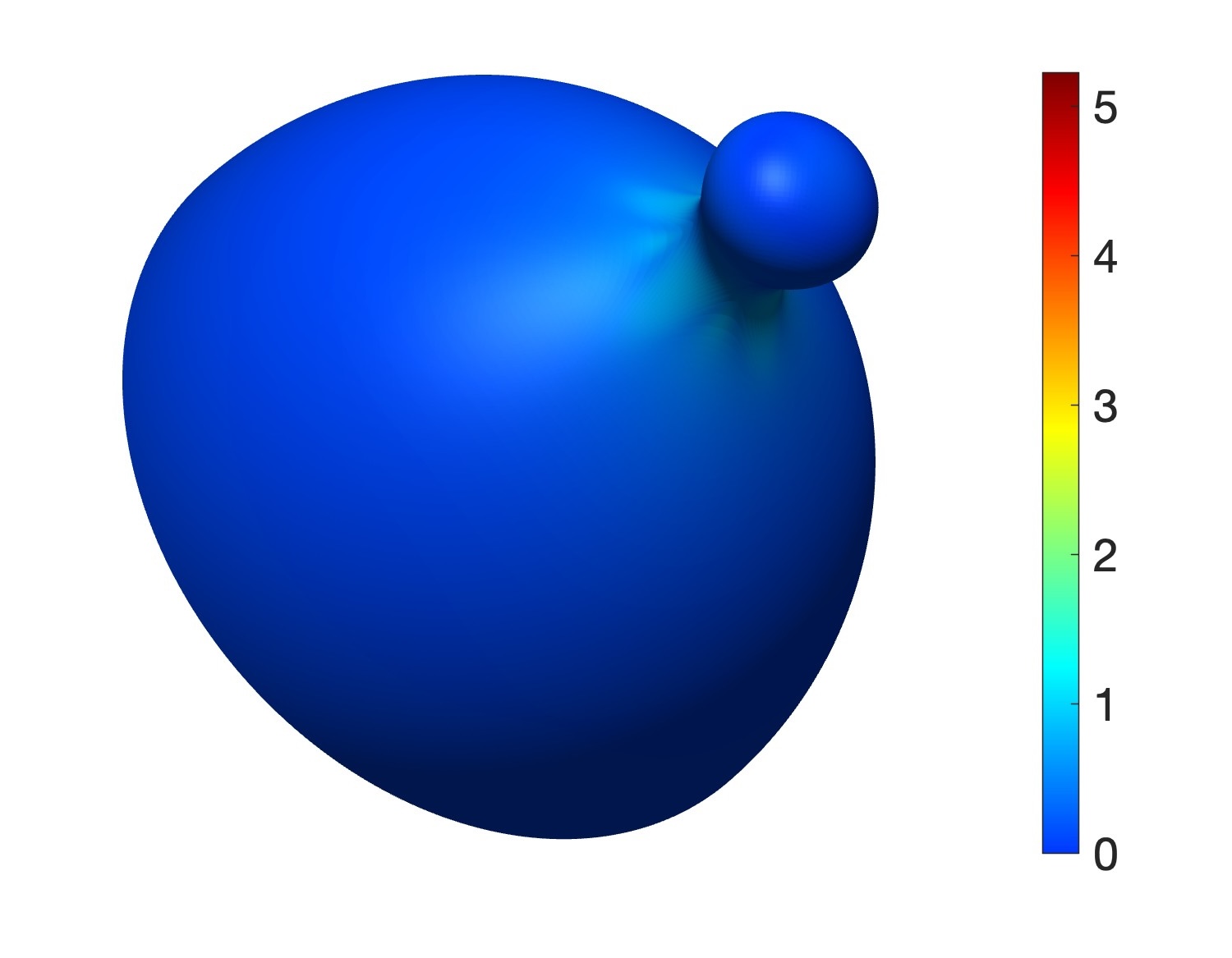}}
\put(7.45,4.95){\scriptsize $\norm{\bv}$}
\put(7.41,3.46){\scriptsize $\big[\!\frac{\mu\mrm}{\mrs}\!\big]$}
\put(-8.25,-.4){\includegraphics[height=31mm]{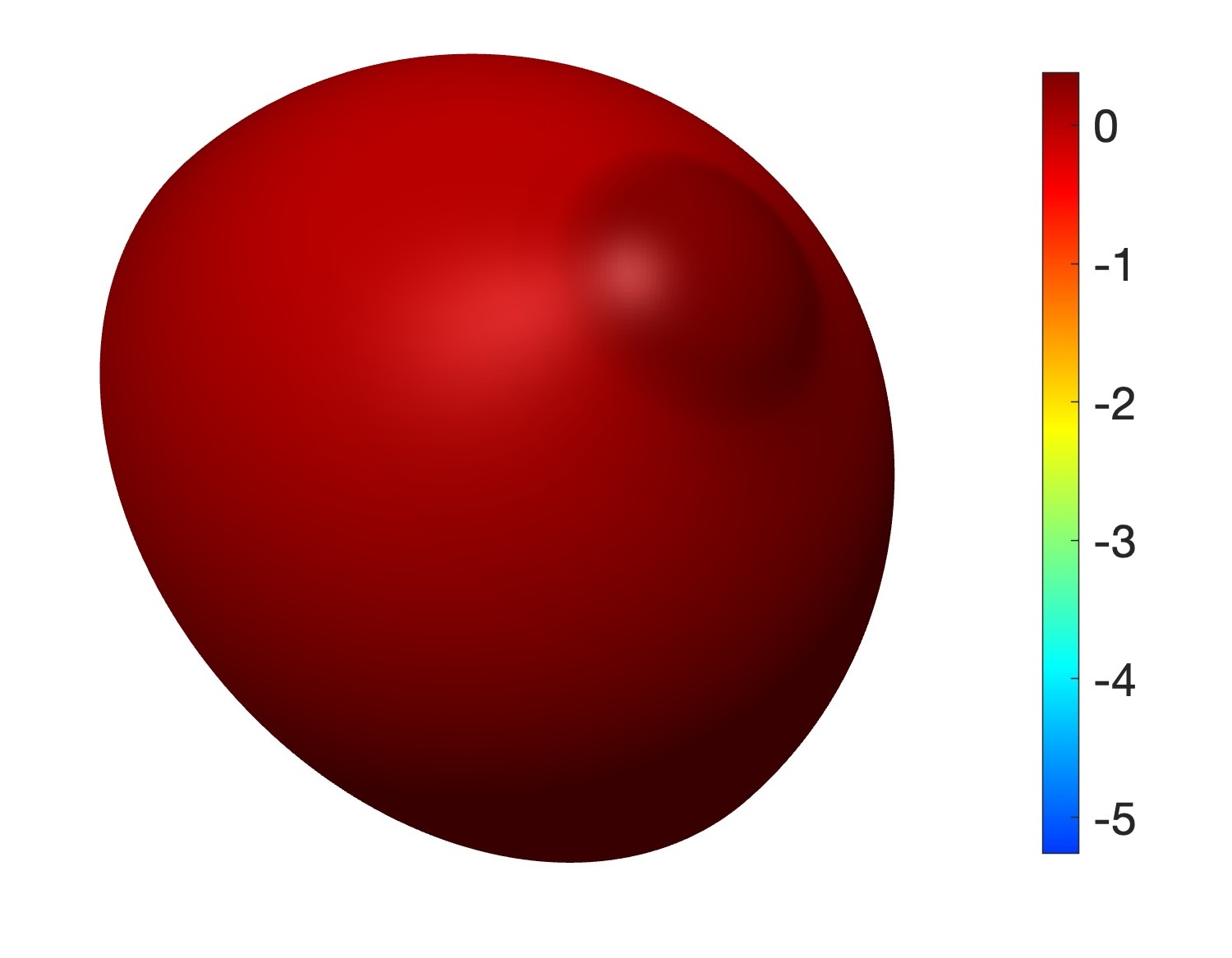}}
\put(-5.2,-.4){\includegraphics[height=31mm]{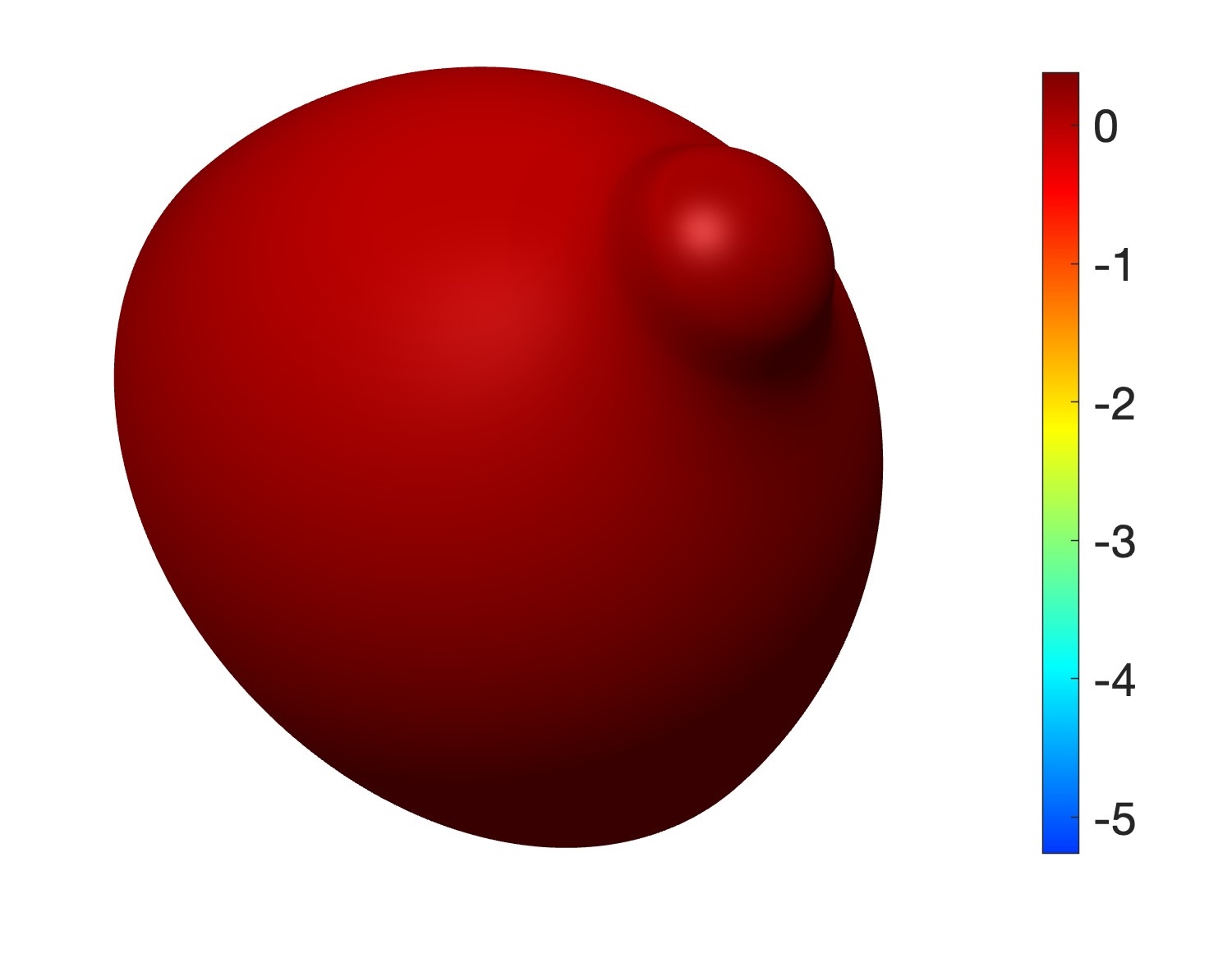}}
\put(-2.22,-.4){\includegraphics[height=31mm]{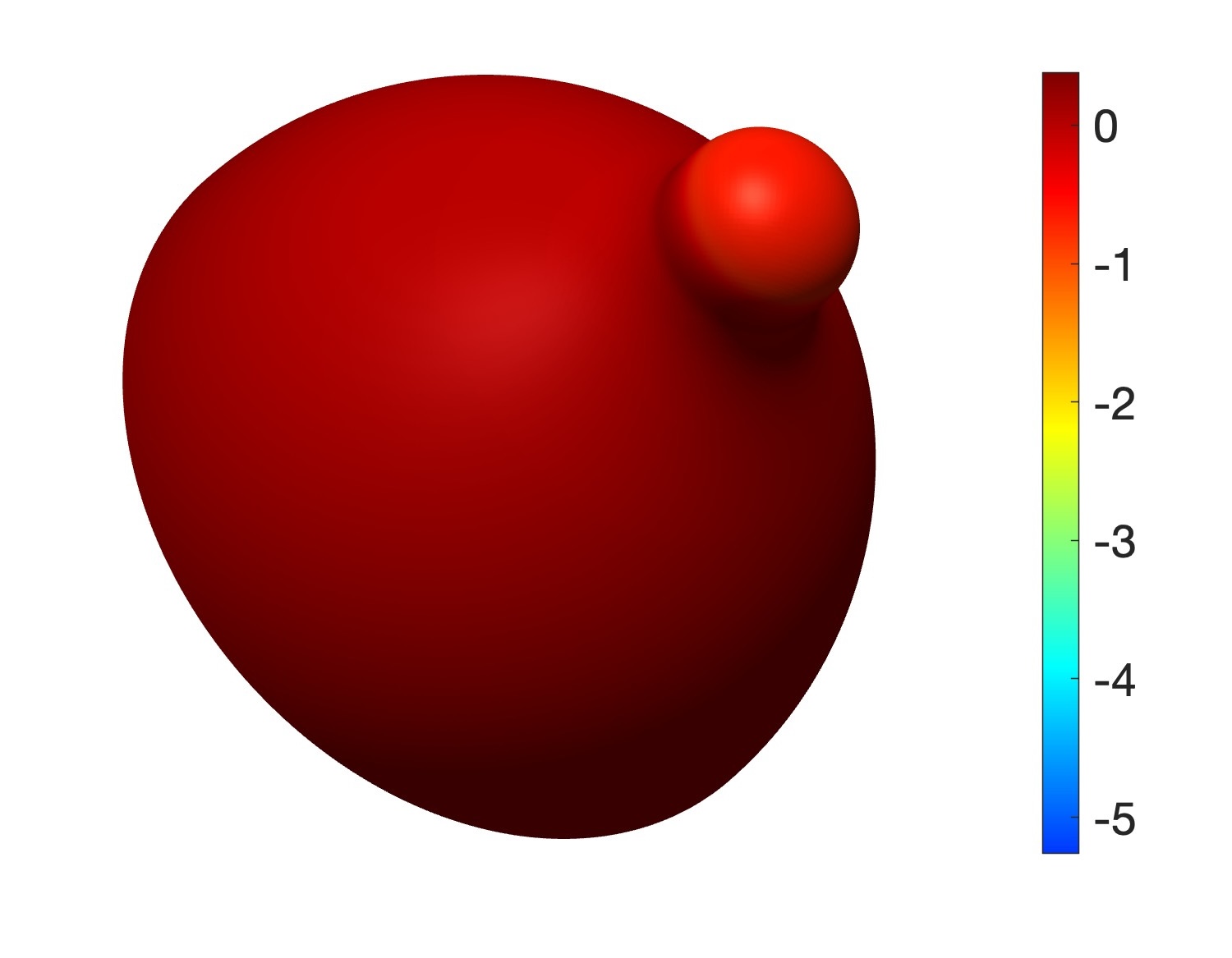}}
\put(0.76,-.4){\includegraphics[height=31mm]{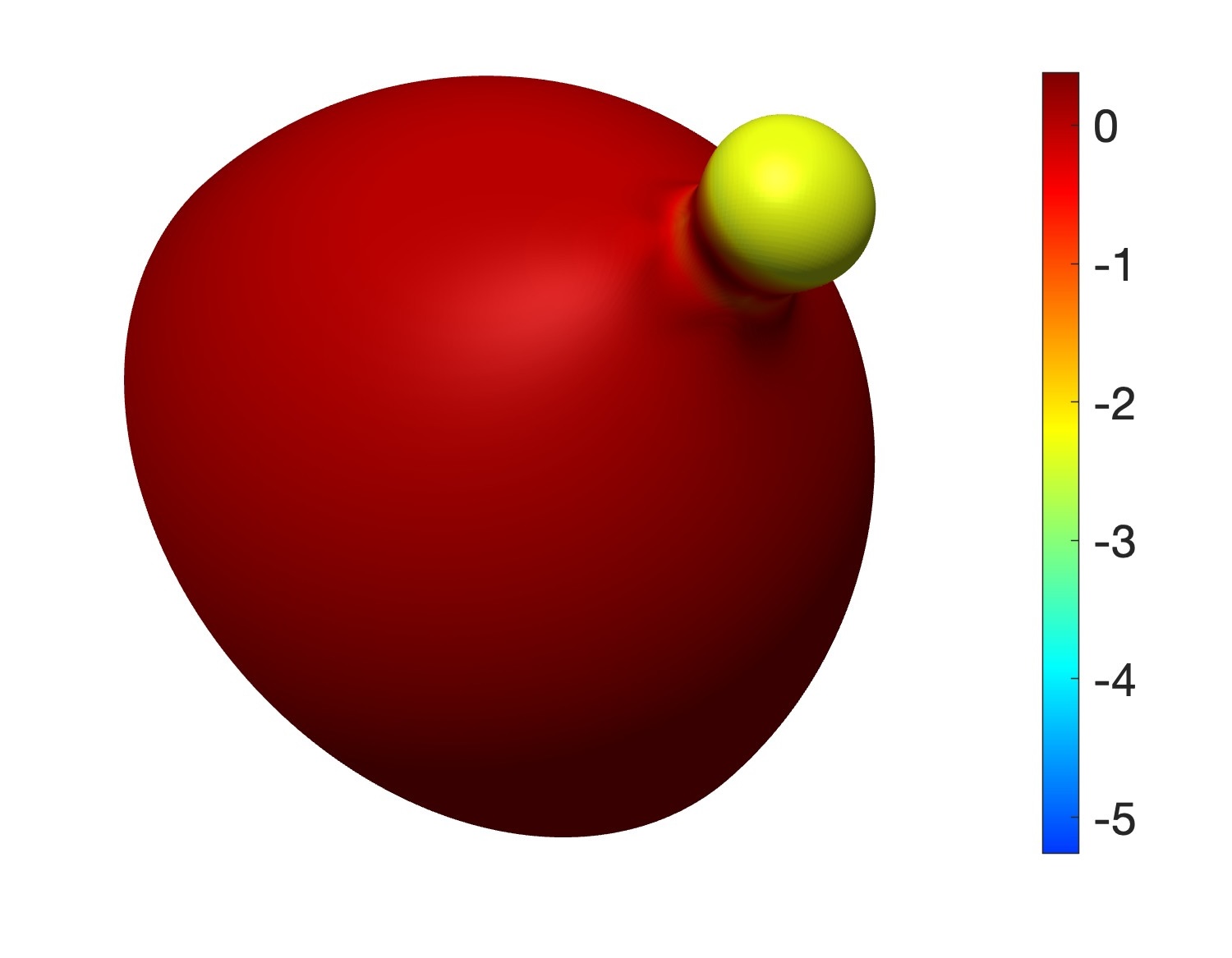}}
\put(3.74,-.4){\includegraphics[height=31mm]{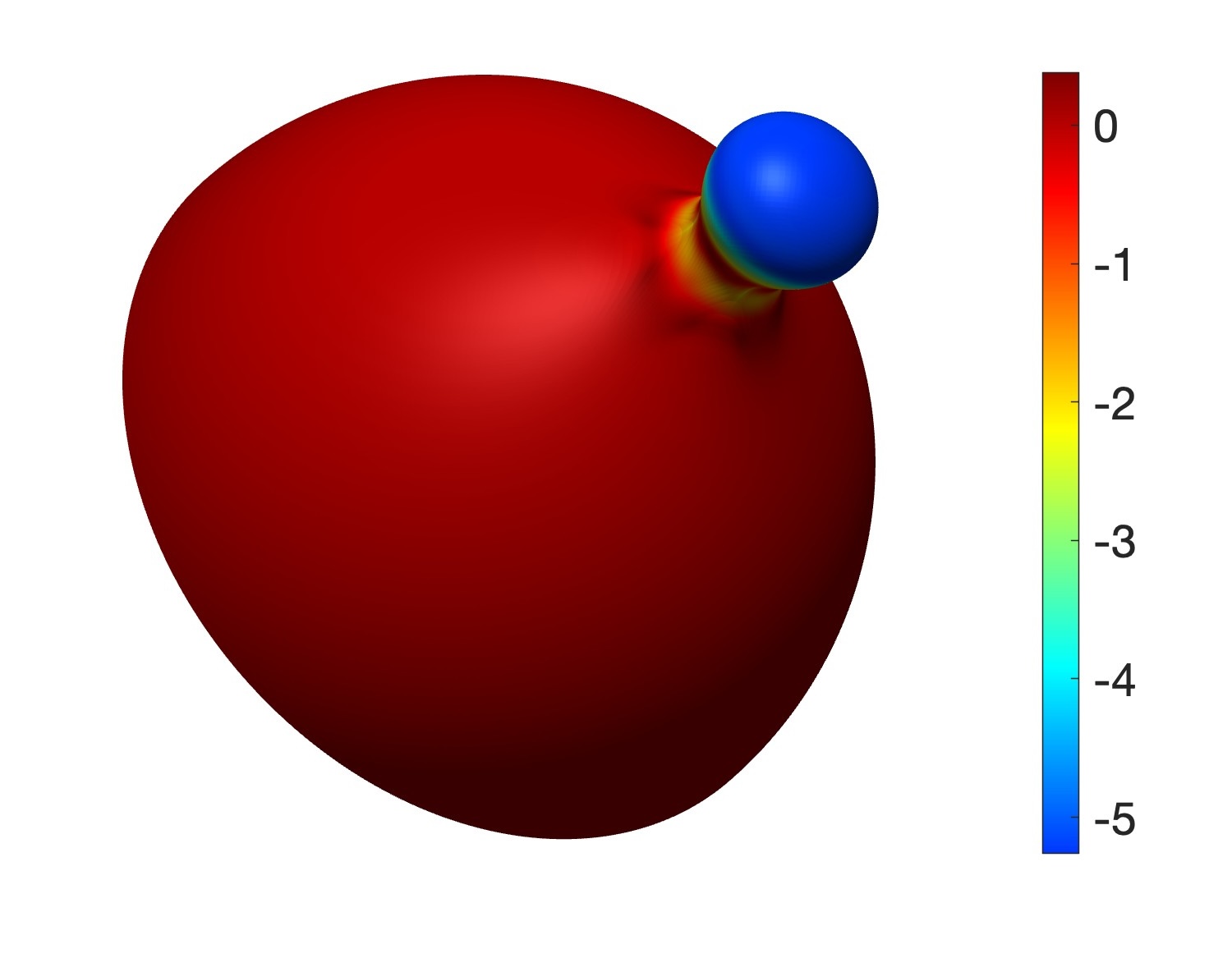}}
\put(7.55,2.05){\scriptsize $q$}
\put(7.43,0.62){\scriptsize $\big[\!\frac{\mu\mathrm{N}}{\mrm}\!\big]$}
\end{picture}
\caption{Vesicle budding: 
Accurate reference solution for the fluid velocity $\norm{\bv}$ (top row) and surface tension $q$ (bottom row) at time $t=[40,\,80,\,120,\,160,\,200]$ms (left to right), 
which corresponds to prescribed curvature $H_0 = -[2,\,4,\,6,\,8,\,10]/R$.
The result is obtained with 49152 NURBS finite elements.\\[-6mm]
}
\label{f:budflow}
\end{center}
\end{figure}

As a final numerical example, the budding of a spherical vesicle is shown.
The example uses the Helfrich bending model of Sec.~\ref{s:bending} and the hemispherical finite element setup from \citet{liquidshell}. 
The initial vesicle radius is taken as $R=1\mu$m, its bending stiffness and viscosity are taken from \citet{omar19}, i.e.~$k = 0.1236$\,nNnm, $k_\mrg = -0.83\,k$ and $\eta = 10^{-8}$Ns/m, and the mesh stiffness is chosen as $\mu_\mrm = 10^{-4}k/R^2$.
As inertia is expected to play no role here, steady-state Stokes flow is considered. 
Budding is induced by prescribing the spontaneous 
curvature $H_0$ at a constant rate $\dot H_0=-0.05$/(nm\,s) within a spherical region centered around the vesicle tip.\footnote{The region has radius $r=0.4R$; full $H_0$ is applied within $2r/3$ and ramped-down to zero between $2r/3$ and~$r$.}
Contrary to \citet{liquidshell}, who consider this region Lagrangian (i.e.~fixed in the reference configuration), it is treated Eulerian here (i.e.~fixed in the current configuration).
Fig.~\ref{f:budflow} shows the bud formation over time together with the fluid velocity and surface tension $q$.
%-----------------------------------------------------------------
% run pBudFlow.m
%-----------------------------------------------------------------

The local kinematics of bud formation resemble that of bubble inflation.
Therefore, the Eulerian surface formulation can be expected to yield much less accurate results than an ALE surface formulation.
This is confirmed by Fig.~\ref{f:budflow2}:
%-----------------------------------------------------------------
\begin{figure}[h]
\begin{center} \unitlength1cm
\begin{picture}(0,9.3)
\put(-8.45,5.45){\includegraphics[height=40mm]{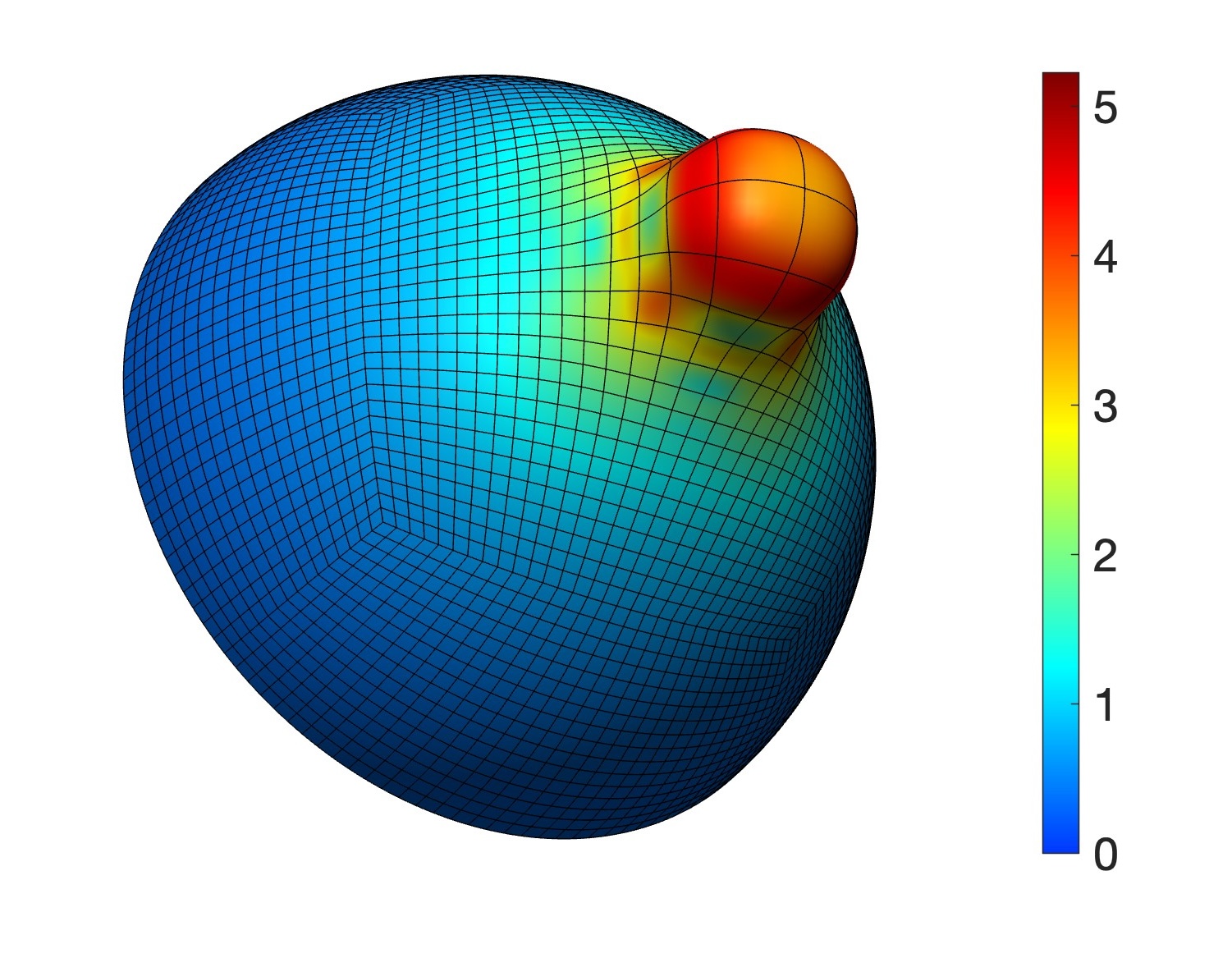}}
\put(-4.75,5.45){\includegraphics[height=40mm]{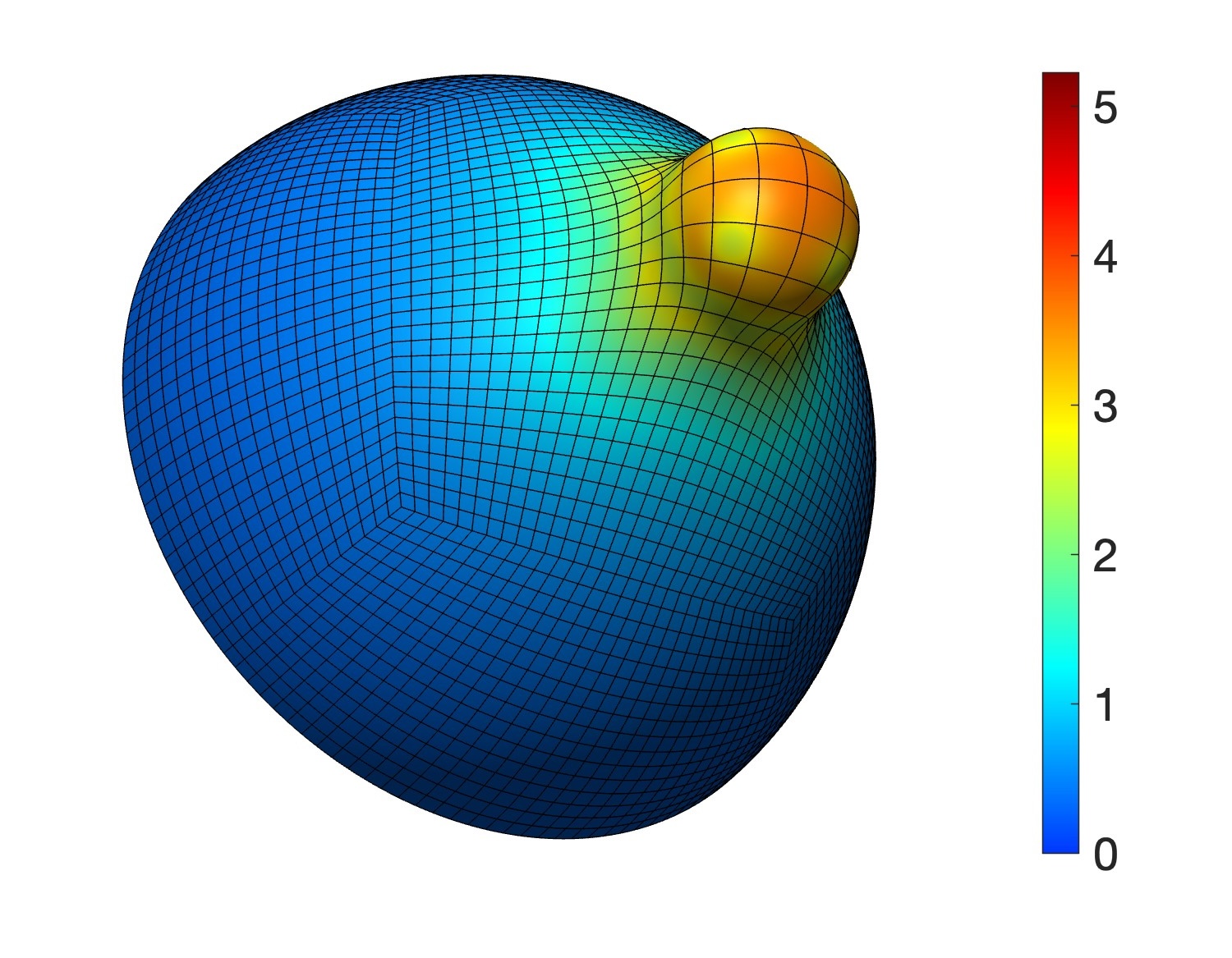}}
\put(-1.05,5.45){\includegraphics[height=40mm]{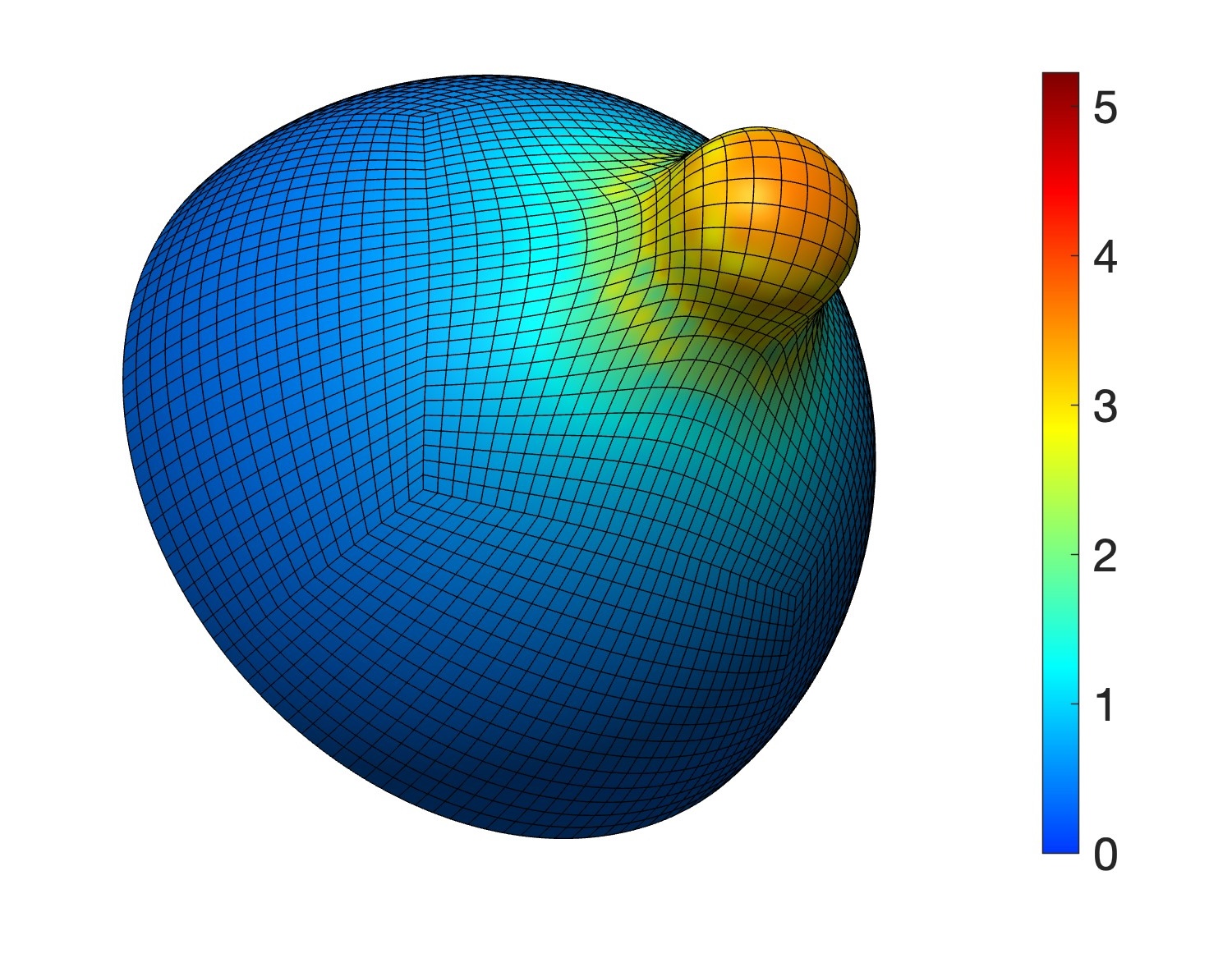}}
\put(2.65,5.45){\includegraphics[height=40mm]{afigs/BudFlow_Lag_m64_H6_DBe1vc.jpg}}
\put(-7.95,-.2){\includegraphics[height=57.5mm]{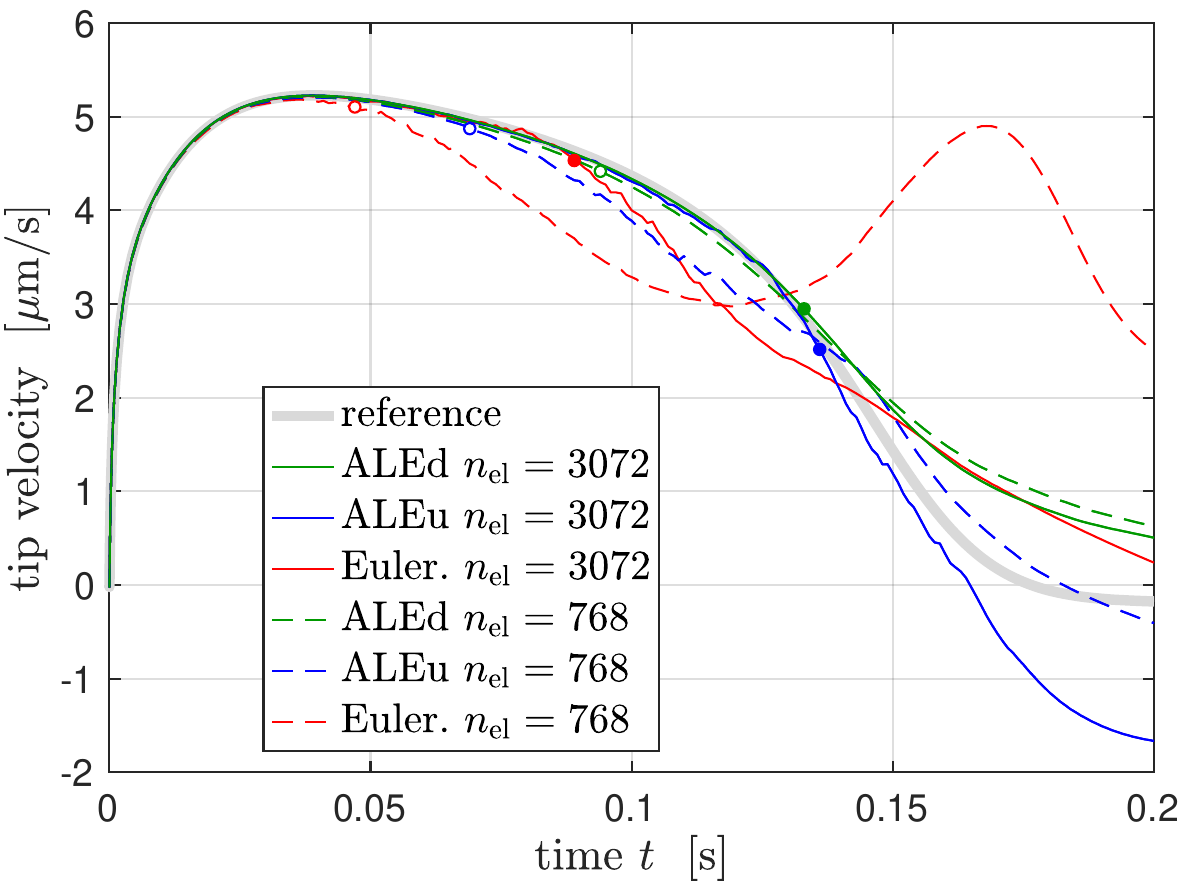}}
\put(0.2,-.2){\includegraphics[height=58mm]{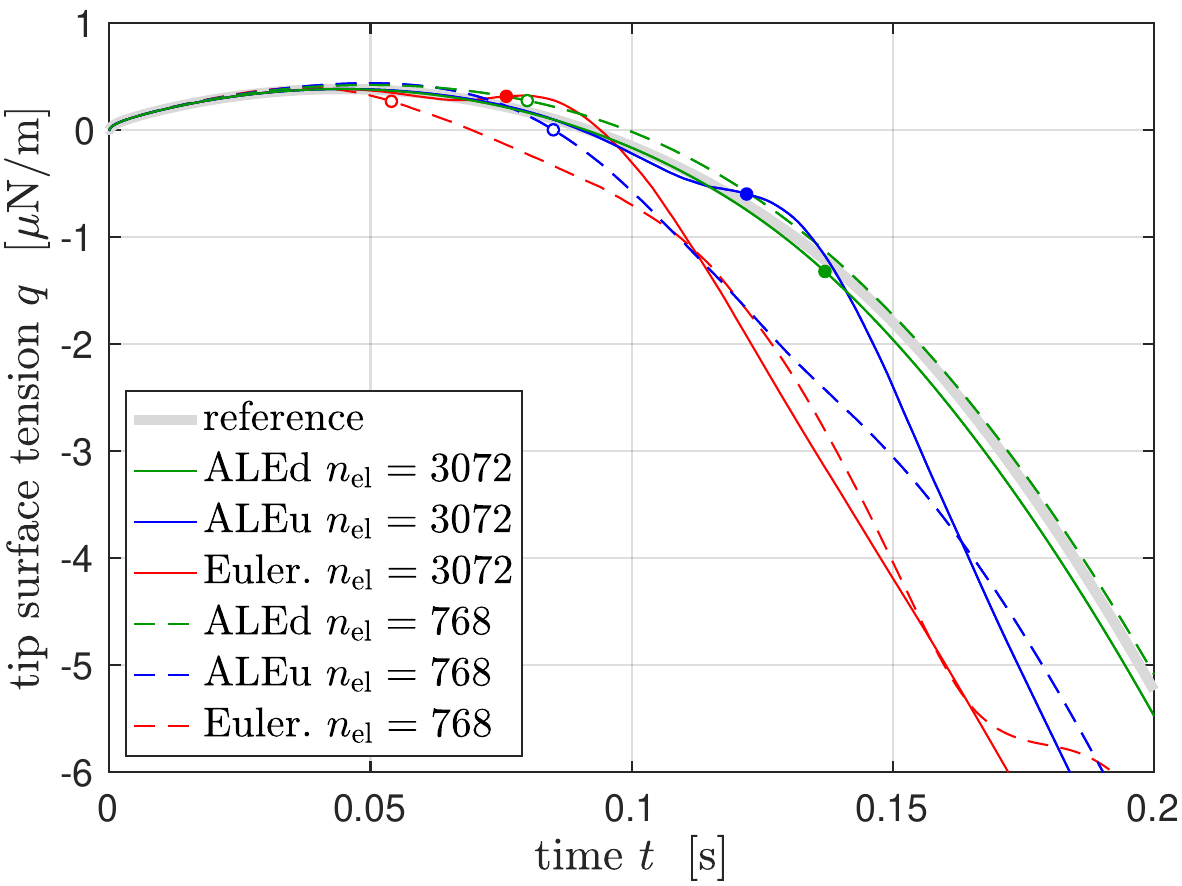}}
\put(-7.95,6.0){\footnotesize (a)}
\put(-4.25,6.0){\footnotesize (b)}
\put(-.55,6.0){\footnotesize (c)}
\put(3.15,6.0){\footnotesize (d)}
\put(7.4,8.6){\scriptsize $\norm{\bv}$}
\put(7.35,6.73){\scriptsize $\big[\!\frac{\mu\mrm}{\mrs}\!\big]$}
\put(-7.95,-.1){\footnotesize (e)}
\put(0.2,-.1){\footnotesize (f)}
\end{picture}
\caption{Vesicle budding: 
Eulerian (a), ALEu (b) and ALEd solution (c) for $n_\mathrm{el} = 3072$ in comparison to the accurate reference solution (d), all for $H_0 = -6/R$.
Evolution of the tip velocity $\norm{\bv}$ (e) and tip surface tension $q$ (f) during budding for the different formulations and FE meshes. 
The dots mark the location where the relative error exceeds 2\%.
As seen the two ALE formulations are much more accurate than the Eulerian one.}
\label{f:budflow2}
\end{center}
\end{figure}
% run pBudFlow.m & pBudFlow2.m
%-----------------------------------------------------------------
The Eulerian formulation becomes inaccurate at the bud due to large mesh distortion. 
These distortions are much smaller for the proposed ALE formulation, resulting in much higher accuracy.
Two cases are shown:
Uniform mesh elasticity (ALEu) and distributed mesh elasticity (ALEd -- where $\mu_\mrm$ is lowered to 10\% for points initially beyond 0.2R from the tip).
The latter allows to reduce mesh distortions at the bud even further and hence allows to obtain highly accurate results as Fig.~\ref{f:budflow2} shows.

The example demonstrates that the proposed ALE formulation also works well in the presence of bending elasticity and large local surface deformations, and it can be tailored to them.
It is interesting to note that here the chosen Eulerian prescription of $H_0$ ensures that the bud remains axisymmetric, which is not the case for a Lagrangian prescription as is seen in \citet{liquidshell}. 
A more detailed study and analysis of vesicle budding, also for non-axisymmetric flows, is planned for future work.

\section{Conclusion}\label{s:concl}

This work presents a general arbitrary Lagrangian-Eulerian surface formulation suitable for transient and steady Navier-Stokes flow on self-evolving manifolds.
The formulation is based on a curvilinear surface parameterization associated with the ALE frame of reference. 
This frame is used to define a basis for the objective description of vectors and tensors on the surface.
In-plane surface elasticity is proposed for obtaining stable mesh motion that does not affect the material flow. 
The presented ALE formulation applies to closed as well as open surfaces, that contain evolving inflow boundaries.
This generality distinguishes it from earlier surface ALE descriptions.
The theory and analytical solutions presented here serve as a basis for the development of advanced finite element formulations in \citet{ALEcomp}.
%future work.

The main theoretical, numerical and physical insights gained from the present work are:
\\[-8mm]
\begin{itemize}
\item [T1.] The material time derivative of the ALE coordinate, $\dot\zeta^\alpha$, that defines the relative in-plane velocity of the material w.r.t.~the mesh, is the essential object of the surface ALE formulation, in particular for understanding Eq.~\eqref{e:ALE} and why $\bv_{,\alpha}\neq\dot\ba_\alpha$. \\[-7mm]
\item [T2.] The out-of-plane mesh motion is constrained to follow the surface motion. This constraint can be eliminated in the weak form without introducing a so-called \textit{mesh pressure} in the framework of Lagrange multiplier or penalty methods. \\[-7mm]
\item [T3.] The weak form and its variation should be derived for fixed ALE coordinate $\zeta^\alpha$. This coordinate then directly and naturally follows from a local finite element parametrization.
\\[-7mm]
\item [N1.] Lagrangian mesh motion is known to be inappropriate for fluid flows.
Here it is shown that also in-plane Eulerian mesh motion can become very inaccurate, especially for large surface motion.
\\[-7mm] 
\item [N2.] A surface ALE formulation based on elastic mesh motion, on the other hand, is much more accurate, and it achieves optimal convergence rates.
The velocity convergence rate for free surface motion is one order lower than for fixed surfaces. \\[-7mm] 
\item [N3.] The mesh elasticity parameters can be tailored to the problem: Using larger mesh stiffness in regions where large mesh distortions are expected allows for further accuracy gains. \\[-7mm]
\item[N4.] A stable way to control bubble inflation is to prescribe the interior volume. Prescribing the inflow velocity or the interior pressure tends to be numerically unstable. \\[-7mm] 
\item [P1.] The coupling between surface geometry, surface flow, mesh velocity and surface tension within the governing field equations is highlighted. 
\\[-7mm] 
\item [P2.] Non-trivial, varying body forces are required to keep a flowing surface in shape. \\[-7mm] 
\item [P3.] Otherwise the shape changes: In the examples of Sec.~\ref{s:shear} and \ref{s:4tex} the surface bulges out, where the velocity is large, and in, where the velocity is small.
The outward bulges are shifted in flow direction, which agrees with intuition. \\[-7mm] 
\item [P4.] Vesicle budding remain axisymmetric if the spontaneous curvature of the Helfrich model is prescribed in an Eulerian instead of a Lagrangian manner.
\\[-7mm] 
\end{itemize}

The formulation here already contains the cases of area-compressibility and bending elasticity, but analytical solutions for these cases are still missing.
Further extensions are evolving surfaces with changing thickness and out-of-plane shear.
The analytical solutions presented here all have uniform surface curvature. 
Also interesting would be analytical solutions with varying curvature.

\bigskip

{\Large{\bf Declaration of Interests}} 

The author reports no conflict of interest.\\[-2mm]

\bigskip

{\Large{\bf Acknowledgements}}

The author is grateful to Kranthi K.~Mandadapu and Thang X.~Duong for discussions on the topic.

\appendix

\section{Basis transformation rules}\label{s:trans}

This section presents formulae for the transformation between bases $\ba_\alpha$ and $\ba_{\hat\alpha}$.
Those are required to adapt a Lagrangian description, as used in \citet{shelltheo,sahu17,FeFi}, to an ALE description.
Applying the chain rule to \eqref{e:ba} gives
\eqb{l}
\ba_\alpha = \ba_{\hat\gamma}\,\ds\pa{\xi^\gamma}{\zeta^\alpha} \quad $and$ \quad
\ba_{\hat\alpha} = \ba_\gamma\,\ds\pa{\zeta^\gamma}{\xi^\alpha}\,.
\label{e:aa1}\eqe
The dual basis must thus satisfy
\eqb{l}
\ba^\alpha = \ds\pa{\zeta^\alpha}{\xi^\gamma}\,\ba^{\hat\gamma} \quad $and$ \quad
\ba^{\hat\alpha} = \ds\pa{\xi^\alpha}{\zeta^\gamma}\,\ba^\gamma\,,
\label{e:aa2}\eqe
to ensure orthonormality between covariant and contravariant basis vectors.
Applying these formulae to the tensor
\eqb{l}
\bc = c_{\alpha\beta}\, \ba^{\alpha} \otimes \ba^{\beta} 
= c^{\alpha\beta} \ba_{\alpha} \otimes \ba_{\beta} 
= c_{\hat\alpha\hat\beta}\, \ba^{\hat\alpha} \otimes \ba^{\hat\beta}
= c^{\hat\alpha\hat\beta}\, \ba_{\hat\alpha} \otimes \ba_{\hat\beta}
\eqe
yields the transformation rules
\eqb{llllll}
c_{\alpha\beta} \is \ds\pa{\xi^\gamma}{\zeta^\alpha}\pa{\xi^\delta}{\zeta^\beta}\,c_{\hat\gamma\hat\delta}\,,~~ & 
c_{\hat\alpha\hat\beta} \is \ds\pa{\zeta^\gamma}{\xi^\alpha}\pa{\zeta^\delta}{\xi^\beta}\,c_{\gamma\delta}\,, \\[3.5mm]
c^{\alpha\beta} \is \ds\pa{\zeta^\alpha}{\xi^\gamma}\pa{\zeta^\beta}{\xi^\delta}\,c^{\hat\gamma\hat\delta}\,,~~ & 
c^{\hat\alpha\hat\beta} \is \ds\pa{\xi^\alpha}{\zeta^\gamma}\pa{\xi^\beta}{\zeta^\delta}\,c^{\gamma\delta}\,.
\label{e:cc}\eqe
From equations \eqref{e:aa1}--\eqref{e:cc} corresponding transformation rules for time derivatives can be derived via the product rule.
From (\ref{e:aa1}.2) follows for example 
\eqb{l}
\dot\ba_{\hat\alpha} = \big(\dot\ba_\gamma + \dot\zeta^\beta_{,\gamma}\,\ba_\beta\big)\,\ds\pa{\zeta^\gamma}{\xi^\alpha}\,,
\label{e:dah}\eqe
while (\ref{e:cc}.2) gives
\eqb{lll}
\dot c_{\hat\alpha\hat\beta}  = \ds\big( \dot c_{\gamma\delta} + \dot\zeta^\lambda_{,\gamma}\,c_{\lambda\delta} +  c_{\gamma\lambda}\,\dot\zeta^\lambda_{,\delta}  \big) \pa{\zeta^\gamma}{\xi^\alpha}\pa{\zeta^\delta}{\xi^\beta}\,.
\eqe
Here, we have used
\eqb{l}
\ds\pa{}{t}\bigg(\pa{\zeta^\gamma}{\xi^\alpha}\bigg)_{\!\!\xi^\beta} = \pa{\dot\zeta^\gamma}{\xi^\alpha}
= \pa{\dot\zeta^\gamma}{\zeta^\lambda}\pa{\zeta^\lambda}{\xi^\alpha}\,,
\eqe
since temporal and spatial differentiation can be exchanged in this case.

\section{Relation between $\bv_{\!,\alpha}$ and $\dot\ba_\alpha$}\label{s:va}

There are three ways to derive relation \eqref{e:badot}.
The simplest is to use 
\eqb{l}
\bv_{\!,\hat\alpha} = \ds\pa{\zeta^\gamma}{\xi^\alpha}\,\bv_{\!,\gamma}\,,
\eqe
together with $\bv_{\!,\hat\alpha}=\dot\ba_{\hat\alpha}$ and \eqref{e:dah}.

Another is to apply the parametric derivative 
to Eq.~\eqref{e:bv}.
This gives
\eqb{l}
\bv_{\!,\alpha} = \ds\pa{}{\zeta^\alpha}\bigg(\pa{\bx}{t}\bigg)_{\!\!\zeta^\beta} + \dot\zeta^\gamma\,\ba_{\gamma,\alpha} + \dot\zeta^\gamma_{,\alpha}\,\ba_\gamma\,. 
\label{e:bvap}\eqe
In the first term, the temporal and spatial differentiation can be exchanged. 
Likewise, $\ba_{\gamma,\alpha} = \ba_{\alpha,\gamma}$.
Further, applying the fundamental ALE equation \eqref{e:ALE} to $\ba_\alpha$ gives
\eqb{l}
\dot\ba_\alpha = \ds\pa{\ba_\alpha}{t}\Big|_{\zeta^\beta} + \ba_{\alpha,\gamma}\,\dot\zeta^\gamma\,.
\label{e:dotaa}\eqe 
Combining \eqref{e:bvap} and \eqref{e:dotaa} then immediately leads to \eqref{e:badot}.

Another alternative derivation is to follow Appendix A.2 of \citet{sahu17}. 
There is has been shown that
\eqb{l}
\dot\ba_\alpha = \ds\pa{}{t}\bigg(\pa{\hat\bx(\xi^\epsilon,t)}{\xi^\gamma}\bigg)_{\!\!\xi^\beta}\pa{\xi^\gamma}{\zeta^\alpha} + 
\pa{\hat\bx(\xi^\epsilon,t)}{\xi^\gamma} \pa{}{t}\bigg(\pa{\xi^\gamma}{\zeta^\alpha}\bigg)_{\!\!\xi^\beta}\,,
\label{e:dotaa2}\eqe
where the fixed-surface coordinate $\theta^\alpha$, originally appearing in \citet{sahu17}, has been replaced by the more general ALE coordinate $\zeta^\alpha$. 
As noted in \citet{sahu17}, temporal and spatial differentiation commute in the first term, such that the first term turns out to be equal to $\bv_{\!,\alpha}$. 
However, the time derivative in the second term is generally not zero.
Only for special choices of $\zeta^\alpha$, such as $\zeta^\alpha=\xi^\alpha$, does the second term vanish.
Eq.~\eqref{e:dotaa2} thus gives
\eqb{l}
\bv_{\!,\alpha} = \dot\ba_\alpha - \ds\pa{}{t}\bigg(\pa{\xi^\gamma}{\zeta^\alpha}\bigg)_{\!\!\xi^\beta}\,\ba_{\hat\gamma}\,.
\label{e:badot2}\eqe
Using the fundamental ALE equation \eqref{e:ALE} we find
\eqb{l}
\ds\pa{}{t}\bigg(\pa{\xi^\gamma}{\zeta^\alpha}\bigg)_{\!\!\xi^\beta} = \ds\pa{}{t}\bigg(\pa{\xi^\gamma}{\zeta^\alpha}\bigg)_{\!\!\zeta^\beta} 
+ \paqq{\xi^\gamma}{\zeta^\alpha}{\zeta^\beta}\dot\zeta^\beta\,,
\eqe
which can be rewritten into
\eqb{l}
\ds\pa{}{t}\bigg(\pa{\xi^\gamma}{\zeta^\alpha}\bigg)_{\!\!\xi^\beta} = \ds\pa{}{\zeta^\alpha}\bigg(\pa{\xi^\gamma}{t}\bigg|_{\zeta^\beta}
+ \pa{\xi^\gamma}{\zeta^\beta}\dot\zeta^\beta \bigg)  - \pa{\xi^\gamma}{\zeta^\beta}\pa{\dot\zeta^\beta}{\zeta^\alpha}\,.
\label{e:dxdzdot}\eqe
Using the fundamental ALE equation, the term in brackets simply becomes $\dot\xi^\gamma$, which is zero by definition.
Inserting \eqref{e:dxdzdot} into \eqref{e:badot2} and using (\ref{e:aa1}.1) then yields \eqref{e:badot}.

\begin{remark}
\textit{Second order counterpart to \eqref{e:badot}.}
Applying derivative $\partial .../\partial\zeta^\beta$ to \eqref{e:badot} and using
\eqb{l}
(\dot\ba_\alpha)_{,\beta} = \dot{(\ba_{\alpha,\beta})} + \ba_{\alpha,\gamma}\,\dot\zeta^\gamma_{,\beta}\,,
\eqe
which follows from applying \eqref{e:ALE} to $\ba_\alpha$ and $\ba_{\alpha,\beta}$,
gives
\eqb{l} 
\bv_{,\alpha\beta} = \dot{(\ba_{\alpha,\beta})} 
+ \dot\zeta^\gamma_{,\alpha}\,\ba_{\gamma,\beta} + \ba_{\alpha,\gamma}\,\dot\zeta^\gamma_{,\beta} + \dot\zeta^\gamma_{,\alpha\beta}\,\ba_\gamma\,.
\label{e:bvab}\eqe
\end{remark}

\section{Constitutive laws in the ALE frame}\label{s:2law}

In \citet{sahu17} and \citet{cism} it was shown that in the Lagrangian frame, the second law of thermodynamics leads to the constitutive relations
\eqb{l}
\sig_\mathrm{el}^{\hat\alpha\hat\gamma} = \ds\frac{2}{J}\pa{\Psi_0}{a_{\hat\alpha\hat\gamma}}\,,
\label{a:2lawcrit_el}\eqe
\eqb{l}
M^{\hat\alpha\hat\gamma} = \ds\frac{1}{J}\pa{\Psi_0}{b_{\hat\alpha\hat\gamma}}
\label{a:2lawcrit_M}\eqe
and
\eqb{l}
\sig_\mathrm{inel}^{\hat\alpha\hat\gamma}\,\dot a_{\hat\alpha\hat\gamma}\geq 0\,.
\label{a:2lawcrit_visc}\eqe 
Using $\dot a_{\hat\alpha\hat\gamma} = 2d_{\hat\alpha\hat\gamma}$, which follows from \eqref{e:dab} since $\dot\zeta^\gamma_{,\alpha}=0$ for the Lagrangian case $\zeta^\alpha=\xi^\alpha$, the last relation becomes
\eqb{l}
\sig_\mathrm{inel}^{\hat\alpha\hat\gamma}\,d_{\hat\alpha\hat\gamma}\geq 0 \,.
\label{a:2lawcrit_visc2}\eqe  
By applying the transformation rules of \eqref{e:cc} to \eqref{a:2lawcrit_el}, \eqref{a:2lawcrit_M} and \eqref{a:2lawcrit_visc2} one immediately arrives at \eqref{e:2lawcrit_el}--\eqref{e:2lawcrit_visc}.

\section{Mechanical weak form derivation}\label{s:wfd}

The derivation of the mechanical weak form is analogous to the derivation of the mechanical power balance.
In \citet{shelltheo} it was thus shown that in the Lagrangian frame the weak form for fluid films with bending resistance is given by
\eqb{l}
\hat G := \hat G_\mathrm{in} + \hat G_\mathrm{int} - \hat G_\mathrm{ext} = 0 \quad\forall\,\delta\hat\bx\in\sV\,,
\label{e:WFhatv}\eqe
with
\eqb{lll}
\hat G_\mathrm{in}
\dis \ds\int_{\sS} \delta\hat\bx\cdot\rho\,\dot\bv\,\dif\hat a\,, \\[4mm]
\hat G_\mathrm{int} 
\dis \ds\frac{1}{2}\int_{\sS}  \sig^{\hat\alpha\hat\gamma} \, \delta\hat a_{\hat\alpha\hat\gamma} \, \dif\hat a 
+ \int_{\sS} M^{\hat\alpha\hat\gamma} \, \delta\hat b_{\hat\alpha\hat\gamma} \, \dif\hat a\,, \\[4mm]
\hat G_\mathrm{ext}
\dis \ds\int_{\sS}\delta\hat\bx\cdot\bff\,\dif\hat a
+ \int_{\partial\sS}\delta\hat\bx\cdot\bT\,\dif\hat s
+ \int_{\partial\sS}\delta\hat\bn\cdot\bM\,\dif\hat s\,,
\label{e:Giiehat}\eqe
and
\eqb{l}
\delta\hat a_{\hat\alpha\hat\gamma}=\delta\hat\ba_{\hat\alpha}\cdot\ba_{\hat\gamma} + \ba_{\hat\alpha}\cdot\delta\hat\ba_{\hat\gamma}\,,\quad
\delta\hat b_{\hat\alpha\hat\gamma} = (\delta\hat\ba_{\hat\alpha,\hat\gamma} + \Gamma^{\hat\mu}_{\hat\alpha\hat\gamma}\,\delta\hat\ba_{\hat\mu})\cdot\bn\,.
\eqe
Here $\dif\hat a$ and $\dif\hat s$ denote differential area and line elements associated with the Lagrangian surface parametrization $\bx=\hat\bx(\xi^\alpha,t)$.
The variation $\delta\hat\bx$ is considered at fixed Lagrangian coordinate $\xi^\alpha$, such that variation $\delta(...)$ and differentiation $(...)_{,\hat\alpha}$ are exchangeable, i.e.~$\delta\hat\bx_{,\hat\alpha} = (\delta\hat\bx)_{,\hat\alpha} = \delta(\hat\bx_{,\hat\alpha}) = \delta\hat\ba_{\hat\alpha}$ and 
$\delta\hat\bx_{,\hat\alpha\hat\gamma} = (\delta\hat\bx)_{,\hat\alpha\hat\gamma} = \delta(\hat\bx_{,\hat\alpha\hat\gamma}) = \delta\hat\ba_{\hat\alpha,\hat\gamma}$.
Variation $\delta\hat\bx$ is thus analogous to time derivative $\dot\bx$.
Exploiting the symmetry of $ \sig^{\hat\alpha\hat\gamma}$ then leads to
\eqb{l}
\hat G_\mathrm{int} 
= \ds\int_{\sS} \sig^{\hat\alpha\hat\gamma}\,\delta\hat\bx_{,\hat\alpha}\cdot\ba_{\hat\gamma} \, \dif\hat a 
+ \int_{\sS} M^{\hat\alpha\hat\gamma}\,\delta\hat\bx_{;\hat\alpha\hat\gamma}\cdot\bn \, \dif\hat a\,,
\label{e:Ginthat}\eqe
with
\eqb{l}
\delta\hat\bx_{;\hat\alpha\hat\gamma} := \delta\hat\bx_{,\hat\alpha\hat\gamma} - \Gamma^{\hat\mu}_{\hat\alpha\hat\gamma}\,\delta\hat\bx_{,\hat\mu}\,.
\label{e:dxabhat}\eqe

The derivation in \citet{shelltheo} works in the same way when variation and integration are associated with the ALE frame.
One thus considers the differential elements $\dif a$ and $\dif s$ of the ALE surface parametrization $\bx=\bx(\zeta^\alpha,t)$ together with the  surface variation $\delta\bx$ at fixed ALE coordinate $\zeta^\alpha$, such that variation $\delta(...)$ and differentiation $(...)_{,\alpha}$ are exchangeable, i.e.~$\delta\bx_{,\alpha} = (\delta\bx)_{,\alpha} = \delta(\bx_{,\alpha}) = \delta\ba_{\alpha}$ and 
$\delta\bx_{,\alpha\gamma} = (\delta\bx)_{,\alpha\gamma} = \delta(\bx_{,\alpha\gamma}) = \delta\ba_{\alpha,\gamma}$.
Variation $\delta\bx$ is thus analogous to time derivative $\bx'$.
This then leads to the weak form \eqref{e:WFv}.

It is important to note that weak forms \eqref{e:WFv} and \eqref{e:WFhatv} are not equal, but both valid weak forms.
If desired they can be related, which is discussed briefly for the case that the ALE frame coincides with the Lagrangian frame initially, i.e.~$\dif A = \dif\hat A$, so that $\dif\hat a = J\,\dif a/J_\mrm$.
Variations $\delta\hat\bx$ and $\delta\bx$ are related in the same way as $\dot\bx$ and $\bx'$ in \eqref{e:bv}, i.e.
\eqb{l}
\delta\hat\bx = \delta\bx + \delta\zeta^\alpha\ba_\alpha\,,
\label{e:dxhat}\eqe
where $\delta\zeta^\alpha$ is a variation of $\zeta^\alpha$ at fixed $\xi^\beta$.
Inserting \eqref{e:dxhat} into
\eqb{l}
\delta\hat\bx_{,\hat\alpha} = (\delta\hat\bx)_{,\hat\alpha} = \ds\pa{\zeta^\gamma}{\xi^\alpha}(\delta\hat\bx)_{,\gamma}
\eqe
gives
\eqb{l}
\delta\hat\bx_{,\hat\alpha} = \ds\pa{\zeta^\gamma}{\xi^\alpha}\Big(\delta\bx_{,\gamma} + \delta\zeta^\beta_{,\gamma}\,\ba_\beta + \delta\zeta^\beta \ba_{\beta,\gamma}\Big)\,,
\eqe
with $\delta\zeta^\beta_{,\gamma} := (\delta\zeta^\beta)_{,\gamma}$.
Using (\ref{e:aa1}.2) and (\ref{e:cc}.3), the first part in $\hat G_\mathrm{int}$ thus becomes
\eqb{l}
\ds\int_{\sS} \sig^{\hat\alpha\hat\gamma}\,\delta\hat\bx_{,\hat\alpha}\cdot\ba_{\hat\gamma} \, \dif\hat a 
= \ds\int_{\sS} \sig^{\alpha\gamma}\,\Big(\delta\bx_{,\alpha}\cdot\ba_\gamma + \delta\zeta^\beta_{,\alpha}\,a_{\beta\gamma} 
+ \delta\zeta^\beta\,\Gamma^\delta_{\alpha\beta}\,a_{\delta\gamma}\Big) \frac{J}{J_\mrm}  \, \dif a\,.
\eqe
This, however, has no apparent advantage over the first term in (\ref{e:Giie}.2).
It is just for illustrating that it makes a difference whether the weak form is based on a variation of the Lagrangian surface description or the ALE surface description.
The latter is considered here, as it is the one required for a computational description in the ALE frame.

\section{Basis derivatives}\label{s:basis}

\subsection{Soap bubble example}

The Cartesian basis used in Sec.~\ref{s:bubble2} has the ALE derivatives
\eqb{llllll}
\be_{r,1} \is \psi_\mro\,\be_\psi\,, ~~& \be_{r,2} \is \mathbf{0}\,, \\[1mm]
\be_{\psi,1} \is -\psi_\mro\,\be_r\,, & \be_{\psi,2} \is \mathbf{0}\,,
\label{e3:bezeta}\eqe
and the time derivatives
\eqb{llllll}
\dot\be_r \is \dot\psi\,\be_\psi\,, ~~& \be_r' \is \psi'\,\be_\psi\,, \\[1mm]
\dot\be_\psi \is -\dot\psi\,\be_r\,, & \be_\psi' \is -\psi'\,\be_r\,.
\label{e3:bedot}\eqe

\subsection{Shear flow examples}

\begin{figure}[h]
\begin{center} \unitlength1cm
\begin{picture}(0,4.6)
\put(-2.5,-.1){\includegraphics[height=48mm]{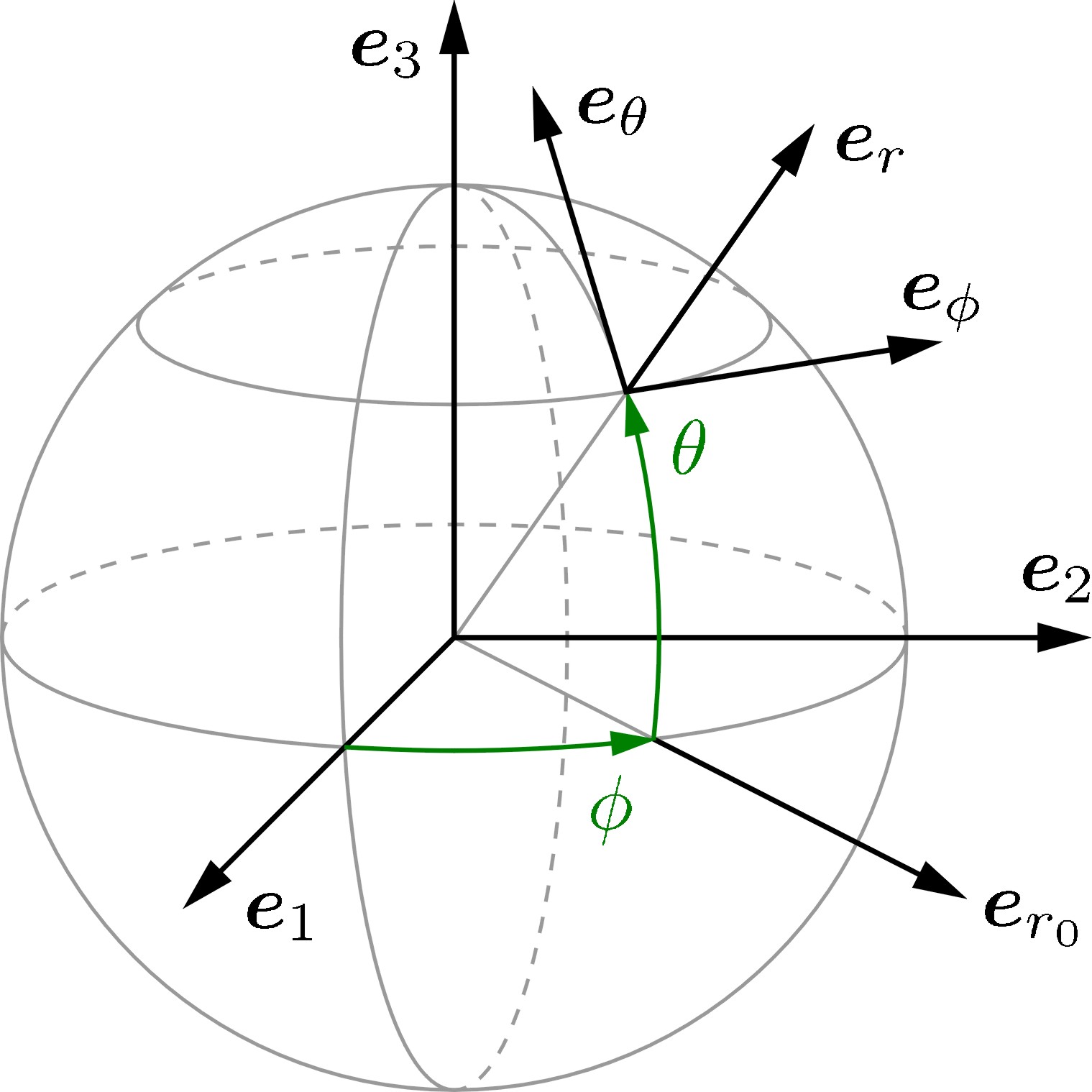}}
\end{picture}
\caption{Basis vectors used in the sphere examples of Secs.~\ref{s:shear}-\ref{s:4tex}.}
\label{f:basis3}
\end{center}
\end{figure}

The sphere examples in Secs.~\ref{s:shear}-\ref{s:4tex} use the basis vectors defined in Eqs.~\eqref{e4:bero} \& \eqref{e4:bezeta} and illustrated in Fig.~\ref{f:basis3}.
%-----------------------------------------------------------------
%-----------------------------------------------------------------
They satisfy
\eqb{lllllll}
\be_{r_0} \is \cos\theta\,\be_r \mi \sin\theta\,\be_\theta\,, \\[1mm]
\be_3 \is \cos\theta\,\be_\theta \plus \sin\theta\,\be_r\,,
\label{e4:be3}\eqe
and have the ALE derivatives
\eqb{llllll}
\be_{r,1} \is \cos\theta\,\be_\phi\,, ~~& \be_{r,2} \is \be_\theta\,, \\[1mm]
\be_{r_0,1} \is \be_\phi\,, & \be_{r_0,2} \is \mathbf{0}\,, \\[1mm]
\be_{\phi,1} \is -\be_{r_0}\,, & \be_{\phi,2} \is \mathbf{0}\,, \\[1mm]
\be_{\theta,1} \is -\sin\theta\,\be_\phi\,, & \be_{\theta,2} \is -\be_r\,,
\label{e:bezeta}\eqe
and the time derivatives
\eqb{llllll}
\dot\be_r \is \dot\zeta^1\cos\theta\,\be_\phi + \dot\zeta^2\,\be_\theta\,, ~~& \be_r' \is \mathbf{0}\,, \\[1mm]
\dot\be_{r_0} \is \dot\zeta^1\,\be_\phi\,, & \be_{r_0}' \is \mathbf{0}\,, \\[1mm]
\dot\be_\phi \is -\dot\zeta^1\,\be_{r_0}\, & \be_\phi' \is \mathbf{0}\,, \\[1mm]
\dot\be_\theta \is -\dot\zeta^1\sin\theta\,\be_\phi - \dot\zeta^2\,\be_r\,, & \be_\theta' \is \mathbf{0}\,.
\label{e:bedot}\eqe
The time derivatives simplify when $\dot\zeta^2=0$, as in Secs.~\ref{s:shear} and \ref{s:spin}.
The Lagrangian derivatives used in Sec.~\ref{s:shear} are
\eqb{llllll}
\be_{r,\hat1} \is \cos\theta\,\be_\phi\,, ~~& \be_{r,\hat2} \is \omega t\cos^2\theta\,\be_\phi + \be_\theta\,, \\[1mm]
\be_{r_0,\hat1} \is \be_\phi\,, & \be_{r_0,\hat2} \is \omega t\cos\theta\,\be_\phi\,,  \\[1mm]
\be_{\phi,\hat1} \is -\be_{r_0}\,, & \be_{\phi,\hat2} \is -\omega t\cos\theta\,\be_{r_0}\,, \\[1mm]
\be_{\theta,\hat1} \is -\sin\theta\,\be_\phi\,, & \be_{\theta,\hat2} \is  -\omega t\sin\theta\cos\theta\,\be_\phi -\be_r\,.
\label{e:bexi}\eqe

\section{Finite element formulation}\label{s:FE}

The three primary fields $\bv$, $q$ and $\bx$ are discretized by the finite element (FE) approximations
\eqb{lll}
\bv(\zeta^\alpha,t) \ais \ds\sum_{I=1}^{n_e} N_I(\zeta^\alpha)\,\bv_I(t)\,,\\[4.5mm]
q(\zeta^\alpha,t) \ais \ds\sum_{I=1}^{n_e} N_I(\zeta^\alpha)\,q_I(t)\,,\\[4.5mm]
\bx(\zeta^\alpha,t) \ais \ds\sum_{I=1}^{n_e} N_I(\zeta^\alpha)\,\bx_I(t)\,,
\label{e:FE}\eqe
where the latter leads to the FE mesh velocity
\eqb{l}
\bv_\mrm = \bx' \approx \ds\sum_{I=1}^{n_e} N_I\,\bx'_I = \ds\sum_{I=1}^{n_e} N_I\,\bv_{\mrm I}\,.
\label{e:FEm}\eqe
Here $N_I = N_I(\zeta^\alpha)$ are the FE shape functions and $\bv_I$, $q_I$, $\bx_I$ and $\bv_{\mrm I} = \bx'_I$ the corresponding nodal values.
Quadratic shape functions are used for all fields -- quadratic Lagrange for the examples in Sec.~\ref{s:bubble2}-\ref{s:4tex} and quadratic NURBS for the example in Sec.~\ref{s:bud}.
The polynomial pressure projection scheme of \citet{dohrmann04} is used to stabilize surface tension $q$. 
The local FE parameterization $\zeta^\alpha\in[-1,1]$ naturally provides an ALE frame of reference.  
Hence all the kinematic quantities and constitutive laws from Secs.~\ref{s:surf} and \ref{s:consti} can be evaluated using \eqref{e:FE}-\eqref{e:FEm}, and then  used in turn to discretize the weak forms of Sec.~\ref{s:wf}.
The three discretized weak forms from Secs.~\ref{s:wfv}, \ref{s:wfq} and \ref{s:wfm} 
then yield three coupled ordinary differential equations for the three fields $\bv_I(t)$, $q_I(t)$ and $\bx_I(t)$ or $\bv_{\mrm I}(t)$.
They are solved in time using the implicit trapezoidal rule and the Newton-Raphson method at every time step.
This requires the full linearization of the coupled three field problem.
%Details on this will be reported in a forthcoming publication.
Details on this are reported in \citet{ALEcomp}.

\bibliographystyle{apalike}
%\bibliography{../../Tex/bibliography}
\bibliography{bibliography}

\end{document}

%% file: neco.tex
\newcommand{\bitm}{\begin{itemize}}
\newcommand{\eitm}{\end{itemize}}
\newcommand{\bnumr}{\begin{enumerate}}
\newcommand{\enumr}{\end{enumerate}}

\newcommand {\eqb}[1]{\begin{equation}\begin{array}{#1}}
\newcommand {\eqe}{\end{array}\end{equation}}

\newcommand {\esb}[1]{\begin{equation*}\begin{array}{#1}}
\newcommand {\ese}{\end{array}\end{equation*}}
\newcommand {\ds}{\displaystyle}

\newcommand {\pa}[2]{\frac{\partial{#1}}{\partial{#2}}}

\newcommand {\paqq}[3]{\frac{\partial^2{#1}}{\partial{#2}\,\partial{#3}}}
\newcommand {\back}{\! \! \!}
\newcommand {\is}{\back &=& \back}
\newcommand {\dis}{\back &:=& \back}

\newcommand {\ais}{\back &\approx& \back}

\newcommand {\plus}{\back &+& \back}
\newcommand {\mi}{\back &-& \back}

\newcommand {\norm}[1]{\|#1\|}
\newcommand {\tr}{\mathrm{tr}\,}

\newcommand {\dif}{\mathrm{d}}

\newcommand {\II}{{I\kern-.3em I}}
\newcommand {\III}{{I\kern-.3em I\kern-.3em I}}

\newcommand {\mra}{\mathrm{a}}
\newcommand {\mrb}{\mathrm{b}}
\newcommand {\mrc}{\mathrm{c}}

\newcommand {\mre}{\mathrm{e}}

\newcommand {\mrg}{\mathrm{g}}

\newcommand {\mri}{\mathrm{i}}

\newcommand {\mrm}{\mathrm{m}}
\newcommand {\mrn}{\mathrm{n}}
\newcommand {\mro}{\mathrm{o}}

\newcommand {\mrs}{\mathrm{s}}

\newcommand{\mrT}{\mathrm{T}}

\newcommand {\ba}{\boldsymbol{a}}

\newcommand {\bc}{\boldsymbol{c}}
\newcommand {\bd}{\boldsymbol{d}}
\newcommand {\be}{\boldsymbol{e}}
\newcommand {\bff}{\boldsymbol{f}}

\newcommand {\bi}{\boldsymbol{i}}

\newcommand {\bn}{\boldsymbol{n}}

\newcommand {\bv}{\boldsymbol{v}}
\newcommand {\bw}{\boldsymbol{w}}
\newcommand {\bx}{\boldsymbol{x}}

\newcommand {\bA}{\boldsymbol{A}}

\newcommand {\bM}{\boldsymbol{M}}

\newcommand {\bT}{\boldsymbol{T}}

\newcommand {\bX}{\boldsymbol{X}}

\newcommand {\sig}{\sigma}

\newcommand {\bsig}{\mbox{\boldmath$\sigma$}}

\newcommand {\IR}{{\rm\kern.24em
   \vrule width.02em height1.53ex depth-.05ex
   \kern-.3em R}}
\newcommand {\ic}{{\rm\kern.20em
   \vrule width.02em height1.0ex depth-.05ex
   \kern-.22em c}}
\newcommand {\ia}{{\rm\kern.20em
   \vrule width.02em height1.05ex depth-.0ex
   \kern-.25em a}}
\newcommand {\IC}{{\rm\kern.24em
   \vrule width.02em height1.4ex depth-.05ex
   \kern-.26em C}}
\newcommand {\ID}{{\rm\kern.34em
   \vrule width.02em height1.5ex depth-.05ex
   \kern-.36em D}}
\newcommand {\IS}{{\rm\kern.24em
   \vrule width.02em height1.6ex depth.05ex
   \kern-.26em S}}
\newcommand {\IT}{{\rm\kern.50em
   \vrule width.02em height1.55ex depth-.05ex
   \kern-.52em T}}

\newcommand {\IE}{{\rm\kern.24em
   \vrule width.02em height1.55ex depth-.05ex
   \kern-.33em E}}
\newcommand {\IEa}{{\rm\kern.24em
   \vrule width.02em height1.55ex depth-.05ex
   \kern-.33em E}^{1}_{ijkl}}
\newcommand {\IEb}{{\rm\kern.24em
   \vrule width.02em height1.55ex depth-.05ex
   \kern-.33em E}^{2}_{ijkl}}

\newcommand {\sP}{\mathcal{P}}
\newcommand {\sQ}{\mathcal{Q}}

\newcommand {\sS}{\mathcal{S}}

\newcommand {\sV}{\mathcal{V}}
\newcommand {\sW}{\mathcal{W}}

\newcommand{\nablas}{\nabla_{\!\mrs}}
\newcommand{\divs}{\mathrm{div_s}\,}

\newcommand {\Ass}[2]{\kern 0.9ex \vrule width0.45em height0.2ex depth0ex \kern -2.1ex \bigwedge_{#1}^{#2}}
\newcommand {\ASS}[2]{\kern 1.45ex \vrule width0.5em height0.2ex depth0ex \kern -2.65ex \bigwedge_{#1}^{#2}}